\documentclass[a4paper,12pt]{article}

\usepackage{amssymb} 
\usepackage{amsmath}

      \def\RR{{\mathchoice {\mathrm{I
    \hspace{-0.2em} R}} {\mathrm{I \hspace{-0.2em} R}} {\mathrm{I
    \hspace{-0.14em} R}} {\mathrm{I \hspace{-0.14em} R}}}}

\newtheorem{theorem}{Theorem} \newtheorem{lemma}{Lemma}

\newenvironment{proof}[1][{}]{\noindent \textbf{Proof #1} \par}{\hfill
$\blacksquare$ \par} \newcounter{rqlfig}

\newcommand{\hc}{\mathrm{h. c.}}

\title{Rigorous Schrieffer-Wolff transformation for the Anderson impurity
model} 
\author{Gilles Poirot}
\date{}

\begin{document}
\maketitle

\begin{abstract}
With the help of computer algebra, I devise an exact unitary transformation
for the Anderson impurity model which allows to kill the hybridization term in
the slightly simplified case of zero chemical potential.  Then I compute
explicitly the outcome of this transformation. This is a rigorous version of
the well known Schrieffer-Wolff transformation. It should be possible to treat
the general case at the price of increased computation time.
\end{abstract}

\section{Introduction}
The Anderson impurity model describes a single magnetic impurity coupled to a
conduction band of electrons (for a recent review see \cite{Hew}).  The
reduced Hamiltonian with chemical potential $\mu$ is \emph{formally} given by
\begin{eqnarray}
H' &=& H -\mu N \\ &=& H'_{0} + (\varepsilon_{d}-\mu) n_{d} + U n_{d,
  \uparrow} n_{d, \downarrow} + g \sum_{\sigma \in \{\uparrow, \downarrow\}}
  \left[ c_{\sigma}^{*}(V) d_{\sigma} + \mathrm{h.c.} \right]
  \label{EQdefmodel} \\
H'_{0} &=& \sum_{\sigma \in \{\uparrow, \downarrow\}} \int \!\! dk \,
  [\varepsilon(k)-\mu] c_{\sigma}^{*}(k) c_{\sigma}(k) \\ c_{\sigma}^{*}(V)
  &=& \int \!\! dk \, V(k) c_{\sigma}^{*}(k) \; = \; \left[c_{\sigma}
  (V)\right]^{*} \\ n_{d, \sigma} &=& d_{\sigma}^{*} d_{\sigma} \\ n_{d} &=&
  n_{d, \uparrow} + n_{d, \downarrow} \\ \| V \|_{L^{2}}^{2} &=& 1
\end{eqnarray}
where in fact one should pass to \emph{particle-hole} representation and
define a new $H'$ (by adding a diverging constant) bounded from below in the
infinite volume limit. The $c$'s and the $d$'s are standard Fermionic
annihilation operators, we assume furthermore
\begin{equation}
\{c_{\sigma}(k), d_{\sigma'} \} = \{c_{\sigma}(k), d_{\sigma'}^{*} \} = 0
\end{equation}

(\emph{In the following, I will note $H$ instead of $H'$.})

Using a Bethe-ansatz, the model was exactly solved in dimension $d=1$ with
linear dispersion relation (without UV-cutoff) and a constant $V$
\cite{TW}. But in the general case, apart from numerical renormalization
\cite{KWW}, the common way to study this model consists in trying to eliminate
the hybridization term. This was first approximatively done by Schrieffer and
Wolff \cite{SW}. This Schrieffer-Wolff transformation leads to a
renormalization of the energy impurity and of the repulsive
interaction. Furthermore it generates an anti-ferromagnetic spin-spin
interaction between the impurity and conduction band electrons. Thus, it maps
the Anderson impurity model to an effective s-d type model exhibiting Kondo
effect.

But this transformation is equivalent to a second order perturbation treatment
and its validity is unclear. Furthermore, it becomes singular when the energy
of the impurity lies in the conduction band. In particular, it is not at all
suitable in the intermediate valence regime (of interest for certain rare
earth compounds) where the impurity energy is close to the Fermi level.

The Schrieffer-Wolff transformation was refined by Kehrein and Mielke by the
use of infinitesimal unitary transformations \cite{KM}. This leads to a
smoother result but it still relies on uncontrolled approximations.

In this paper, I want to study the possibility of an \emph{exact} elimination
of the hybridization term through unitary transformation with the help of
\emph{computer algebra}.  As a first step, I will consider the slightly
simplified case $\mu=0$, \emph{i.e.} there are only particles and no holes. In
this case, I prove the existence of such a generalized Schrieffer-Wolff
transformation and I compute the resulting Hamiltonian as a s-d type model (up
to irrelevant terms) with an effective spin-spin interaction which is
anti-ferromagnetic in one channel.

Then I explain how the computation would be changed in the general case and
how one can then partly use the $\mu=0$ computation. This suggests that it
should be possible to treat also the general case.


\section{Main result}
\begin{theorem} 
\label{MainThm}
Let $H$ be as in (\ref{EQdefmodel}) with $U>0$, $\mu=0$ and
$\frac{V}{\varepsilon^{1/2}} \not \in L^{2}(\RR^{d})$, then there exist
\begin{itemize}
\item $(t_{1}, t_{2}) \in \RR^{2}$;
\item $(f_{1}, f_{2}) \in L^{2}(\RR^{d})^{2}$ with $\| f_{1} \|_{L^{2}}^{2} =
\|f_{2}\|_{L^{2}}^{2} = 1$;
\item and some unitary operators
\begin{eqnarray}
U_{1} &=& \exp \left\{ t_{1} \sum_{\sigma} [c_{\sigma}^{*}(f_{1}) d_{\sigma}
  -\hc] \right\} \\ U_{2} &=& \exp \left\{ t_{2} \sum_{\sigma}
  [c_{\sigma}^{*}(f_{2}) d_{-\sigma}^{*} d_{-\sigma} d_{\sigma} -\hc] \right\}
\end{eqnarray} 
\end{itemize}
such that
\begin{equation}
H_{2} = U_{2} U_{1} H U_{1}^{*} U_{2}^{*}
\end{equation}
contains no term linear in $c_{\sigma}$ or $c_{\sigma}^{*}$.
\end{theorem}

The proof is through brute force. Using the above expressions as ansatz for
$U_{1}$ and $U_{2}$, I compute (with the help of computer algebra) the effect
of the two unitary transformations as a function of $(t_{1}, f_{1}; t_{2},
f_{2})$ then I impose that the terms linear in $c_{\sigma}$ or
$c_{\sigma}^{*}$ vanish.

After simplification, I obtain the following expected result
\begin{theorem}
  $H$ is unitarily equivalent to an s-d type Hamiltonian
\begin{eqnarray}
H_{2} &=& H_{0} + \varepsilon_{R} n_{d} + U_{R} n_{d, \uparrow} n_{d,
  \downarrow} + K_{1} \left[c_{\uparrow}^{*}(f_{1}) c_{\downarrow}^{*}(f_{1})
  d_{\downarrow} d_{\uparrow} + \hc \right] \nonumber \\ && + \sum_{{\sigma
  \in \{\uparrow, \downarrow\}} \atop {i \in \{+, -\}}} \lambda_{i}
  c_{\sigma}^{*}(\varphi_{i}) c_{\sigma}(\varphi_{i}) - \sum_{{\sigma \in
  \{\uparrow, \downarrow\}} \atop {x \in \{AF, F\}}} \lambda_{x}
  c_{\sigma}^{*}(\varphi_{x}) c_{\sigma}(\varphi_{x}) n_{d} \nonumber \\ && +
  4 \sum_{x \in \{AF, F\}} \lambda_{x} \vec{S}_{c}(\varphi_{x} \otimes
  \varphi_{x}) \textbf{.} \vec{S}_{d} + H_{\mathit{Irr}} \\
  S_{c}^{(i)}(\varphi \otimes \psi) &=& \sum_{(\tau, \tau') \in \{\uparrow,
  \downarrow\}^{2}} c_{\tau}^{*}(\varphi)
  \left(\frac{\sigma_{i}}{2}\right)_{\tau \tau'} c_{\tau'}(\psi), \quad i \in
  \{x, y, z\} \\ S_{d}^{(i)} &=& \sum_{(\tau, \tau') \in \{\uparrow,
  \downarrow\}^{2}} d_{\tau}^{*} \left(\frac{\sigma_{i}}{2}\right)_{\tau
  \tau'} d_{\tau'}
\end{eqnarray}
where $H_{\mathit{Irr}}$ is an \emph{irrelevant} part (in the Renormalization
Group sense) which consists in a (large) number of terms of order at least 3
in $c$ or $c^{*}$, and the $\sigma_{i}$ are Pauli matrices (thus the
$\vec{S}$'s are spin operators).

Furthermore, for values of the parameters $(\varepsilon_{d}, U,g)$ in the
so-called \emph{local moment} or \emph{intermediate valence} regimes, one has
\begin{equation}
\lambda_{AF} > 0 \quad \textrm{and} \quad \lambda_{F} < 0
\end{equation}
\emph{i.e.} the resulting spin-spin interaction is anti-ferromagnetic in one
channel.
\end{theorem}

\subsection*{Remarks}
\begin{itemize}
\item Let us note that $U_{1}$ and $U_{2}$ preserve the original
\emph{electron number} and \emph{total spin} conservation as well as the $U(1)
\times SU(2)$ symmetry. Therefore, the only possible terms linear in
$c_{\sigma}$ or $c_{\sigma}^{*}$ are
\begin{displaymath}
\sum_{\sigma} [c_{\sigma}^{*}(.) d_{\sigma} + \hc] \quad \mathrm{and} \quad
  \sum_{\sigma} [c_{\sigma}^{*}(.) d_{-\sigma}^{*} d_{-\sigma} d_{\sigma} +
  \hc]
\end{displaymath}
this means that I perform a kind of minimal transformation (2 adjustable
functions to meet 2 constraints).
\item The Schrieffer Wolff transformation amounts to the particular choice
\begin{eqnarray}
f_{1} &\propto& \frac{V}{\varepsilon-\varepsilon_{d}} \\ f_{2} &\propto&
\frac{V}{(\varepsilon-\varepsilon_{d}) \left[\varepsilon -(U+\varepsilon_{d})
\right]}
\end{eqnarray}
We will see that this is not so bad provided one replaces $\varepsilon_{d}$
and $U$ by their \emph{renormalized value}. Indeed I find
\begin{eqnarray}
f_{1} &\propto& \frac{V}{\varepsilon-\varepsilon_{R}} \\ f_{2} &\propto& a_{2}
\frac{V}{\varepsilon-\varepsilon_{R}} + b_{2}
\frac{V}{(\varepsilon-\varepsilon_{R}) (\varepsilon+\mu_{2})} \quad
\textrm{where } \mu_{2} \sim -(U_{R}+\varepsilon_{R})
\end{eqnarray}
\item Up to irrelevant terms, one can see that in $H_{2}$ the channels
  $n_{d}=1$ and $n_{d} \in \{0, 2\}$ are decoupled.  Yet
  $U_{R}+\varepsilon_{R} \sim -\mu_{2}$ is negative but small (in the local
  moment regime, I find that $\mu_{2} \sim e^{- C
  |\varepsilon_{d}|/g^{2}}$). Thus it is a priori not clear in which channel
  will be the ground state of $H_{2}$.
\item Finally an important remark is that strictly speaking eliminating the
  hybridization term is not enough because there remains a dangerous term in
  $H_{2}$, namely
\begin{equation}
\mathcal{W} = K_{1} \left[c_{\uparrow}^{*}(f_{1}) c_{\downarrow}^{*}(f_{1})
  d_{\downarrow} d_{\uparrow} + \hc \right]
\end{equation}
The reason is that when one performs some Renormalization Group analysis on
$H_{2}$, $\mathcal{W}$ combined with some cubic term in $H_{\mathit{Irr}}$ can
generate back some terms which are linear in $c_{\sigma}$ or
$c_{\sigma}^{*}$. Thus one should perform a third transformation
\begin{equation}
U_{3} = \exp \left\{ t_{3} [c_{\uparrow}^{*}(f_{3}) c_{\downarrow}^{*}(f_{3})
  d_{\downarrow} d_{\uparrow} -\hc] \right\}
\end{equation}
to kill also this term. This should a priori be possible, one then would have
to meet three constraints with three adjustable functions.It is easy to see
that there are no other \emph{dangerous} term which could generate some
part linear in $c_{\sigma}$ or $c_{\sigma}^{*}$. 

This problem was totally overlooked up to know because people usually do not
worry about irrelevant terms. This is not too dramatic in the \emph{local
moment} regime since in that case, one has
\begin{equation}
K_{1} \sim e^{- C |\varepsilon_{d}|/g^{2}}
\end{equation}
which means that the terms linear in $c_{\sigma}$ or $c_{\sigma}^{*}$ that are
generated will in fact remain small at least up to the Kondo Temperature
scale. 
\end{itemize}


\section{Proof of theorem \protect \ref{MainThm}}
\subsection{Principle of the computation}
The unitary transformations that I consider in this paper are of the form
  $U(t) = e^{t \Gamma}$, and the general problem is, given an observable $O$,
  to compute
\begin{equation}
O(t) = e^{t \Gamma} O e^{-t \Gamma}
\end{equation}
$O(t)$ is solution of the following differential equation
\begin{eqnarray}
\frac{d}{dt} O(t) &=& e^{t \Gamma} [\Gamma, O] e^{-t \Gamma} = [\Gamma, O](t)
  \\ O(t=0) &=& O
\end{eqnarray}

Thus the problem amounts to find a finite (possibly large) set $\{O_{1}=O,
O_{2}, \ldots, O_{n} \}$ whose span is stable under commutation with $\Gamma$.
\begin{equation}
[\Gamma, O_{i}] = \sum_{j} M_{ij} O_{j}
\end{equation}
Then one has just to integrate the first order linear equation
\begin{equation}
\frac{d {X}_{t}}{dt} = M X_{t} \quad \mathrm{with} \quad X_{t=0} = (O_{1},
  \ldots, O_{n})
\end{equation}

The difficulty comes from the large value of $n$ (for the second
transformation, $n \approx 150$) which makes the problem in practical
impossible to solve by hand, thus the need for computer assistance.

The computation was done on a Unix workstation running \emph{Mathematica},
using a personal implementation of Fermionic operator algebras. The
computation time is hard to assess because computation went in parallel with
code writing and furthermore the whole process is not yet automated and
requires an heavy human participation.

I can nevertheless estimate a kind of effective computation time
(human+computer) which is of order of a few months.  This proves that such
computations are tractable.


\subsection{First Unitary transformation}
Let $H_{1} = U_{1} H U_{1}^{*}$, $\alpha_{1} = \cos(t_{1})$ and $\beta_{1} =
\sin(t_{1})$, this transformation is quite easy to compute since one can check
that
\begin{eqnarray}
c_{\sigma}(f_{1}) &\stackrel{U_{1}}{\longmapsto}& \alpha_{1} c_{\sigma}(f_{1})
  - \beta_{1} d_{\sigma} \\ d_{\sigma} &\stackrel{U_{1}}{\longmapsto}&
  \beta_{1} c_{\sigma}(f_{1}) + \alpha_{1} d_{\sigma} \\ c_{\sigma}(k)
  &\stackrel{U_{1}}{\longmapsto}& c_{\sigma}(k) + f_{1}(k) [(\alpha_{1}-1)
  c_{\sigma}(f_{1}) - \beta_{1} d_{\sigma}]
\end{eqnarray}

An important simplification coming from the case $\mu=0$ is the fact that the
second transformation does not affect the ``$c_{\sigma}^{*}(.) d_{\sigma}$''
part. Therefore I can set it to zero at this stage. The constraint reads
\begin{eqnarray}
(\varepsilon + \lambda_{1}) f_{1} &=& g \frac{\alpha_{1}}{\beta_{1}} V
  \label{EQfun} \\
\lambda_{1} &=& g \frac{\alpha_{1}}{\beta_{1}} (1-\alpha_{1}) \left<f_{1},
  V\right> - (1-\alpha_{1}) \left<f_{1}, \varepsilon f_{1}\right> + g
  \beta_{1} \left<V, f_{1}\right> - \alpha_{1} \varepsilon_{d}
\end{eqnarray}
This leads to the following equation for $\lambda_{1}$
\begin{equation}
\lambda_{1} = g^{2} \left<V, \frac{1}{\varepsilon+\lambda_{1}} V\right>
  -\varepsilon_{d}
\label{EQlambdaun}
\end{equation}
from which one can see that $\lambda_{1}$ is real, just like $\left<V,
  f_{1}\right>$.

Let us consider
\begin{equation}
\varepsilon_{d}(\lambda) = g^{2} \left<V, \frac{1}{\varepsilon+\lambda} V
  \right> -\lambda, \quad \lambda \in (0, +\infty)
\end{equation}
it is easy to see that this is a strictly decreasing function ranging from
$+\infty$ (because $\frac{V}{\varepsilon^{1/2}} \not \in L^{2}$) to
$-\infty$. Therefore it admits an inverse function
$\lambda_{1}(\varepsilon_{d})>0$ giving the unique solution to
(\ref{EQlambdaun}).

Since $\lambda_{1}>0$, $f_{1} \in L^{2}(\RR^{d})$ and the normalization
condition fixes $\alpha_{1}^{2}$ and $\beta_{1}^{2}$.
\begin{eqnarray}
\alpha_{1}^{2} &=& \frac{1}{1+g^{2} M_{1}^{2}} \; \leqslant \; 1
  \label{EQalphaun} \\
M_{1}^{2} &=& \left<V, (\varepsilon+\lambda_{1})^{-2} V\right>
  \label{EQMun}
\end{eqnarray}

Finally, choosing $\alpha_{1}$ and $\beta_{1}$ to be positive, one obtains.
\begin{eqnarray}
H_{1} &=& H_{0} + C_{1} d_{\uparrow}^{*} d_{\downarrow}^{*} d_{\downarrow}
  d_{\uparrow} + C_{3} [c_{\uparrow}^{*}(f_{1}) c_{\downarrow}^{*}(f_{1})
  d_{\downarrow} d_{\uparrow} + \hc] \nonumber \\ && \quad + C_{5}
  c_{\uparrow}^{*}(f_{1}) c_{\downarrow}^{*}(f_{1}) c_{\downarrow}(f_{1})
  c_{\uparrow}(f_{1}) + \sum_{\sigma \in \{\uparrow, \downarrow\}}
  W_{\sigma}^{(1)}\\ W_{\sigma}^{(1)} &=& C_{2} [c_{\sigma}^{*}(f_{1})
  d_{-\sigma}^{*} d_{-\sigma} d_{\sigma} + \hc] - \frac{C_{3}}{2}
  [c_{\sigma}^{*}(f_{1}) c_{-\sigma}(f_{1}) d_{-\sigma}^{*} d_{\sigma} + \hc]
  \nonumber \\ && \quad + C_{3} c_{\sigma}^{*}(f_{1}) c_{\sigma}(f_{1})
  d_{-\sigma}^{*} d_{-\sigma} + C_{4} [c_{\sigma}^{*}(f_{1})
  c_{-\sigma}^{*}(f_{1}) c_{-\sigma}(f_{1}) d_{\sigma} + \hc] \nonumber \\ &&
  \quad + C_{6} [c_{\sigma}^{*}(f_{1}) c_{\sigma}(\varepsilon f_{1}) + \hc] +
  C_{7} d_{\sigma}^{*} d_{\sigma} + C_{8} c_{\sigma}^{*}(f_{1})
  c_{\sigma}(f_{1})
\end{eqnarray}
where the $C$'s are given by
\begin{eqnarray}
x_{1} &=& g M_{1} \\ C_{1} &=& \frac{U}{(1 + x_{1}^{2})^{2}} \\ C_{2} &=&
x_{1} C_{1} \\ C_{3} &=& x_{1}^{2} C_{1} \\ C_{4} &=& x_{1}^{3} C_{1} \\ C_{5}
&=& x_{1}^{4} C_{1} \\ C_{6} &=& -1 + \sqrt{1 + x_{1}^{2}} \\ C_{7} &=& -
\lambda_{1} \\ C_{8} &=& \frac{1}{x_{1}^{2}} \left\{(2+x_{1}^{2})
\varepsilon_{d} + (2-x_{1}^{2}) \lambda_{1} -2 \sqrt{1+x_{1}^{2}} \left[
\varepsilon_{d}+ \lambda_{1} (1-x_{1}^{2}) \right] \right\}
\end{eqnarray}


\subsection{Second Unitary Transform}
By analogy with the first transformation, I will look for a function $f_{2}$
satisfying
\begin{equation}
\forall \, k\in \RR^{d}, \quad \overline{f_{1}(k)} \, f_{2}(k) \in \RR
\end{equation}

The interested reader will find in appendix the differential equations
governing this transformation, from this one can reconstruct the integrated
flow.

I note
\begin{eqnarray}
\alpha_{2} &=& \cos(\sqrt{2} \, t_{2}) \\ \beta_{2} &=& \sin(\sqrt{2} \,
t_{2}) \\ \Gamma_{i} &=& \left<f_{i}, \varepsilon f_{i} \right> \quad i=1,2 \\
\omega_{12} &=& \left<f_{1}, f_{2}\right> \quad (\in \RR) \\ \Gamma_{12} &=&
\left<f_{1}, \varepsilon f_{2}\right> \quad (\in \RR)
\end{eqnarray}

Having the $c_{\sigma}^{*}(.) d_{-\sigma}^{*} d_{-\sigma} d_{\sigma}$ term
vanishing leads to the following constraint
\begin{eqnarray}
(\varepsilon + \mu_{2}) f_{2} &=& -\omega_{12} C_{6} (\varepsilon+\mu_{1})
  f_{1} \label{EQfdeux} \\ -\omega_{12} C_{6} \mu_{1} &=& \sqrt{2} \,
  \frac{\alpha_{2}}{\beta_{2}} C_{2}-2 \omega_{12} C_{3} -\omega_{12} C_{8} -
  \Gamma_{12} C_{6}
  \label{EQmuun} \\
\mu_{2} &=& -\alpha_{2} (C_{1}+C_{7}) -\Gamma_{2} (1-\alpha_{2})+ \sqrt{2} \,
  \frac{\alpha_{2}}{\beta_{2}} \omega_{12} C_{2} (1-\alpha_{2}) + \sqrt{2} \,
  \beta_{2} \omega_{12} C_{2} \nonumber
  \label{EQmudeux} \\ 
&& \quad - \omega_{12}^{2} (2 C_{3}+C_{8})(1-\alpha_{2}) - 2 \Gamma_{12}
  \omega_{12} C_{6} (1-\alpha_{2})
\end{eqnarray}

Using expression (\ref{EQfdeux}) one gets
\begin{eqnarray}
\omega_{12} &=& - \omega_{12} C_{6} \left[1 - (\mu_{2}-\mu_{1}) \left<f_{1},
    \frac{1}{\varepsilon+\mu_{2}} f_{1}\right>\right]
  \label{EQcontraintedeux} \\
\Gamma_{12} &=& -\omega_{12} [C_{6}(\Gamma_{1}+\mu_{1})+\mu_{2}]
  \label{EQGammadouze}\\
\Gamma_{2} &=& \omega_{12}^{2} C_{6} [C_{6} (\Gamma_{1}+\mu_{1})+\mu_{2}
  -\mu_{1}] - \mu_{2} \label{EQGammadeux} \\
\end{eqnarray}
And from (\ref{EQfun}), (\ref{EQlambdaun}) and (\ref{EQalphaun})
\begin{equation}
\Gamma_{1} = \frac{\lambda_{1} + \varepsilon_{d}}{g^{2} M_{1}^{2}} -
  \lambda_{1} \label{EQGammaun}
\end{equation}

Substituting (\ref{EQGammadouze}) into (\ref{EQmuun}) yields
\begin{eqnarray}
\alpha_{2}^{2} &=& \frac{Z^{2}}{Z^{2}+2 C_{2}^{2}} \; < \; 1
  \label{EQdefalphadeux} \\
Z &=& \omega_{12} \left[-2 C_{3}-C_{8}+C_{6} \mu_{2} +C_{6}(C_{6}+1) \mu_{1} +
\Gamma_{1} C_{6}^{2}\right]
\end{eqnarray}
Then I choose ($\theta = \pm 1$)
\begin{eqnarray}
\alpha_{2} &=& \frac{\theta Z}{\sqrt{Z^{2}+2 C_{2}^{2}}} \\ \beta_{2} &=&
\frac{- \theta \sqrt{2} C_{2}}{\sqrt{Z^{2}+2 C_{2}^{2}}}
\end{eqnarray}
With this particular choice, $\omega_{12}$ disappears from equation
(\ref{EQmudeux}) so that I can express $\mu_{1}$ as a function of $\mu_{2}$.
\begin{equation}
\mu_{1} = - \frac{2 C_{2}^{2}+ (C_{1}+C_{7}+\mu_{2}) \left[-2
    C_{3}-C_{8}+C_{6} (\mu_{2}+C_{6} \Gamma_{1})\right]}{C_{6} (C_{6}+1)
    (C_{1}+C_{7}+\mu_{2})} \label{EQdefmuun}
\end{equation}
And from the normalization condition, one gets $\omega_{12}$ as a function of
$\mu_{1}$ and $\mu_{2}$.
\begin{equation}
1 = \left<f_{2}, f_{2}\right> = \omega_{12}^{2} C_{6}^{2} \left[ -
    \frac{C_{6}+2}{C_{6}} + (\mu_{2}-\mu_{1})^{2} \left<f_{1},
    \frac{1}{(\varepsilon+\mu_{2})^{2}} f_{1}\right>\right]
\label{EQdefomegadouze}
\end{equation}

Finally, substituting (\ref{EQdefmuun}) into (\ref{EQcontraintedeux}), we are
left with the following equation
\begin{eqnarray}
(1+C_{6})^{2} (C_{1}+C_{7}+\mu_{2}) &=& \left<f_{1},
    \frac{1}{\varepsilon+\mu_{2}} f_{1}\right> \Big\{ 2 C_{2}^{2} \nonumber \\
    && \hspace{-2cm} +(C_{1}+C_{7}+\mu_{2}) \big[-2 C_{3}-C_{8}+C_{6}
    (C_{6}+2) \mu_{2} + C_{6}^{2} \Gamma_{1} \big] \Big\}
\end{eqnarray}
which after simplification becomes
\begin{eqnarray}
\label{EQphimudeux}
\varphi(\mu_{2}) &=& 0 \\ \varphi(\mu) &=& U + (1+x_{1}^{2})^{2} \,
(\mu-\lambda_{1}) \nonumber \\ && \quad + \frac{x_{1}^{2}}{1+x_{1}^{2}}
(\mu-\lambda_{1}) \big[U - (1+x_{1}^{2})^{2} \, (\mu-\lambda_{1}) \big]
\left<f_{1}, \frac{1}{\varepsilon+\mu} f_{1}\right>
\end{eqnarray}

\begin{lemma} 
\

The equation $\varphi(\mu)=0$ has a unique solution $\mu_{2}$, furthermore
    $\mu_{2} \in (0, \lambda_{1})$.
\end{lemma}
\begin{proof}
\begin{itemize}
\item First, let us note that $\varphi$ is a smooth function with
\begin{eqnarray}
\lim_{\mu \rightarrow 0} \, \varphi(\mu) &=& -\infty \quad \textrm{(since
  $\displaystyle \frac{V}{\varepsilon^{1/2}} \not \in L^{2}$)}\\
  \varphi(\lambda_{1}) &=& U \; > \; 0
\end{eqnarray}
This implies that there is at least one solution $\mu_{2} \in (0,
\lambda_{1})$.
\item For $\mu > \lambda_{1}$ we have $\varphi(\mu) \geqslant U > 0$ at least
  as long as $U - (1+x_{1}^{2})^{2} \, (\mu-\lambda_{1}) \geqslant 0$,
  \emph{i.e.} as long as $\mu \leqslant \lambda_{1} + U (1+x_{1}^{2})^{-2}$.
\item For $\mu > \lambda_{1} + U (1+x_{1}^{2})^{-2}$, we note that
\begin{eqnarray}
\varphi(\mu) &=& U + (1+x_{1}^{2})^{2} \, (\mu-\lambda_{1}) \nonumber \\ &&
\quad - \frac{x_{1}^{2}}{1+x_{1}^{2}} (\mu-\lambda_{1}) \big[(1+x_{1}^{2})^{2}
\, (\mu-\lambda_{1}) - U \big] \left<f_{1}, \frac{1}{\varepsilon+\mu}
f_{1}\right> \\ &\geqslant& U + (1+x_{1}^{2})^{2} \, (\mu-\lambda_{1})
\nonumber \\ && \quad - \frac{x_{1}^{2}}{1+x_{1}^{2}} (\mu-\lambda_{1})
\big[(1+x_{1}^{2})^{2} \, (\mu-\lambda_{1}) - U \big] \, \frac{1}{\mu} \\
&\geqslant& U + (1+x_{1}^{2})^{2} \, (\mu-\lambda_{1}) -x_{1}^{2}
(1+x_{1}^{2}) (\mu-\lambda_{1}) \\ &\geqslant& U + (1+x_{1}^{2}) \,
(\mu-\lambda_{1}) \; > \; 0
\end{eqnarray}
Thus there are no solution in $[\lambda_{1}, +\infty)$.
\item Finally, for $0< \mu < \lambda_{1}$, we have
\begin{eqnarray}
\varphi'(\mu) &=& (1+x_{1})^{2} + \frac{x_{1}^{2}}{1+ x_{1}^{2}} \big[U + 2
  (1+x_{1}^{2})^{2} (\lambda_{1}-\mu) \big] \left<f_{1},
  \frac{1}{\varepsilon+\mu} f_{1}\right> \nonumber \\ && + \frac{x_{1}^{2}}{1+
  x_{1}^{2}} (\lambda_{1}-\mu) \big[U + (1+x_{1}^{2})^{2} (\lambda_{1}-\mu)
  \big] \left<f_{1}, \frac{1}{(\varepsilon+\mu)^{2}} f_{1}\right> \\ &>& 0
\end{eqnarray}
therefore there is a unique $\mu_{2} \in (0, \lambda_{1})$ such that
$\varphi(\mu_{2}) = 0$.
\end{itemize}
\end{proof}

Since $\mu_{2}>0$, $f_{2} \in L^{2}(\RR^{d})$ and I can successively solve
(\ref{EQdefmuun}), (\ref{EQdefomegadouze}) and (\ref{EQdefalphadeux}).
\begin{eqnarray}
\mu_{1} &=& \frac{x_{1}^{2}}{(1+x_{1}^{2})-\sqrt{1+x_{1}^{2}}} \lambda_{1} -
  \frac{\mu_{2}}{\sqrt{1+x_{1}^{2}}} \nonumber \\ && \quad - \frac{2 U
  x_{1}^{2}}{\left[U -(1+x_{1}^{2})^{2} (\lambda_{1}-\mu_{2}) \right]
  \left[(1+x_{1}^{2})-\sqrt{1+x_{1}^{2}} \right]} (\lambda_{1}-\mu_{2})
  \label{EQsolvemuun} \\
\omega_{12}^{2} &=& \left[(\mu_{2}-\mu_{1})^{2} (\sqrt{1+x_{1}^{2}}-1)^{2}
  \left<f_{1},(\varepsilon+\mu_{2})^{-2} f_{1} \right>-x_{1}^{2}\right]^{-1}
  \label{EQsolveomegadouze}\\
\alpha_{2}^{2} &=& \frac{N_{\alpha_{2}}^{2}}{D_{\alpha_{2}}^{2}} \\
N_{\alpha_{2}} &=& (\mu_{2}-\mu_{1}) (\sqrt{1+x_{1}^{2}}-1)-x_{1}^{2} \left[
\frac{2 U}{(1+x_{1}^{2})^{2}} +\lambda_{1}-\mu_{1} \right] \\
D_{\alpha_{2}}^{2} &=& N_{\alpha_{2}}^{2} + \frac{2 U^{2}
x_{1}^{2}}{(1+x_{1}^{2})^{4}} \omega_{12}^{-2}
\end{eqnarray}
Furthermore, $f_{2}$ is given by
\begin{equation}
f_{2} = -\omega_{12} (\sqrt{1+x_{1}^{2}}-1)
  \left(\frac{\varepsilon+\mu_{1}}{\varepsilon+\mu_{2}}\right) f_{1}
\end{equation}
This concludes the proof of theorem \ref{MainThm}.


\section{s-d type Hamiltonian}
After simplification, I find that the initial Anderson impurity Hamiltonian
$H$ is unitarily equivalent to $H_{2}$ whose interacting part, up to
irrelevant terms, consists in a potential scattering term and a s-d exchange
term.
\begin{eqnarray}
H_{2} &=& H_{0} +\varepsilon_{R} n_{d} + U_{R} n_{d, \uparrow} n_{d,
  \downarrow} + K_{1} \left[c_{\uparrow}^{*}(f_{1}) c_{\downarrow}^{*}(f_{1})
  d_{\downarrow} d_{\uparrow} + \hc \right] \nonumber \\ && + \sum_{\sigma}
  W_{\sigma} - 4 W_{\mathrm{sd}} + H_{\mathit{Irr}} \\ W_{\sigma} &=& \Big\{
  K_{5} \left[ c_{\sigma}^{*}(f_{1}) c_{\sigma} (\varepsilon f_{1}) + \hc
  \right] + K_{6} c_{\sigma}^{*}(f_{1}) c_{\sigma}(f_{1}) \Big\} \nonumber \\
  && + \Big\{ K_{2} c_{\sigma}^{*}(f_{1}) c_{\sigma}(f_{1}) + K_{3}
  c_{\sigma}^{*}(F_{2}) c_{\sigma}(F_{2}) \nonumber \\ && \quad \quad + K_{4}
  \left[ c_{\sigma}^{*}(f_{1}) c_{\sigma}(F_{2}) +\hc \right] \Big\} n_{d} \\
  W_{\mathrm{sd}} &=& K_{2} \vec{S}_{c} (f_{1} \otimes
  f_{1})\textbf{.}\vec{S}_{d} + K_{3} \vec{S}_{c}(F_{2} \otimes
  F_{2})\textbf{.}\vec{S}_{d} \nonumber \\ && \quad + K_{4}
  \left[\vec{S}_{c}(f_{1} \otimes F_{2}) +\vec{S}_{c}(F_{2} \otimes
  f_{1})\right]\textbf{.}\vec{S}_{d} \\ F_{2} &=& (\varepsilon+\mu_{2})^{-1}
  f_{1}
\end{eqnarray}

The various coefficients have the following expression
\begin{eqnarray}
\varepsilon_{R} &=& -\lambda_{1} \\ U_{R} &=& \frac{U}{(1+x_{1}^{2})^{2}}
\left[1-2\sqrt{2} x_{1} \alpha_{2} \beta_{2} \omega_{12}-\beta_{2}^{2}(1-2
x_{1}^{2} \omega_{12}^{2})\right] \nonumber \\ && +\beta_{2}^{2}
\left\{\lambda_{1}-\mu_{2} +\omega_{12}^{2} \left[ x_{1}^{2}
(\lambda_{1}-\mu_{1})+(\sqrt{1+x_{1}^{2}}-1) (\mu_{1}-\mu_{2}) \right]\right\}
  \label{EQUrenorm}\\
K_{1} &=& \frac{x_{1}^{2} (\alpha_{2}-\sqrt{2} x_{1} \beta_{2}
  \omega_{12})}{(1+x_{1}^{2})^{2}} U
\end{eqnarray}
\begin{eqnarray}
K_{2} &=& \frac{\omega_{12}^{2}}{4} (1-\alpha_{2}) KC_{2} + \frac{U}{4
  (1+x_{1}^{2})^{5/2}} KU_{2}\\ KC_{2} &=& -x_{1}^{2} \Big\{ \lambda_{1}
  \left[1+\alpha_{2} - x_{1}^{2} \omega_{12}^{2} (1-\alpha_{2})\right]
  -\mu_{1} \left[2-\omega_{12}^{2} (2+x_{1}^{2}) (1-\alpha_{2}) \right]
  \nonumber \\ && \quad \quad +\mu_{2} (1-\alpha_{2}) (1-2 \omega_{12}^{2})
  \Big\} \nonumber \\ && + \left(\sqrt{1+x_{1}^{2}}-1\right) \Big\{ 2
  \lambda_{1} \left[ 1 +\alpha_{2} +x_{1}^{2} -x_{1}^{2} \omega_{12}^{2}
  (1-\alpha_{2}) \right] \nonumber \\ && \quad \quad -\mu_{1} \left[4 +2
  x_{1}^{2} -4 \omega_{12}^{2} (1 -\alpha_{2}) -3 x_{1}^{2} \omega_{12}^{2}
  (1-\alpha_{2}) \right] \nonumber \\ && \quad \quad +\mu_{2} (1-\alpha_{2})
  \left[2 -(4 +x_{1}^{2}) \omega_{12}^{2} \right]\Big\} \\ KU_{2} &=&
  x_{1}^{2} \Big\{ 2 -\beta_{2}^{2} \omega_{12}^{2} -2 \sqrt{2} \beta_{2}
  x_{1} \omega_{12} \left[1-(1-\alpha_{2}) \omega_{12}^{2}\right] \nonumber \\
  && \quad \quad + 2 x_{1}^{2} \omega_{12}^{2} (1-\alpha_{2}) \left[ 2
  -(1-\alpha_{2}) \omega_{12}^{2}\right]\Big\} \nonumber \\ && +
  \left(\sqrt{1+x_{1}^{2}}-1\right) \Big\{ 2 \sqrt{2} \beta_{2} x_{1}
  \omega_{12} \left[1-2(1-\alpha_{2}) \omega_{12}^{2}\right] \nonumber \\ &&
  \quad \quad +x_{1}^{2} \left[2 - (4-4 \alpha_{2}-\beta_{2}^{2})
  \omega_{12}^{2} + 4 \omega_{12}^{4} (1-\alpha_{2})^{2} \right] +2
  \beta_{2}^{2} \omega_{12}^{2} \nonumber \\ && \quad \quad -2 \sqrt{2}
  \beta_{2} x_{1}^{3} \omega_{12}^{3} (1-\alpha_{2}) + 2 x_{1}^{4}
  \omega_{12}^{4} (1-\alpha_{2})^{2} \Big\}
\end{eqnarray}
\begin{eqnarray}
K_{3} &=& \frac{U \omega_{12}^{2}}{4 (1+x_{1}^{2})^{2}} (1-\alpha_{2})
  (\mu_{1}-\mu_{2})^{2} (2+x_{1}^{2}-2\sqrt{1+x_{1}^{2}}) \nonumber \\ &&
  \quad \quad \left[1 +\alpha_{2}- 2 \sqrt{2} \beta_{2} x_{1} \omega_{12} + 2
  x_{1}^{2} \omega_{12}^{2} (1-\alpha_{2})\right] \nonumber \\ && -
  \frac{\omega_{12}^{2}}{4} (1-\alpha_{2}) (\mu_{1}-\mu_{2})^{2} KC_{3} \\
  KC_{3} &=& \lambda_{1} (2+x_{1}^{2}-2 \sqrt{1+x_{1}^{2}})\left[1 +\alpha_{2}
  - x_{1}^{2} \omega_{12}^{2} (1-\alpha_{2}) \right] \nonumber \\ && + \mu_{1}
  \omega_{12}^{2} (1-\alpha_{2}) \left[4+x_{1}^{4}-4 \sqrt{1+x_{1}^{2}}
  +x_{1}^{2}\left(5-3 \sqrt{1+x_{1}^{2}}\right)\right] \nonumber \\ &&
  -\mu_{2} \Big\{ 2 (\sqrt{1+x_{1}^{2}}-1) \left[1+\alpha_{2} +2
  \omega_{12}^{2} (1-\alpha_{2}) \right] \nonumber \\ && \quad \quad +
  x_{1}^{2} \left[1 +\alpha_{2} +\omega_{12}^{2} (1-\alpha_{2})
  (3-\sqrt{1+x_{1}^{2}}) \right]\Big\}
\end{eqnarray}
\begin{eqnarray}
K_{4} &=& -\frac{\omega_{12}^{2}}{4} (1-\alpha_{2}) (\mu_{2}-\mu_{1}) KC_{4} +
  \frac{U \omega_{12} (\mu_{2}-\mu_{1})}{4 (1+x_{1}^{2})^{2}} KU_{4} \\ KC_{4}
  &=& \lambda_{1} \Big\{ 2 \left(\sqrt{1+x_{1}^{2}} -1\right) (1 +\alpha_{2})
  +x_{1}^{4} \omega_{12}^{2} (1-\alpha_{2}) \nonumber \\ && \quad \quad
  -x_{1}^{2} \left[ 2+\alpha_{2} -\omega_{12}^{2} \sqrt{1+x_{1}^{2}} +2
  \omega_{12}^{2} (1-\alpha_{2}) \left(\sqrt{1+x_{1}^{2}} -1\right) \right]
  \Big\} \nonumber \\ && + \mu_{1} \Big\{ 2 (1+x_{1}^{2}) -\omega_{12}^{2} (1
  -\alpha_{2}) (4 +5 x_{1}^{2} +x_{1}^{4}) \nonumber \\ && \quad \quad
  -\sqrt{1+x_{1}^{2}} \left[ 2+x_{1}^{2} -\omega_{12}^{2} (1 -\alpha_{2}) (4+3
  x_{1}^{2}) \right]\Big\} \nonumber \\ && +\mu_{2} \Big\{ \alpha_{2}
  (2+x_{1}^{2}) +\omega_{12}^{2} (1 -\alpha_{2}) (4 +3 x_{1}^{2}) \nonumber \\
  && \quad \quad -\sqrt{1+x_{1}^{2}} \left[ 2 \alpha_{2} + \omega_{12}^{2} (1
  -\alpha_{2}) (4+x_{1}^{2}) \right]\Big\}\\ KU_{4} &=& -x_{1}^{2} \omega_{12}
  (1-\alpha_{2}) \left[ 1+\alpha_{2} -2 \sqrt{2} \beta_{2} x_{1} \omega_{12}
  +2 x_{1}^{2} \omega_{12}^{2} (1-\alpha_{2}) \right] \nonumber \\ && +
  \left(\sqrt{1+x_{1}^{2}} -1\right) \Big\{ 2 \beta_{2}^{2} \omega_{12}
  +\sqrt{2} \beta_{2} x_{1} \left[1-4 (1-\alpha_{2}) \omega_{12}^{2} \right]
  \nonumber \\ && \quad \quad -2 x_{1}^{2} \omega_{12} (1-\alpha_{2}) \left[ 1
  -2 \omega_{12}^{2} (1-\alpha_{2}) \right]\Big\}
\end{eqnarray}
\begin{eqnarray}
K_{5} &=& \sqrt{1+x_{1}^{2}} -1\\ K_{6} &=& \frac{2+x_{1}^{2}-2
\sqrt{1+x_{1}^{2}}}{x_{1}^{2}} \varepsilon_{d} + \frac{2-x_{1}^{2}-2
(1-x_{1}^{2}) \sqrt{1+x_{1}^{2}}}{x_{1}^{2}} \lambda_{1}
\end{eqnarray}

Then, I have to \emph{diagonalize} the various \emph{quadratic forms}.  First
I define
\begin{eqnarray}
F_{3} &=& \varepsilon f_{1} - \Gamma_{1} f_{1} \\ F_{4} &=& F_{2} - \omega_{4}
f_{1} \\ \Gamma_{1} &=& \left<f_{1}, \varepsilon f_{1} \right> \; = \;
\frac{\lambda_{1}+\varepsilon_{d}}{x_{1}^{2}} - \lambda_{1} \\ \omega_{4} &=&
\left<f_{1}, F_{2}\right> \; = \; \left<f_{1}, (\varepsilon+\mu_{2})^{-1}
f_{1} \right> \\ &=& \frac{1}{x_{1}^{2}} (1+x_{1}^{2}) (\lambda_{1}-\mu_{2})
\, \frac{U-(1 +x_{1}^{2})^{2} (\lambda_{1}-\mu_{2})}{U +(1 +x_{1}^{2})^{2}
(\lambda_{1}-\mu_{2})} \\ f_{i} &=& \frac{F_{i}}{\| F_{i}\|}, \quad i \in \{3,
4\} \\ \|F_{3} \|^{2} &=& \frac{1}{x_{1}^{4}} \left[g^{2} x_{1}^{2} -
(\lambda_{1}+\varepsilon_{d})^{2} \right] \\ &=& \frac{\left<V,
(\varepsilon+\lambda_{1})^{-2} V\right> - \left<V,
(\varepsilon+\lambda_{1})^{-1} V\right>^{2}}{\left<V,
(\varepsilon+\lambda_{1})^{-2} V\right>^{2}}\\ \| F_{4} \|^{2} &=& \| F_{2}
\|^{2} - \omega_{4}^{2} \\ \| F_{2} \|^{2} &=& \frac{1 + x_{1}^{2}
\omega_{12}^{2}}{\omega_{12}^{2} (\mu_{1}-\mu_{2})^{2}
\left(\sqrt{1+x_{1}^{2}}-1\right)^{2}}
\end{eqnarray}
This allows to rewrite
\begin{eqnarray}
W_{\sigma} &=& \Big\{ a_{1} c_{\sigma}^{*}(f_{1}) c_{\sigma}(f_{1}) + c_{1}
  \left[ c_{\sigma}^{*}(f_{1}) c_{\sigma} (f_{3}) + \hc \right] \Big\}
  \nonumber \\ && + \Big\{ a_{2} c_{\sigma}^{*}(f_{1}) c_{\sigma}(f_{1}) +
  b_{2} c_{\sigma}^{*}(f_{4}) c_{\sigma}(F_{4}) \nonumber \\ && \quad \quad +
  c_{2} \left[ c_{\sigma}^{*}(f_{1}) c_{\sigma}(f_{4}) +\hc \right] \Big\}
  n_{d} \\ a_{1} &=& \lambda_{1} + \varepsilon_{d} \\ b_{1} &=& 0 \\ c_{1} &=&
  \frac{1}{x_{1}^{2}} \left(\sqrt{1+x_{1}^{2}}-1\right) \sqrt{g^{2} x_{1}^{2}
  -(\lambda_{1}+\varepsilon_{d})^{2}} \\ a_{2} &=& K_{2} +K_{3} \omega_{4}^{2}
  +2 K_{4} \omega_{4} \\ b_{2} &=& K_{3} \|F_{4}\|^{2} \\ c_{2} &=& \|F_{4}\|
  (K_{3} \omega_{4}+K_{4})
\end{eqnarray}

It is then a standard exercise to diagonalize $W_{\sigma}$, and one should
note that it allows also to diagonalize the spin-spin interaction.  I will
note
\begin{eqnarray}
\Delta_{i} &=& (a_{i}-b_{i})^{2}+4 c_{i}^{2} \\ \lambda_{\pm} &=& \frac{1}{2}
(a_{1} \pm \sqrt{\Delta_{1}}) \\ \lambda_{AF} &=& \frac{1}{2}
\left[\sqrt{\Delta_{2}} - (a_{2}+b_{2}) \right]\\ \lambda_{F} &=& -\frac{1}{2}
\left[\sqrt{\Delta_{2}} + (a_{2}+b_{2}) \right]\\ N_{\pm}^{2} &=& 2
(\Delta_{1} \pm a_{1} \sqrt{\Delta_{1}}) \\ N_{AF}^{2} &=& 2 \left[\Delta_{2}
- (a_{2}-b_{2}) \sqrt{\Delta_{2}} \right] \\ N_{F}^{2} &=& 2 \left[\Delta_{2}
+ (a_{2}-b_{2}) \sqrt{\Delta_{2}} \right] \\ \varphi_{\pm} &=&
\frac{1}{N_{\pm}} \left[(\sqrt{\Delta_{1}} \pm a_{1}) f_{1} + 2 c_{1} f_{3}
\right] \\ \varphi_{AF} &=& \frac{1}{N_{AF}} \left\{\left[\sqrt{\Delta_{2}}-
(a_{2}-b_{2})\right] f_{1} - 2 c_{2} f_{4} \right\} \\ \varphi_{F} &=&
\frac{1}{N_{F}} \left\{\left[\sqrt{\Delta_{2}}+(a_{2}-b_{2})\right] f_{1} + 2
c_{2} f_{4} \right\}
\end{eqnarray}

Then I can finally write $H_{2}$ as
\begin{eqnarray}
H_{2} &=& H_{0} + \varepsilon_{R} n_{d} + U_{R} n_{d, \uparrow} n_{d,
  \downarrow} + K_{1} \left[c_{\uparrow}^{*}(f_{1}) c_{\downarrow}^{*}(f_{1})
  d_{\downarrow} d_{\uparrow} + \hc \right] \nonumber \\ && + \sum_{{\sigma
  \in \{\uparrow, \downarrow\}} \atop {i \in \{+, -\}}} \lambda_{i}
  c_{\sigma}^{*}(\varphi_{i}) c_{\sigma}(\varphi_{i}) - \sum_{{\sigma \in
  \{\uparrow, \downarrow\}} \atop {x \in \{AF, F\}}} \lambda_{x}
  c_{\sigma}^{*}(\varphi_{x}) c_{\sigma}(\varphi_{x}) n_{d} \nonumber \\ && +
  4 \sum_{x \in \{ AF, F\}} \lambda_{x} \vec{S}_{c}(\varphi_{x} \otimes
  \varphi_{x}) \textbf{.} \vec{S}_{d} + H_{\mathit{Irr}}
\end{eqnarray}


\section{Asymptotics}
I will now evaluate the various coefficients in different regimes, assuming
\begin{equation}
g \ll 1
\end{equation}

More precisely, I am interested in
\begin{itemize}
\item the \emph{local moment} regime
\begin{itemize}
\item $\varepsilon_{d}<0$, $(U+\varepsilon_{d})>0$,
\item $|\varepsilon_{d}|, (U+\varepsilon_{d}) \gg g^{2} \log g^{-1}$.
\end{itemize}
\item the \emph{intermediate valence} regime
\begin{itemize}
\item $|\varepsilon_{d}| \ll g^{2} \log g^{-1} \ll U$.
\end{itemize}
\end{itemize}
I will note respectively $C^{(lm)}$ and $C^{(iv)}$ the value of $C$ in the
local moment or intermediate valence regimes.  I assume that $V$ is continuous
and non-vanishing at the Fermi surface and that $\varepsilon$ vanishes
linearly (this corresponds to the physical situation) so that the following
asymptotics hold
\begin{eqnarray}
\left<V, (\varepsilon +\lambda)^{-1} V\right> &\sim& a_{1} \log \lambda^{-1}
\\ \left<V, (\varepsilon +\lambda)^{-2} V\right> &\sim& a_{2} \lambda^{-1} \\
\left<V, (\varepsilon +\lambda_{1})^{-2} (\varepsilon+\mu)^{-1} V\right>
  &%
  \begin{array}[t]{c}
    \sim \\[-0.6em] \scriptstyle \mu \ll \lambda_{1}
  \end{array}
  & a_{1} \lambda_{1}^{-2} \log\left(\frac{\lambda_{1}}{\mu}\right) \\
\left<V, (\varepsilon +\lambda_{1})^{-2} (\varepsilon+\mu)^{-2} V\right>
  &%
  \begin{array}[t]{c}
    \sim \\[-0.6em] \scriptstyle \mu \ll \lambda_{1}
  \end{array}
  & a_{2} \lambda_{1}^{-2} \mu^{-1}
\end{eqnarray}
Furthermore, I will assume that $|\varepsilon_{d}|$ is much smaller than the
width of the conduction band (\emph{i.e.} the range of $\varepsilon$) and I
choose units so that $|\varepsilon_{d}| \ll 1$. 

\begin{itemize}
\item $\lambda_{1} = - \varepsilon_{R}$ is given by equation
(\ref{EQlambdaun})
\begin{eqnarray}
\lambda_{1}^{(lm)} &\sim& -\varepsilon_{d} \\ \lambda_{1}^{(lm)} +
\varepsilon_{d} &\sim& a_{1} g^{2} \log |\varepsilon_{d}|^{-1} \\
\lambda_{1}^{(iv)} &\sim& a_{1} g^{2} \log g^{-2}
\end{eqnarray}
\item I will need $x_{1}$ as expansion parameter
\begin{eqnarray}
x_{1}^{2} &=& g^{2} \left<V, (\varepsilon+\lambda_{1})^{-1} V\right> \; \sim
  \; a_{2} \frac{g^{2}}{\lambda_{1}} \\ \left[x_{1}^{(lm)}\right]^{2} &\sim&
  a_{2} \frac{g^{2}}{|\varepsilon_{d}|} \; \ll \; 1 \\
  \left[x_{1}^{(iv)}\right]^{2} &\sim& \frac{a_{2}}{a_{1} \log g^{-2}} \; \ll
  \; 1
\end{eqnarray}
\item $\mu_{2}$ is given by equation (\ref{EQphimudeux})
\begin{eqnarray}
\log \left(\frac{\lambda_{1}}{\mu_{2}}\right) &\sim& a_{1}^{-1}
  \left(\frac{U-\lambda_{1}}{U+\lambda_{1}}\right) \frac{\lambda_{1}}{g^{2}}
  \; \gg \; 1 \\ \mu_{2}^{(lm)} &\sim& \lambda_{1} e^{-a_{1}^{-1}
  \left(\frac{U-\lambda_{1}}{U+\lambda_{1}}\right) \frac{\lambda_{1}}{g^{2}}}
  \; = \; O\left(x_{1}^{\infty}, \lambda_{1}^{\infty} \right)\\ \mu_{2}^{(iv)}
  &\sim& g^{2} \lambda_{1} \; \sim \; a_{1} g^{4} \log g^{-2} \; = \;
  O\left(x_{1}^{\infty}, \lambda_{1}^{2}\right)
\end{eqnarray}
\item $\mu_{1}$ is given by equation (\ref{EQsolvemuun})
\begin{eqnarray}
\mu_{1} &\sim& -2 \lambda_{1} \left(\frac{U+\lambda_{1}}{U-\lambda_{1}}
  \right) \\ \mu_{1}^{(lm)} &\sim& -2 |\varepsilon_{d}| \left(\frac{U
  +|\varepsilon_{d}|}{U-|\varepsilon_{d}|} \right) \\ \mu_{1}^{(iv)} &\sim& -2
  \lambda_{1}
\end{eqnarray}
\item $\omega_{12}$ is also an important expansion parameter, it is given by
equation (\ref{EQsolveomegadouze})
\begin{eqnarray}
\omega_{12}^{2} &\sim& \frac{1}{a_{2}} \left(\frac{2
    \lambda_{1}}{\mu_{1}}\right)^{2} \frac{\mu_{2}}{g^{2} x_{1}^{2}} \\
    \left[\omega_{12}^{(lm)}\right]^{2} &\sim& \frac{1}{a_{2}^{2}}
    \left(\frac{U -|\varepsilon_{d}|}{U+|\varepsilon_{d}|} \right)^{2}
    \frac{|\varepsilon_{d}|}{g^{4}} \mu_{2}^{(lm)} \; = \;
    O\left(x_{1}^{\infty}, \lambda_{1}^{\infty}\right) \\
    \left[\omega_{12}^{(iv)}\right]^{2} &\sim& \frac{a_{1}^{2}}{a_{2}^{2}}
    g^{2} \left(\log g^{-2}\right)^{2} \; = \; O(x_{1}^{\infty}, \lambda_{1})
\end{eqnarray}
\item $U_{R}$ is given by equation (\ref{EQUrenorm})
\begin{equation}
U_{R} = \lambda_{1} - \mu_{2} +O\left(\frac{\lambda_{1}^{2} x_{1}^{2}
    \omega_{12}^{2}}{U}\right) = -\varepsilon_{R} -\mu_{2} + o(\mu_{2})
\end{equation}
\item For the scattering potential, I find
\begin{eqnarray}
K_{1} &\sim& -\theta \frac{(2\lambda_{1}-\mu_{1})}{4 \sqrt{2}} x_{1}^{3}
  \omega_{12} \\ \lambda_{\pm} &\sim& \pm \frac{g x_{1}}{2} \\
  \lambda_{\pm}^{(lm)} &\sim& \pm \frac{\sqrt{a_{2}}}{2} \,
  \frac{g^{2}}{|\varepsilon_{d}|^{1/2}} \\ \lambda_{\pm}^{(iv)} &\sim& \pm
  \frac{1}{2} \sqrt{\frac{a_{2}}{a_{1}}} \, \frac{g}{(\log g^{-2})^{1/2}} \\
  \varphi_{\pm} &\sim& \frac{f_{1} \pm f_{3}}{\sqrt{2}}
\end{eqnarray}
\item Finally, for the spin-spin part
\begin{eqnarray}
\lambda_{AF}^{(lm)} &\sim& \frac{a_{2}g^{2}}{2} \,
  \left(\frac{U}{U+\varepsilon_{d}}\right) \; > \; 0 \\ \lambda_{F}^{(lm)}
  &\sim& - \frac{(U+\varepsilon_{d})}{4} \; < \; 0\\ \varphi_{AF}^{(lm)}
  &\sim& f_{1} - \theta \sqrt{2 a_{2}} \frac{g}{|\varepsilon_{d}|^{1/2}}
  \left(\frac{U}{U+\varepsilon_{d}}\right) f_{4} \\ \varphi_{F}^{(lm)} &\sim&
  \sqrt{2 a_{2}} \frac{g}{|\varepsilon_{d}|^{1/2}}
  \left(\frac{U}{U+\varepsilon_{d}}\right) f_{1} + \theta f_{4} \\
  \lambda_{AF}^{(iv)} &\sim& \frac{a_{2}g^{2}}{2} \; > \; 0 \\
  \lambda_{F}^{(iv)} &\sim& - \frac{U}{4} \; < \; 0 \\ \varphi_{AF} & \sim&
  f_{1} -\theta \sqrt{\frac{a_{2}}{a_{1} \log g^{-1}}} f_{4} \\ \varphi_{F} &
  \sim& \sqrt{\frac{a_{2}}{a_{1} \log g^{-1}}} f_{1} +\theta f_{4}
\end{eqnarray}
\end{itemize}

Thus I find a strong ferro-magnetic interaction in one channel and a weak
anti-ferromagnetic one in another channel. But one has also to take into
account the potential scattering term
\begin{equation}
- \sum_{{\sigma \in \{\uparrow, \downarrow\}} \atop {x \in \{ AF, F\}}}
  \lambda_{x} c_{\sigma}^{*}(\varphi_{x}) c_{\sigma}(\varphi_{x}) n_{d}
\end{equation}
whose effect is to \emph{suppress} the ferro-magnetic channel. Indeed putting
an electron in the orbital $\varphi_{F}$ with its spin aligned to the impurity
one will result in an energy gain (from this part of the interaction) 
\begin{equation}
\Delta E_{F} = -\lambda_{F} + 4 \lambda_{F} \times \frac{1}{4} = 0
\end{equation}
On the other hand, putting an electron in the orbital $\varphi_{AF}$ with its
spin opposite to the impurity one will result in an energy gain
\begin{equation}
\Delta E_{AF} = -\lambda_{AF} + 4 \lambda_{AF} \times \left( -\frac{1}{4}
  \right) = - 2 \lambda_{AF} < 0
\end{equation}

Therefore at low temperature one will see only the anti-ferromagnetic channel
and the model will exhibit a Kondo effect.


\section{Computation in the general $\mu \neq 0$ case}
In the general case, as $\varepsilon-\mu$ is no longer positive one has to go
to the so-called \emph{particle-hole} representation. Formally, this amounts
to perform the change
\begin{equation}
c_{\sigma}(k) \longmapsto a_{\sigma}(k) \theta \left[\varepsilon(k)
  -\mu\right] + b_{\sigma}^{*}(k) \theta \left[\mu - \varepsilon(k) \right]
\end{equation}
and then do some normal ordering with respect to the $a^{*}$'s and $b^{*}$'s.

Now one would like to kill the terms which are linear in $a_{\sigma}$,
$a_{\sigma}^{*}$, $b_{\sigma}$ or $b_{\sigma}^{*}$. One can still apply the
two general unitary transformations $U_{1}$ and $U_{2}$ but then one must
perform some normal ordering. Thus terms of order 3, 5, \emph{etc.} in
$c_{\sigma}$ or $c_{\sigma}^{*}$ will give some contribution to the terms
linear in the $a$'s or $b$'s.

One can still use my computation of the flow of $U_{1}$ and
$U_{2}$, change the $c$'s into $a$'s and $b$'s, do the normal ordering and set
the linear terms to zero. Once again one will have two constraints to meet
with two adjustable functions so this should be a priori possible, the only
point is that the constraints have now much more complicated expressions.


\section{Conclusion}
In this paper, I showed that with the help of computer algebra it is possible
to compute explicitly non trivial unitary transformations in Quantum Field
Theory. For the Anderson impurity model, I succeeded in eliminating the
hybridization term in the slightly simplified case of zero chemical
potential. This is a rigorous version of the well known Schrieffer-Wolff
transformation. Furthermore it should be possible to treat the general case in
the same way.

More generally, such exact unitary transformations should be useful when one
wants to decouple some small system with finitely many degrees of freedom from
a background field, \emph{e.g.} in dissipative systems. I expect also that one
could use computer algebra to perform \emph{Hamiltonian conditioning}. For
instance, in the case of the Kondo problem, I am investigating the possibility
of deriving explicitly the effective low-temperature Hamiltonian for the s-d
model. This would enable some Renormalization Group study of the model in the
low-temperature phase.



\clearpage
\appendix
\section{Differential Flow of $\boldmath{O \mapsto U_{2}(t) \, O \, 
    U_{2}^{*}(t)}$}
The interested reader will find here the raw material to compute the flow of
my second unitary transformation. I put only the differential flow since the
integrated one would be too long. 




\newcommand{\dispSFPrintmath}[1]{#1 \\}
\newcommand{\Mstring}[1]{\textrm{#1}}

%
\begin{eqnarray*}
\dispSFPrintmath{{P_1}  &=&  {{({d_{\sigma }})}^*}\, {d_{\sigma }}}
\dispSFPrintmath{{P_2}  &=&  {{{a_{\sigma }}[{f_2}]}^*}\, {a_{\sigma }}[{f_2}]}
\dispSFPrintmath{{P_3}  &=&  {{{a_{\sigma }}[{f_2}]}^*}\, {d_{\sigma }}}
\dispSFPrintmath{{P_4}  &=&  {a_{\sigma }}[{f_2}]\, {{({d_{\sigma }})}^*}}
\dispSFPrintmath{{P_5}  &=&  {{{a_{\sigma }}[{f_2}]}^*}\, {d_{\sigma }}\, {{({d_{-\sigma }})}^*}\, {d_{-\sigma
}}}
\dispSFPrintmath{{P_6}  &=&  {a_{\sigma }}[{f_2}]\, {{({d_{\sigma }})}^*}\, {{({d_{-\sigma }})}^*}\, {d_{-\sigma
}}}
\dispSFPrintmath{{P_7}  &=&  {{({d_{\sigma }})}^*}\, {d_{\sigma }}\, {{({d_{-\sigma }})}^*}\, {d_{-\sigma
}}}
\dispSFPrintmath{{P_8}  &=&  {{{a_{\sigma }}[{f_2}]}^*}\, {a_{-\sigma }}[{f_2}]\, {d_{\sigma }}\, {{({d_{-\sigma
}})}^*}}
\dispSFPrintmath{{P_9}  &=&  {{{a_{\sigma }}[{f_2}]}^*}\, {a_{\sigma }}[{f_2}]\, {{({d_{-\sigma }})}^*}\, {d_{-\sigma
}}}
\dispSFPrintmath{{P_{10}}  &=&  {a_{\sigma }}[{f_2}]\, {{{a_{-\sigma }}[{f_2}]}^*}\, {{({d_{\sigma }})}^*}\, {d_{-\sigma
}}}
\dispSFPrintmath{{P_{11}}  &=&  {{{a_{-\sigma }}[{f_2}]}^*}\, {{({d_{\sigma }})}^*}\, {d_{\sigma }}\, {d_{-\sigma
}}}
\dispSFPrintmath{{P_{12}}  &=&  {a_{-\sigma }}[{f_2}]\, {{({d_{\sigma }})}^*}\, {d_{\sigma }}\, {{({d_{-\sigma
}})}^*}}
\dispSFPrintmath{{P_{13}}  &=&  {{{a_{-\sigma }}[{f_2}]}^*}\, {a_{-\sigma }}[{f_2}]\, {{({d_{\sigma }})}^*}\, {d_{\sigma
}}}
\dispSFPrintmath{{P_{14}}  &=&  {{{a_{\sigma }}[{f_2}]}^*}\, {{{a_{-\sigma }}[{f_2}]}^*}\, {d_{\sigma }}\, {d_{-\sigma
}}}
\dispSFPrintmath{{P_{15}}  &=&  {a_{\sigma }}[{f_2}]\, {a_{-\sigma }}[{f_2}]\, {{({d_{\sigma }})}^*}\, {{({d_{-\sigma
}})}^*}}
\dispSFPrintmath{{P_{16}}  &=&  {{{a_{\sigma }}[{f_2}]}^*}\, {a_{\sigma }}[{f_2}]\, {{{a_{-\sigma }}[{f_2}]}^*}\, {d_{-\sigma
}}}
\dispSFPrintmath{{P_{17}}  &=&  {{{a_{\sigma }}[{f_2}]}^*}\, {a_{\sigma }}[{f_2}]\, {a_{-\sigma }}[{f_2}]\, {{({d_{-\sigma
}})}^*}}
\dispSFPrintmath{{P_{18}}  &=&  {{{a_{\sigma }}[{f_2}]}^*}\, {{{a_{-\sigma }}[{f_2}]}^*}\, {a_{-\sigma }}[{f_2}]\, {d_{\sigma
}}}
\dispSFPrintmath{{P_{19}}  &=&  {a_{\sigma }}[{f_2}]\, {{{a_{-\sigma }}[{f_2}]}^*}\, {a_{-\sigma }}[{f_2}]\, {{({d_{\sigma
}})}^*}}
\dispSFPrintmath{{P_{20}}  &=&  {{{a_{\sigma }}[{f_2}]}^*}\, {a_{\sigma }}[{f_2}]\, {a_{-\sigma }}[{f_2}]\, {{({d_{\sigma
}})}^*}\, {d_{\sigma }}\, {{({d_{-\sigma }})}^*}}
\dispSFPrintmath{{P_{21}}  &=&  {{{a_{\sigma }}[{f_2}]}^*}\, {{{a_{-\sigma }}[{f_2}]}^*}\, {a_{-\sigma }}[{f_2}]\, {d_{\sigma
}}\, {{({d_{-\sigma }})}^*}\, {d_{-\sigma }}}
\dispSFPrintmath{{P_{22}}  &=&  {{{a_{\sigma }}[{f_2}]}^*}\, {a_{\sigma }}[{f_2}]\, {{{a_{-\sigma }}[{f_2}]}^*}\, {{({d_{\sigma
}})}^*}\, {d_{\sigma }}\, {d_{-\sigma }}}
\dispSFPrintmath{{P_{23}}  &=&  {a_{\sigma }}[{f_2}]\, {{{a_{-\sigma }}[{f_2}]}^*}\, {a_{-\sigma }}[{f_2}]\, {{({d_{\sigma
}})}^*}\, {{({d_{-\sigma }})}^*}\, {d_{-\sigma }}}
\dispSFPrintmath{{P_{24}}  &=&  {{{a_{\sigma }}[{f_2}]}^*}\, {a_{\sigma }}[{f_2}]\, {{({d_{\sigma }})}^*}\, {d_{\sigma
}}\, {{({d_{-\sigma }})}^*}\, {d_{-\sigma }}}
\dispSFPrintmath{{P_{25}}  &=&  {{{a_{\sigma }}[{f_2}]}^*}\, {a_{\sigma }}[{f_2}]\, {{{a_{-\sigma }}[{f_2}]}^*}\, {a_{-\sigma
}}[{f_2}]\, {{({d_{\sigma }})}^*}\, {d_{\sigma }}}
\dispSFPrintmath{{P_{26}}  &=&  {{{a_{-\sigma }}[{f_2}]}^*}\, {a_{-\sigma }}[{f_2}]\, {{({d_{\sigma }})}^*}\, {d_{\sigma
}}\, {{({d_{-\sigma }})}^*}\, {d_{-\sigma }}}
\dispSFPrintmath{{P_{27}}  &=&  {{{a_{\sigma }}[{f_2}]}^*}\, {a_{\sigma }}[{f_2}]\, {{{a_{-\sigma }}[{f_2}]}^*}\, {a_{-\sigma
}}[{f_2}]\, {{({d_{-\sigma }})}^*}\, {d_{-\sigma }}}
\end{eqnarray*}
%
%
\clearpage
\begin{eqnarray*}
\dispSFPrintmath{\Mstring{$\partial_{t}$}
 {P_1}  &=&  {P_5}-{P_6}}
\dispSFPrintmath{\Mstring{$\partial_{t}$}
 {P_2}  &=&  -{P_5}+{P_6}}
\dispSFPrintmath{\Mstring{$\partial_{t}$}
 {P_3}  &=&  -{P_7}+{P_8}+{P_9}+{P_{14}}}
\dispSFPrintmath{\Mstring{$\partial_{t}$}
 {P_4}  &=&  {P_7}-{P_9}-{P_{10}}-{P_{15}}}
\dispSFPrintmath{\Mstring{$\partial_{t}$}
 {P_5}  &=&  -{P_7}+{P_8}+{P_9}}
\dispSFPrintmath{\Mstring{$\partial_{t}$}
 {P_6}  &=&  {P_7}-{P_9}-{P_{10}}}
\dispSFPrintmath{\Mstring{$\partial_{t}$}
 {P_7}  &=&  {P_5}-{P_6}+{P_{11}}-{P_{12}}}
\dispSFPrintmath{\Mstring{$\partial_{t}$}
 {P_8}  &=&  -{P_5}+{P_{12}}-{P_{20}}+{P_{21}}}
\dispSFPrintmath{\Mstring{$\partial_{t}$}
 {P_9}  &=&  -{P_5}+{P_6}-{P_{20}}+{P_{22}}}
\dispSFPrintmath{\Mstring{$\partial_{t}$}
 {P_{10}}  &=&  {P_6}-{P_{11}}+{P_{22}}-{P_{23}}}
\dispSFPrintmath{\Mstring{$\partial_{t}$}
 {P_{11}}  &=&  -{P_7}+{P_{10}}+{P_{13}}}
\dispSFPrintmath{\Mstring{$\partial_{t}$}
 {P_{12}}  &=&  {P_7}-{P_8}-{P_{13}}}
\dispSFPrintmath{\Mstring{$\partial_{t}$}
 {P_{13}}  &=&  -{P_{11}}+{P_{12}}+{P_{21}}-{P_{23}}}
\dispSFPrintmath{\Mstring{$\partial_{t}$}
 {P_{14}}  &=&  -{P_{16}}-{P_{18}}+{P_{21}}+{P_{22}}}
\dispSFPrintmath{\Mstring{$\partial_{t}$}
 {P_{15}}  &=&  {P_{17}}+{P_{19}}-{P_{20}}-{P_{23}}}
\dispSFPrintmath{\Mstring{$\partial_{t}$}
 {P_{16}}  &=&  {P_{14}}-{P_{24}}+{P_{25}}}
\dispSFPrintmath{\Mstring{$\partial_{t}$}
 {P_{17}}  &=&  -{P_{15}}+{P_{24}}-{P_{25}}}
\dispSFPrintmath{\Mstring{$\partial_{t}$}
 {P_{18}}  &=&  {P_{14}}-{P_{26}}+{P_{27}}}
\dispSFPrintmath{\Mstring{$\partial_{t}$}
 {P_{19}}  &=&  -{P_{15}}+{P_{26}}-{P_{27}}}
\dispSFPrintmath{\Mstring{$\partial_{t}$}
 {P_{20}}  &=&  {P_{24}}-{P_{25}}}
\dispSFPrintmath{\Mstring{$\partial_{t}$}
 {P_{21}}  &=&  -{P_{26}}+{P_{27}}}
\dispSFPrintmath{\Mstring{$\partial_{t}$}
 {P_{22}}  &=&  -{P_{24}}+{P_{25}}}
\dispSFPrintmath{\Mstring{$\partial_{t}$}
 {P_{23}}  &=&  {P_{26}}-{P_{27}}}
\dispSFPrintmath{\Mstring{$\partial_{t}$}
 {P_{24}}  &=&  -{P_{20}}+{P_{22}}}
\dispSFPrintmath{\Mstring{$\partial_{t}$}
 {P_{25}}  &=&  {P_{20}}-{P_{22}}}
\dispSFPrintmath{\Mstring{$\partial_{t}$}
 {P_{26}}  &=&  {P_{21}}-{P_{23}}}
\dispSFPrintmath{\Mstring{$\partial_{t}$}
 {P_{27}}  &=&  -{P_{21}}+{P_{23}}}
\end{eqnarray*}
%
%
\begin{eqnarray*}
\dispSFPrintmath{{R_1}  &=&  {{{a_{\sigma }}[{f_1}]}^*}\, {a_{\sigma }}[{f_1}]}
\dispSFPrintmath{{R_2}  &=&  {{{a_{\sigma }}[{f_1}]}^*}\, {d_{\sigma }}\, {{({d_{-\sigma }})}^*}\, {d_{-\sigma
}}-{a_{\sigma }}[{f_1}]\, {{({d_{\sigma }})}^*}\, {{({d_{-\sigma }})}^*}\, {d_{-\sigma }}}
\dispSFPrintmath{
{R_3}  &=&  {{{a_{\sigma }}[{f_1}]}^*}\, {a_{-\sigma }}[{f_2}]\, {d_{\sigma }}\, {{({d_{-\sigma }})}^*}+{{{a_{\sigma
}}[{f_1}]}^*}\, {a_{\sigma }}[{f_2}]\, {{({d_{-\sigma }})}^*}\, {d_{-\sigma }}+  
\\ &&
\hspace{2.em} {{{a_{\sigma }}[{f_2}]}^*}\, {a_{\sigma }}[{f_1}]\, {{({d_{-\sigma }})}^*}\, {d_{-\sigma }}+{a_{\sigma }}[{f_1}]\, {{{a_{-\sigma }}[{f_2}]}^*}\, {{({d_{\sigma
}})}^*}\, {d_{-\sigma }}
}
\dispSFPrintmath{
{R_4}  &=&  -{{{a_{\sigma }}[{f_1}]}^*}\, {a_{\sigma }}[{f_2}]\, {a_{-\sigma }}[{f_2}]\, {{({d_{\sigma }})}^*}\, {d_{\sigma
}}\, {{({d_{-\sigma }})}^*}+  
\\ &&
\hspace{2.em} {{{a_{\sigma }}[{f_2}]}^*}\, {a_{\sigma }}[{f_1}]\, {{{a_{-\sigma }}[{f_2}]}^*}\, {{({d_{\sigma }})}^*}\, {d_{\sigma }}\, {d_{-\sigma }}
}
\dispSFPrintmath{
{R_5}  &=&  {{{a_{\sigma }}[{f_1}]}^*}\, {{{a_{-\sigma }}[{f_2}]}^*}\, {a_{-\sigma }}[{f_2}]\, {d_{\sigma
}}\, {{({d_{-\sigma }})}^*}\, {d_{-\sigma }}+  
\\ &&
\hspace{2.em} {{{a_{\sigma }}[{f_1}]}^*}\, {a_{\sigma }}[{f_2}]\, {{{a_{-\sigma }}[{f_2}]}^*}\, {{({d_{\sigma }})}^*}\, {d_{\sigma }}\, {d_{-\sigma }}-
\\ &&
\hspace{2.em} {{{a_{\sigma }}[{f_2}]}^*}\, {a_{\sigma }}[{f_1}]\, {a_{-\sigma }}[{f_2}]\, {{({d_{\sigma }})}^*}\, {d_{\sigma }}\, {{({d_{-\sigma }})}^*}-
\\ &&
\hspace{2.em} {a_{\sigma }}[{f_1}]\, {{{a_{-\sigma }}[{f_2}]}^*}\, {a_{-\sigma }}[{f_2}]\, {{({d_{\sigma }})}^*}\, {{({d_{-\sigma }})}^*}\, {d_{-\sigma
}}
}
\dispSFPrintmath{
{R_6}  &=&  -{{{a_{\sigma }}[{f_1}]}^*}\, {a_{\sigma }}[{f_2}]\, {{({d_{\sigma }})}^*}\, {d_{\sigma }}\, {{({d_{-\sigma
}})}^*}\, {d_{-\sigma }}+  
\\ &&
\hspace{2.em} {{{a_{\sigma }}[{f_1}]}^*}\, {a_{\sigma }}[{f_2}]\, {{{a_{-\sigma }}[{f_2}]}^*}\, {a_{-\sigma }}[{f_2}]\, {{({d_{\sigma }})}^*}\, {d_{\sigma
}}-  
\\ &&
\hspace{2.em} {{{a_{\sigma }}[{f_2}]}^*}\, {a_{\sigma }}[{f_1}]\, {{({d_{\sigma }})}^*}\, {d_{\sigma }}\, {{({d_{-\sigma }})}^*}\, {d_{-\sigma }}+  
\\ &&
\hspace{2.em} {{{a_{\sigma }}[{f_2}]}^*}\, {a_{\sigma }}[{f_1}]\, {{{a_{-\sigma }}[{f_2}]}^*}\, {a_{-\sigma }}[{f_2}]\, {{({d_{\sigma }})}^*}\, {d_{\sigma
}}
}
\dispSFPrintmath{
{R_7}  &=&  {{{a_{\sigma }}[{f_2}]}^*}\, {{{a_{\sigma }}[{f_1}]}^*}\, {a_{\sigma }}[{f_2}]\, {a_{-\sigma }}[{f_2}]\, {d_{\sigma
}}\, {{({d_{-\sigma }})}^*}+  
\\ &&
\hspace{2.em} {{{a_{\sigma }}[{f_2}]}^*}\, {a_{\sigma }}[{f_1}]\, {a_{\sigma }}[{f_2}]\, {{{a_{-\sigma }}[{f_2}]}^*}\, {{({d_{\sigma }})}^*}\, {d_{-\sigma
}}
}
\dispSFPrintmath{
{R_8}  &=&  {{{a_{\sigma }}[{f_1}]}^*}\, {a_{\sigma }}[{f_2}]\, {{{a_{-\sigma }}[{f_2}]}^*}\, {a_{-\sigma
}}[{f_2}]\, {{({d_{-\sigma }})}^*}\, {d_{-\sigma }}+  
\\ &&
\hspace{2.em} {{{a_{\sigma }}[{f_2}]}^*}\, {a_{\sigma }}[{f_1}]\, {{{a_{-\sigma }}[{f_2}]}^*}\, {a_{-\sigma }}[{f_2}]\, {{({d_{-\sigma }})}^*}\, {d_{-\sigma
}}
}
\dispSFPrintmath{
{R_9}  &=&  -{{{a_{\sigma }}[{f_1}]}^*}\, {a_{\sigma }}[{f_2}]\, {{{a_{-\sigma }}[{f_2}]}^*}\, {{({d_{\sigma
}})}^*}\, {d_{\sigma }}\, {d_{-\sigma }}+  
\\ &&
\hspace{2.em} {{{a_{\sigma }}[{f_2}]}^*}\, {a_{\sigma }}[{f_1}]\, {a_{-\sigma }}[{f_2}]\, {{({d_{\sigma }})}^*}\, {d_{\sigma }}\, {{({d_{-\sigma }})}^*}
}
\dispSFPrintmath{
{R_{10}}  &=&  -{{{a_{\sigma }}[{f_2}]}^*}\, {{{a_{\sigma }}[{f_1}]}^*}\, {a_{\sigma }}[{f_2}]\, {d_{\sigma
}}\, {{({d_{-\sigma }})}^*}\, {d_{-\sigma }}+  
\\ &&
\hspace{2.em} {{{a_{\sigma }}[{f_2}]}^*}\, {a_{\sigma }}[{f_1}]\, {a_{\sigma }}[{f_2}]\, {{({d_{\sigma }})}^*}\, {{({d_{-\sigma }})}^*}\, {d_{-\sigma
}}
}
\dispSFPrintmath{
{R_{11}}  &=&    
{{{a_{\sigma }}[{f_2}]}^*}\, {{{a_{\sigma }}[{f_1}]}^*}\, {a_{\sigma }}[{f_2}]\, {{{a_{-\sigma }}[{f_2}]}^*}\, {a_{-\sigma }}[{f_2}]\, {d_{\sigma
}}\, {{({d_{-\sigma }})}^*}\, {d_{-\sigma }}-  
\\ &&
\hspace{2.em} {{{a_{\sigma }}[{f_2}]}^*}\, {a_{\sigma }}[{f_1}]\, {a_{\sigma }}[{f_2}]\, {{{a_{-\sigma }}[{f_2}]}^*}\, {a_{-\sigma }}[{f_2}]\, {{({d_{\sigma
}})}^*}\, {{({d_{-\sigma }})}^*}\, {d_{-\sigma }}
}
\dispSFPrintmath{
{R_{12}}  &=&  {{{a_{\sigma }}[{f_1}]}^*}\, {a_{\sigma }}[{f_2}]\, {{{a_{-\sigma }}[{f_2}]}^*}\, {a_{-\sigma
}}[{f_2}]\, {{({d_{\sigma }})}^*}\, {d_{\sigma }}+  
\\ &&
\hspace{2.em} {{{a_{\sigma }}[{f_2}]}^*}\, {a_{\sigma }}[{f_1}]\, {{{a_{-\sigma }}[{f_2}]}^*}\, {a_{-\sigma }}[{f_2}]\, {{({d_{\sigma }})}^*}\, {d_{\sigma
}}
}
\dispSFPrintmath{
{R_{13}}  &=&  {{{a_{\sigma }}[{f_1}]}^*}\, {a_{\sigma }}[{f_2}]\, {{{a_{-\sigma }}[{f_2}]}^*}\, {a_{-\sigma
}}[{f_2}]\, {{({d_{\sigma }})}^*}\, {d_{\sigma }}\, {{({d_{-\sigma }})}^*}\, {d_{-\sigma }}+  
\\ &&
\hspace{2.em} {{{a_{\sigma }}[{f_2}]}^*}\, {a_{\sigma }}[{f_1}]\, {{{a_{-\sigma }}[{f_2}]}^*}\, {a_{-\sigma }}[{f_2}]\, {{({d_{\sigma }})}^*}\, {d_{\sigma
}}\, {{({d_{-\sigma }})}^*}\, {d_{-\sigma }}
}
\dispSFPrintmath{{R_{14}}  &=&  {{{a_{\sigma }}[{f_2}]}^*}\, {a_{\sigma }}[{f_2}]\, {{{a_{-\sigma }}[{f_2}]}^*}\, {a_{-\sigma
}}[{f_2}]\, {{({d_{\sigma }})}^*}\, {d_{\sigma }}\, {{({d_{-\sigma }})}^*}\, {d_{-\sigma }}}
\dispSFPrintmath{{R_{15}}  &=&  {{{a_{\sigma }}[{f_1}]}^*}\, {a_{\sigma }}[\epsilon [{f_1}]]+{{{a_{\sigma
}}[\epsilon [{f_1}]]}^*}\, {a_{\sigma }}[{f_1}]}
\dispSFPrintmath{{R_{16}}  &=&  {{{a_{\sigma }}[\epsilon [{f_1}]]}^*}\, {d_{\sigma }}\, {{({d_{-\sigma }})}^*}\, {d_{-\sigma
}}-{a_{\sigma }}[\epsilon [{f_1}]]\, {{({d_{\sigma }})}^*}\, {{({d_{-\sigma }})}^*}\, {d_{-\sigma }}}
\dispSFPrintmath{
{R_{17}}  &=& {{{a_{\sigma }}[{f_2}]}^*}\, {a_{\sigma }}[\epsilon [{f_1}]]\, {{({d_{-\sigma }})}^*}\, {d_{-\sigma }}+{{{a_{\sigma }}[\epsilon [{f_1}]]}^*}\, {a_{-\sigma
}}[{f_2}]\, {d_{\sigma }}\, {{({d_{-\sigma }})}^*}+  
\\ &&
\hspace{2.em} {{{a_{\sigma }}[\epsilon [{f_1}]]}^*}\, {a_{\sigma }}[{f_2}]\, {{({d_{-\sigma }})}^*}\, {d_{-\sigma }}+{a_{\sigma }}[\epsilon [{f_1}]]\, {{{a_{-\sigma
}}[{f_2}]}^*}\, {{({d_{\sigma }})}^*}\, {d_{-\sigma }}
}
\dispSFPrintmath{
{R_{18}}  &=&  {{{a_{\sigma }}[{f_2}]}^*}\, {a_{\sigma }}[\epsilon [{f_1}]]\, {{{a_{-\sigma }}[{f_2}]}^*}\, {{({d_{\sigma
}})}^*}\, {d_{\sigma }}\, {d_{-\sigma }}-  
\\ &&
\hspace{2.em} {{{a_{\sigma }}[\epsilon [{f_1}]]}^*}\, {a_{\sigma }}[{f_2}]\, {a_{-\sigma }}[{f_2}]\, {{({d_{\sigma }})}^*}\, {d_{\sigma }}\, {{({d_{-\sigma
}})}^*}
}
\end{eqnarray*}
\begin{eqnarray*}
\dispSFPrintmath{
{R_{19}}  &=&  -{{{a_{\sigma }}[{f_2}]}^*}\, {a_{\sigma }}[\epsilon [{f_1}]]\, {a_{-\sigma }}[{f_2}]\, {{({d_{\sigma
}})}^*}\, {d_{\sigma }}\, {{({d_{-\sigma }})}^*}+  
\\ &&
\hspace{2.em} {{{a_{\sigma }}[\epsilon [{f_1}]]}^*}\, {{{a_{-\sigma }}[{f_2}]}^*}\, {a_{-\sigma }}[{f_2}]\, {d_{\sigma }}\, {{({d_{-\sigma }})}^*}\, {d_{-\sigma
}}+  
\\ &&
\hspace{2.em} {{{a_{\sigma }}[\epsilon [{f_1}]]}^*}\, {a_{\sigma }}[{f_2}]\, {{{a_{-\sigma }}[{f_2}]}^*}\, {{({d_{\sigma }})}^*}\, {d_{\sigma }}\, {d_{-\sigma
}}-  
\\ &&
\hspace{2.em} {a_{\sigma }}[\epsilon [{f_1}]]\, {{{a_{-\sigma }}[{f_2}]}^*}\, {a_{-\sigma }}[{f_2}]\, {{({d_{\sigma }})}^*}\, {{({d_{-\sigma }})}^*}\, {d_{-\sigma
}}
}
\dispSFPrintmath{
{R_{20}}  &=&  {{{a_{\sigma }}[{f_2}]}^*}\, {{{a_{\sigma }}[\epsilon [{f_1}]]}^*}\, {a_{\sigma }}[{f_2}]\, {a_{-\sigma
}}[{f_2}]\, {d_{\sigma }}\, {{({d_{-\sigma }})}^*}+  
\\ &&
\hspace{2.em} {{{a_{\sigma }}[{f_2}]}^*}\, {a_{\sigma }}[\epsilon [{f_1}]]\, {a_{\sigma }}[{f_2}]\, {{{a_{-\sigma }}[{f_2}]}^*}\, {{({d_{\sigma }})}^*}\, {d_{-\sigma
}}
}
\dispSFPrintmath{
{R_{21}}  &=&  {{{a_{\sigma }}[{f_2}]}^*}\, {a_{\sigma }}[\epsilon [{f_1}]]\, {{({d_{\sigma }})}^*}\, {d_{\sigma
}}\, {{({d_{-\sigma }})}^*}\, {d_{-\sigma }}+  
\\ &&
\hspace{2.em} {{{a_{\sigma }}[\epsilon [{f_1}]]}^*}\, {a_{\sigma }}[{f_2}]\, {{({d_{\sigma }})}^*}\, {d_{\sigma }}\, {{({d_{-\sigma }})}^*}\, {d_{-\sigma
}}
}
\dispSFPrintmath{
{R_{22}}  &=&  {{{a_{\sigma }}[{f_2}]}^*}\, {a_{\sigma }}[\epsilon [{f_1}]]\, {{{a_{-\sigma }}[{f_2}]}^*}\, {a_{-\sigma
}}[{f_2}]\, {{({d_{\sigma }})}^*}\, {d_{\sigma }}+  
\\ &&
\hspace{2.em} {{{a_{\sigma }}[\epsilon [{f_1}]]}^*}\, {a_{\sigma }}[{f_2}]\, {{{a_{-\sigma }}[{f_2}]}^*}\, {a_{-\sigma }}[{f_2}]\, {{({d_{\sigma }})}^*}\, {d_{\sigma
}}
}
\dispSFPrintmath{
{R_{23}}  &=&  {{{a_{\sigma }}[{f_2}]}^*}\, {a_{\sigma }}[\epsilon [{f_1}]]\, {{{a_{-\sigma }}[{f_2}]}^*}\, {a_{-\sigma
}}[{f_2}]\, {{({d_{-\sigma }})}^*}\, {d_{-\sigma }}+  
\\ &&
\hspace{2.em} {{{a_{\sigma }}[\epsilon [{f_1}]]}^*}\, {a_{\sigma }}[{f_2}]\, {{{a_{-\sigma }}[{f_2}]}^*}\, {a_{-\sigma }}[{f_2}]\, {{({d_{-\sigma }})}^*}\, {d_{-\sigma
}}
}
\dispSFPrintmath{
{R_{24}}  &=&  -{{{a_{\sigma }}[{f_2}]}^*}\, {{{a_{\sigma }}[\epsilon [{f_1}]]}^*}\, {a_{\sigma }}[{f_2}]\, {d_{\sigma
}}\, {{({d_{-\sigma }})}^*}\, {d_{-\sigma }}+  
\\ &&
\hspace{2.em} {{{a_{\sigma }}[{f_2}]}^*}\, {a_{\sigma }}[\epsilon [{f_1}]]\, {a_{\sigma }}[{f_2}]\, {{({d_{\sigma }})}^*}\, {{({d_{-\sigma }})}^*}\, {d_{-\sigma
}}
}
\dispSFPrintmath{
{R_{25}}  &=& {{{a_{\sigma }}[{f_2}]}^*}\, {{{a_{\sigma }}[\epsilon [{f_1}]]}^*}\, {a_{\sigma }}[{f_2}]\, {{{a_{-\sigma }}[{f_2}]}^*}\, {a_{-\sigma }}[{f_2}]\, {d_{\sigma
}}\, {{({d_{-\sigma }})}^*}\, {d_{-\sigma }}-  
\\ &&
\hspace{2.em} {{{a_{\sigma }}[{f_2}]}^*}\, {a_{\sigma }}[\epsilon [{f_1}]]\, {a_{\sigma }}[{f_2}]\,   
\\ &&
\hspace{3.em} {{{a_{-\sigma }}[{f_2}]}^*}\, {a_{-\sigma }}[{f_2}]\, {{({d_{\sigma }})}^*}\, {{({d_{-\sigma }})}^*}\, {d_{-\sigma }}
}
\dispSFPrintmath{
{R_{26}}  &=&  {{{a_{\sigma }}[\epsilon [{f_1}]]}^*}\, {{{a_{-\sigma }}[{f_2}]}^*}\, {a_{-\sigma }}[{f_2}]\, {d_{\sigma
}}\, {{({d_{-\sigma }})}^*}\, {d_{-\sigma }}-  
\\ &&
\hspace{2.em} {a_{\sigma }}[\epsilon [{f_1}]]\, {{{a_{-\sigma }}[{f_2}]}^*}\, {a_{-\sigma }}[{f_2}]\, {{({d_{\sigma }})}^*}\, {{({d_{-\sigma }})}^*}\, {d_{-\sigma
}}
}
\dispSFPrintmath{
{R_{27}}  &=& {{{a_{\sigma }}[{f_2}]}^*}\, {a_{\sigma }}[\epsilon [{f_1}]]\, {{{a_{-\sigma }}[{f_2}]}^*}\, {a_{-\sigma }}[{f_2}]\, {{({d_{\sigma }})}^*}\, {d_{\sigma
}}\, {{({d_{-\sigma }})}^*}\, {d_{-\sigma }}+  
\\ &&
\hspace{2.em} {{{a_{\sigma }}[\epsilon [{f_1}]]}^*}\, {a_{\sigma }}[{f_2}]\, {{{a_{-\sigma }}[{f_2}]}^*}\, {a_{-\sigma }}[{f_2}]\, {{({d_{\sigma }})}^*}\, {d_{\sigma
}}\, {{({d_{-\sigma }})}^*}\, {d_{-\sigma }}
}
\dispSFPrintmath{{R_{28}}  &=&  {{{a_{\sigma }}[{f_1}]}^*}\, {a_{\sigma }}[{f_1}]\, {{{a_{-\sigma }}[{f_1}]}^*}\, {a_{-\sigma
}}[{f_1}]}
\dispSFPrintmath{
{R_{29}}  &=&  -{{{a_{\sigma }}[{f_1}]}^*}\, {{{a_{-\sigma }}[{f_1}]}^*}\, {a_{-\sigma }}[{f_1}]\, {d_{\sigma
}}\, {{({d_{-\sigma }})}^*}\, {d_{-\sigma }}-  
\\ &&
\hspace{2.em} {{{a_{\sigma }}[{f_1}]}^*}\, {a_{\sigma }}[{f_1}]\, {{{a_{-\sigma }}[{f_1}]}^*}\, {{({d_{\sigma }})}^*}\, {d_{\sigma }}\, {d_{-\sigma }}+
 \\ &&
\hspace{2.em} {{{a_{\sigma }}[{f_1}]}^*}\, {a_{\sigma }}[{f_1}]\, {a_{-\sigma }}[{f_1}]\, {{({d_{\sigma }})}^*}\, {d_{\sigma }}\, {{({d_{-\sigma }})}^*}+
 \\ &&
\hspace{2.em} {a_{\sigma }}[{f_1}]\, {{{a_{-\sigma }}[{f_1}]}^*}\, {a_{-\sigma }}[{f_1}]\, {{({d_{\sigma }})}^*}\, {{({d_{-\sigma }})}^*}\, {d_{-\sigma
}}
}
\dispSFPrintmath{
{R_{30}}  &=&  {{{a_{-\sigma }}[{f_1}]}^*}\, {a_{-\sigma }}[{f_1}]\, {{({d_{\sigma }})}^*}\, {d_{\sigma }}\, {{({d_{-\sigma
}})}^*}\, {d_{-\sigma }}+  
\\ &&
\hspace{2.em} {{{a_{\sigma }}[{f_1}]}^*}\, {a_{\sigma }}[{f_1}]\, {{({d_{\sigma }})}^*}\, {d_{\sigma }}\, {{({d_{-\sigma }})}^*}\, {d_{-\sigma }}
}
\dispSFPrintmath{
{R_{31}}  &=&  {{{a_{\sigma }}[{f_1}]}^*}\, {{{a_{-\sigma }}[{f_1}]}^*}\, {a_{-\sigma }}[{f_1}]\, {a_{-\sigma
}}[{f_2}]\, {d_{\sigma }}\, {{({d_{-\sigma }})}^*}+  
\\ &&
\hspace{2.em} {{{a_{\sigma }}[{f_1}]}^*}\, {a_{\sigma }}[{f_1}]\, {{{a_{-\sigma }}[{f_1}]}^*}\, {a_{-\sigma }}[{f_2}]\, {{({d_{\sigma }})}^*}\, {d_{\sigma
}}+  
\\ &&
\hspace{2.em} {{{a_{\sigma }}[{f_1}]}^*}\, {a_{\sigma }}[{f_1}]\, {{{a_{-\sigma }}[{f_2}]}^*}\, {a_{-\sigma }}[{f_1}]\, {{({d_{\sigma }})}^*}\, {d_{\sigma
}}+  
\\ &&
\hspace{2.em} {{{a_{\sigma }}[{f_1}]}^*}\, {a_{\sigma }}[{f_1}]\, {a_{\sigma }}[{f_2}]\, {{{a_{-\sigma }}[{f_1}]}^*}\, {{({d_{\sigma }})}^*}\, {d_{-\sigma
}}+  
\\ &&
\hspace{2.em} {{{a_{\sigma }}[{f_1}]}^*}\, {a_{\sigma }}[{f_2}]\, {{{a_{-\sigma }}[{f_1}]}^*}\, {a_{-\sigma }}[{f_1}]\, {{({d_{-\sigma }})}^*}\, {d_{-\sigma
}}+  
\\ &&
\hspace{2.em} {{{a_{\sigma }}[{f_2}]}^*}\, {{{a_{\sigma }}[{f_1}]}^*}\, {a_{\sigma }}[{f_1}]\, {a_{-\sigma }}[{f_1}]\, {d_{\sigma }}\, {{({d_{-\sigma
}})}^*}+  
\\ &&
\hspace{2.em} {{{a_{\sigma }}[{f_2}]}^*}\, {a_{\sigma }}[{f_1}]\, {{{a_{-\sigma }}[{f_1}]}^*}\, {a_{-\sigma }}[{f_1}]\, {{({d_{-\sigma }})}^*}\, {d_{-\sigma
}}+  
\\ &&
\hspace{2.em} {a_{\sigma }}[{f_1}]\, {{{a_{-\sigma }}[{f_2}]}^*}\, {{{a_{-\sigma }}[{f_1}]}^*}\, {a_{-\sigma }}[{f_1}]\, {{({d_{\sigma }})}^*}\, {d_{-\sigma
}}
}
\end{eqnarray*}
\begin{eqnarray*}
\dispSFPrintmath{
{R_{32}}  &=&  -{{{a_{-\sigma }}[{f_1}]}^*}\, {a_{-\sigma }}[{f_1}]\, {a_{-\sigma }}[{f_2}]\, {{({d_{\sigma
}})}^*}\, {d_{\sigma }}\, {{({d_{-\sigma }})}^*}+  
\\ &&
\hspace{2.em} {{{a_{-\sigma }}[{f_2}]}^*}\, {{{a_{-\sigma }}[{f_1}]}^*}\, {a_{-\sigma }}[{f_1}]\, {{({d_{\sigma }})}^*}\, {d_{\sigma }}\, {d_{-\sigma
}}+  
\\ &&
\hspace{2.em} {{{a_{\sigma }}[{f_1}]}^*}\, {a_{\sigma }}[{f_1}]\, {{{a_{-\sigma }}[{f_2}]}^*}\, {{({d_{\sigma }})}^*}\, {d_{\sigma }}\, {d_{-\sigma }}-
 \\ &&
\hspace{2.em} {{{a_{\sigma }}[{f_1}]}^*}\, {a_{\sigma }}[{f_1}]\, {a_{-\sigma }}[{f_2}]\, {{({d_{\sigma }})}^*}\, {d_{\sigma }}\, {{({d_{-\sigma }})}^*}-
 \\ &&
\hspace{2.em} {{{a_{\sigma }}[{f_1}]}^*}\, {a_{\sigma }}[{f_1}]\, {a_{\sigma }}[{f_2}]\, {{({d_{\sigma }})}^*}\, {{({d_{-\sigma }})}^*}\, {d_{-\sigma
}}+  
\\ &&
\hspace{2.em} {{{a_{\sigma }}[{f_2}]}^*}\, {{{a_{-\sigma }}[{f_1}]}^*}\, {a_{-\sigma }}[{f_1}]\, {d_{\sigma }}\, {{({d_{-\sigma }})}^*}\, {d_{-\sigma
}}+  
\\ &&
\hspace{2.em} {{{a_{\sigma }}[{f_2}]}^*}\, {{{a_{\sigma }}[{f_1}]}^*}\, {a_{\sigma }}[{f_1}]\, {d_{\sigma }}\, {{({d_{-\sigma }})}^*}\, {d_{-\sigma }}-
 \\ &&
\hspace{2.em} {a_{\sigma }}[{f_2}]\, {{{a_{-\sigma }}[{f_1}]}^*}\, {a_{-\sigma }}[{f_1}]\, {{({d_{\sigma }})}^*}\, {{({d_{-\sigma }})}^*}\, {d_{-\sigma
}}
}
\dispSFPrintmath{
{R_{33}}  &=&  {{{a_{\sigma }}[{f_1}]}^*}\, {{{a_{-\sigma }}[{f_1}]}^*}\, {a_{-\sigma }}[{f_2}]\, {d_{\sigma
}}\, {{({d_{-\sigma }})}^*}\, {d_{-\sigma }}+  
\\ &&
\hspace{2.em} {{{a_{\sigma }}[{f_1}]}^*}\, {a_{\sigma }}[{f_2}]\, {{{a_{-\sigma }}[{f_1}]}^*}\, {{({d_{\sigma }})}^*}\, {d_{\sigma }}\, {d_{-\sigma }}-
 \\ &&
\hspace{2.em} {{{a_{\sigma }}[{f_2}]}^*}\, {a_{\sigma }}[{f_1}]\, {a_{-\sigma }}[{f_1}]\, {{({d_{\sigma }})}^*}\, {d_{\sigma }}\, {{({d_{-\sigma }})}^*}-
 \\ &&
\hspace{2.em} {a_{\sigma }}[{f_1}]\, {{{a_{-\sigma }}[{f_2}]}^*}\, {a_{-\sigma }}[{f_1}]\, {{({d_{\sigma }})}^*}\, {{({d_{-\sigma }})}^*}\, {d_{-\sigma
}}
}
\dispSFPrintmath{
{R_{34}}  &=&  {{{a_{\sigma }}[{f_1}]}^*}\, {a_{\sigma }}[{f_1}]\, {a_{\sigma }}[{f_2}]\, {{{a_{-\sigma }}[{f_1}]}^*}\, {a_{-\sigma }}[{f_2}]\, {{({d_{\sigma
}})}^*}\, {{({d_{-\sigma }})}^*}\, {d_{-\sigma }}+  
\\ &&
\hspace{2.em} {{{a_{\sigma }}[{f_1}]}^*}\, {a_{\sigma }}[{f_2}]\, {{{a_{-\sigma }}[{f_1}]}^*}\, {a_{-\sigma }}[{f_1}]\, {a_{-\sigma }}[{f_2}]\, {{({d_{\sigma
}})}^*}\, {d_{\sigma }}\, {{({d_{-\sigma }})}^*}-  
\\ &&
\hspace{2.em} {{{a_{\sigma }}[{f_2}]}^*}\, {{{a_{\sigma }}[{f_1}]}^*}\, {a_{\sigma }}[{f_1}]\, {{{a_{-\sigma }}[{f_2}]}^*}\, {a_{-\sigma }}[{f_1}]\, {d_{\sigma
}}\, {{({d_{-\sigma }})}^*}\, {d_{-\sigma }}-  
\\ &&
\hspace{2.em} {{{a_{\sigma }}[{f_2}]}^*}\, {a_{\sigma }}[{f_1}]\, {{{a_{-\sigma }}[{f_2}]}^*}\, {{{a_{-\sigma }}[{f_1}]}^*}\, {a_{-\sigma }}[{f_1}]\, {{({d_{\sigma
}})}^*}\, {d_{\sigma }}\, {d_{-\sigma }}
}
\dispSFPrintmath{
{R_{35}}  &=& {{{a_{\sigma }}[{f_1}]}^*}\, {{{a_{-\sigma }}[{f_2}]}^*}\, {{{a_{-\sigma }}[{f_1}]}^*}\, {a_{-\sigma }}[{f_1}]\, {a_{-\sigma }}[{f_2}]\, {d_{\sigma
}}\, {{({d_{-\sigma }})}^*}\, {d_{-\sigma }}-  
\\ &&
\hspace{2.em} {{{a_{\sigma }}[{f_1}]}^*}\, {a_{\sigma }}[{f_1}]\, {a_{\sigma }}[{f_2}]\, {{{a_{-\sigma }}[{f_2}]}^*}\, {a_{-\sigma }}[{f_1}]\, {{({d_{\sigma
}})}^*}\, {{({d_{-\sigma }})}^*}\, {d_{-\sigma }}+  
\\ &&
\hspace{2.em} {{{a_{\sigma }}[{f_1}]}^*}\, {a_{\sigma }}[{f_2}]\, {{{a_{-\sigma }}[{f_2}]}^*}\, {{{a_{-\sigma }}[{f_1}]}^*}\, {a_{-\sigma }}[{f_1}]\, {{({d_{\sigma
}})}^*}\, {d_{\sigma }}\, {d_{-\sigma }}+  
\\ &&
\hspace{2.em} {{{a_{\sigma }}[{f_2}]}^*}\, {{{a_{\sigma }}[{f_1}]}^*}\, {a_{\sigma }}[{f_1}]\, {{{a_{-\sigma }}[{f_1}]}^*}\, {a_{-\sigma }}[{f_2}]\, {d_{\sigma
}}\, {{({d_{-\sigma }})}^*}\, {d_{-\sigma }}+  
\\ &&
\hspace{2.em} {{{a_{\sigma }}[{f_2}]}^*}\, {{{a_{\sigma }}[{f_1}]}^*}\, {a_{\sigma }}[{f_1}]\, {a_{\sigma }}[{f_2}]\, {{{a_{-\sigma }}[{f_1}]}^*}\, {{({d_{\sigma
}})}^*}\, {d_{\sigma }}\, {d_{-\sigma }}-  
\\ &&
\hspace{2.em} {{{a_{\sigma }}[{f_2}]}^*}\, {{{a_{\sigma }}[{f_1}]}^*}\, {a_{\sigma }}[{f_1}]\, {a_{\sigma }}[{f_2}]\, {a_{-\sigma }}[{f_1}]\, {{({d_{\sigma
}})}^*}\, {d_{\sigma }}\, {{({d_{-\sigma }})}^*}-  
\\ &&
\hspace{2.em} {{{a_{\sigma }}[{f_2}]}^*}\, {a_{\sigma }}[{f_1}]\, {{{a_{-\sigma }}[{f_1}]}^*}\, {a_{-\sigma }}[{f_1}]\, {a_{-\sigma }}[{f_2}]\, {{({d_{\sigma
}})}^*}\, {d_{\sigma }}\, {{({d_{-\sigma }})}^*}-  
\\ &&
\hspace{2.em} {a_{\sigma }}[{f_1}]\, {{{a_{-\sigma }}[{f_2}]}^*}\, {{{a_{-\sigma }}[{f_1}]}^*}\, {a_{-\sigma }}[{f_1}]\, {a_{-\sigma }}[{f_2}]\, {{({d_{\sigma
}})}^*}\, {{({d_{-\sigma }})}^*}\, {d_{-\sigma }}
}
\dispSFPrintmath{
{R_{36}}  &=&  {{{a_{-\sigma }}[{f_2}]}^*}\, {{{a_{-\sigma }}[{f_1}]}^*}\, {a_{-\sigma }}[{f_1}]\, {a_{-\sigma
}}[{f_2}]\, {{({d_{\sigma }})}^*}\, {d_{\sigma }}+  
\\ &&
\hspace{2.em} {{{a_{\sigma }}[{f_1}]}^*}\, {a_{\sigma }}[{f_1}]\, {{{a_{-\sigma }}[{f_2}]}^*}\, {a_{-\sigma }}[{f_2}]\, {{({d_{\sigma }})}^*}\, {d_{\sigma
}}+  
\\ &&
\hspace{2.em} {{{a_{\sigma }}[{f_1}]}^*}\, {a_{\sigma }}[{f_1}]\, {a_{\sigma }}[{f_2}]\, {{{a_{-\sigma }}[{f_2}]}^*}\, {{({d_{\sigma }})}^*}\, {d_{-\sigma
}}+  
\\ &&
\hspace{2.em} {{{a_{\sigma }}[{f_2}]}^*}\, {{{a_{-\sigma }}[{f_1}]}^*}\, {a_{-\sigma }}[{f_1}]\, {a_{-\sigma }}[{f_2}]\, {d_{\sigma }}\, {{({d_{-\sigma
}})}^*}+  
\\ &&
\hspace{2.em} {{{a_{\sigma }}[{f_2}]}^*}\, {{{a_{\sigma }}[{f_1}]}^*}\, {a_{\sigma }}[{f_1}]\, {a_{-\sigma }}[{f_2}]\, {d_{\sigma }}\, {{({d_{-\sigma
}})}^*}+  
\\ &&
\hspace{2.em} {{{a_{\sigma }}[{f_2}]}^*}\, {{{a_{\sigma }}[{f_1}]}^*}\, {a_{\sigma }}[{f_1}]\, {a_{\sigma }}[{f_2}]\, {{({d_{-\sigma }})}^*}\, {d_{-\sigma
}}+  
\\ &&
\hspace{2.em} {{{a_{\sigma }}[{f_2}]}^*}\, {a_{\sigma }}[{f_2}]\, {{{a_{-\sigma }}[{f_1}]}^*}\, {a_{-\sigma }}[{f_1}]\, {{({d_{-\sigma }})}^*}\, {d_{-\sigma
}}+  
\\ &&
\hspace{2.em} {a_{\sigma }}[{f_2}]\, {{{a_{-\sigma }}[{f_2}]}^*}\, {{{a_{-\sigma }}[{f_1}]}^*}\, {a_{-\sigma }}[{f_1}]\, {{({d_{\sigma }})}^*}\, {d_{-\sigma
}}
}
\dispSFPrintmath{
{R_{37}}  &=&  {{{a_{-\sigma }}[{f_1}]}^*}\, {a_{-\sigma }}[{f_2}]\, {{({d_{\sigma }})}^*}\, {d_{\sigma }}\, {{({d_{-\sigma
}})}^*}\, {d_{-\sigma }}+  
\\ &&
\hspace{2.em} {{{a_{-\sigma }}[{f_2}]}^*}\, {a_{-\sigma }}[{f_1}]\, {{({d_{\sigma }})}^*}\, {d_{\sigma }}\, {{({d_{-\sigma }})}^*}\, {d_{-\sigma }}
}
\end{eqnarray*}
\begin{eqnarray*}
\dispSFPrintmath{
{R_{38}}  &=&  {{{a_{\sigma }}[{f_1}]}^*}\, {a_{\sigma }}[{f_2}]\, {{{a_{-\sigma }}[{f_1}]}^*}\, {a_{-\sigma
}}[{f_2}]\, {{({d_{-\sigma }})}^*}\, {d_{-\sigma }}+  
\\ &&
\hspace{2.em} {{{a_{\sigma }}[{f_1}]}^*}\, {a_{\sigma }}[{f_2}]\, {{{a_{-\sigma }}[{f_1}]}^*}\, {a_{-\sigma }}[{f_2}]\, {{({d_{\sigma }})}^*}\, {d_{\sigma
}}+  
\\ &&
\hspace{2.em} {{{a_{\sigma }}[{f_2}]}^*}\, {a_{\sigma }}[{f_1}]\, {{{a_{-\sigma }}[{f_2}]}^*}\, {a_{-\sigma }}[{f_1}]\, {{({d_{-\sigma }})}^*}\, {d_{-\sigma
}}+  
\\ &&
\hspace{2.em} {{{a_{\sigma }}[{f_2}]}^*}\, {a_{\sigma }}[{f_1}]\, {{{a_{-\sigma }}[{f_2}]}^*}\, {a_{-\sigma }}[{f_1}]\, {{({d_{\sigma }})}^*}\, {d_{\sigma
}}
}
\dispSFPrintmath{
{R_{39}}  &=& {{{a_{\sigma }}[{f_1}]}^*}\, {a_{\sigma }}[{f_1}]\, {a_{\sigma }}[{f_2}]\, {{{a_{-\sigma }}[{f_2}]}^*}\, {{{a_{-\sigma }}[{f_1}]}^*}\, {a_{-\sigma
}}[{f_2}]\, {{({d_{\sigma }})}^*}\, {d_{-\sigma }}+  
\\ &&
\hspace{2.em} {{{a_{\sigma }}[{f_2}]}^*}\, {{{a_{\sigma }}[{f_1}]}^*}\, {a_{\sigma }}[{f_1}]\, {{{a_{-\sigma }}[{f_2}]}^*}\, {a_{-\sigma }}[{f_1}]\, 
\\ &&
\hspace{3.em} {a_{-\sigma }}[{f_2}]\, {d_{\sigma }}\, {{({d_{-\sigma }})}^*}+{{{a_{\sigma }}[{f_2}]}^*}\, {{{a_{\sigma }}[{f_1}]}^*}\, {a_{\sigma }}[{f_2}]\, 
 \\ &&
\hspace{3.em} {{{a_{-\sigma }}[{f_1}]}^*}\, {a_{-\sigma }}[{f_1}]\, {a_{-\sigma }}[{f_2}]\, {d_{\sigma }}\, {{({d_{-\sigma }})}^*}+  
\\ &&
\hspace{2.em} {{{a_{\sigma }}[{f_2}]}^*}\, {a_{\sigma }}[{f_1}]\, {a_{\sigma }}[{f_2}]\, {{{a_{-\sigma }}[{f_2}]}^*}\, {{{a_{-\sigma }}[{f_1}]}^*}\, {a_{-\sigma
}}[{f_1}]\, {{({d_{\sigma }})}^*}\, {d_{-\sigma }}
}
\dispSFPrintmath{
{R_{40}}  &=&  {{{a_{\sigma }}[{f_1}]}^*}\, {a_{\sigma }}[{f_1}]\, {{{a_{-\sigma }}[{f_1}]}^*}\, {a_{-\sigma
}}[{f_2}]\, {{({d_{\sigma }})}^*}\, {d_{\sigma }}\, {{({d_{-\sigma }})}^*}\, {d_{-\sigma }}+  
\\ &&
\hspace{2.em} {{{a_{\sigma }}[{f_1}]}^*}\, {a_{\sigma }}[{f_1}]\, {{{a_{-\sigma }}[{f_2}]}^*}\, {a_{-\sigma }}[{f_1}]\, {{({d_{\sigma }})}^*}\, {d_{\sigma
}}\, {{({d_{-\sigma }})}^*}\, {d_{-\sigma }}+  
\\ &&
\hspace{2.em} {{{a_{\sigma }}[{f_1}]}^*}\, {a_{\sigma }}[{f_2}]\, {{{a_{-\sigma }}[{f_1}]}^*}\, {a_{-\sigma }}[{f_1}]\, {{({d_{\sigma }})}^*}\, {d_{\sigma
}}\, {{({d_{-\sigma }})}^*}\, {d_{-\sigma }}-  
\\ &&
\hspace{2.em} {{{a_{\sigma }}[{f_1}]}^*}\, {a_{\sigma }}[{f_2}]\, {{{a_{-\sigma }}[{f_2}]}^*}\, {{{a_{-\sigma }}[{f_1}]}^*}\, {a_{-\sigma }}[{f_1}]\, 
 \\ &&
\hspace{3.em} {a_{-\sigma }}[{f_2}]\, {{({d_{\sigma }})}^*}\, {d_{\sigma }}-{{{a_{\sigma }}[{f_2}]}^*}\, {{{a_{\sigma }}[{f_1}]}^*}\, {a_{\sigma }}[{f_1}]\, {a_{\sigma
}}[{f_2}]\,   
\\ &&
\hspace{3.em} {{{a_{-\sigma }}[{f_1}]}^*}\, {a_{-\sigma }}[{f_2}]\, {{({d_{-\sigma }})}^*}\, {d_{-\sigma }}-{{{a_{\sigma }}[{f_2}]}^*}\, {{{a_{\sigma
}}[{f_1}]}^*}\,   
\\ &&
\hspace{3.em} {a_{\sigma }}[{f_1}]\, {a_{\sigma }}[{f_2}]\, {{{a_{-\sigma }}[{f_2}]}^*}\, {a_{-\sigma }}[{f_1}]\, {{({d_{-\sigma }})}^*}\, {d_{-\sigma
}}+  
\\ &&
\hspace{2.em} {{{a_{\sigma }}[{f_2}]}^*}\, {a_{\sigma }}[{f_1}]\, {{{a_{-\sigma }}[{f_1}]}^*}\, {a_{-\sigma }}[{f_1}]\, {{({d_{\sigma }})}^*}\, {d_{\sigma
}}\, {{({d_{-\sigma }})}^*}\, {d_{-\sigma }}-  
\\ &&
\hspace{2.em} {{{a_{\sigma }}[{f_2}]}^*}\, {a_{\sigma }}[{f_1}]\, {{{a_{-\sigma }}[{f_2}]}^*}\, {{{a_{-\sigma }}[{f_1}]}^*}\, {a_{-\sigma }}[{f_1}]\, {a_{-\sigma
}}[{f_2}]\, {{({d_{\sigma }})}^*}\, {d_{\sigma }}
}
\dispSFPrintmath{
{R_{41}}  &=&  {{{a_{\sigma }}[{f_1}]}^*}\, {a_{\sigma }}[{f_2}]\, {{{a_{-\sigma }}[{f_1}]}^*}\, {a_{-\sigma
}}[{f_2}]\, {{({d_{\sigma }})}^*}\, {d_{\sigma }}\, {{({d_{-\sigma }})}^*}\, {d_{-\sigma }}+  
\\ &&
\hspace{2.em} {{{a_{\sigma }}[{f_2}]}^*}\, {a_{\sigma }}[{f_1}]\, {{{a_{-\sigma }}[{f_2}]}^*}\, {a_{-\sigma }}[{f_1}]\, {{({d_{\sigma }})}^*}\, {d_{\sigma
}}\, {{({d_{-\sigma }})}^*}\, {d_{-\sigma }}
}
\dispSFPrintmath{
{R_{42}}  &=&  {{{a_{\sigma }}[{f_1}]}^*}\, {a_{\sigma }}[{f_2}]\, {{{a_{-\sigma }}[{f_2}]}^*}\,   
\\ &&
\hspace{3.em} {{{a_{-\sigma }}[{f_1}]}^*}\, {a_{-\sigma }}[{f_1}]\, {a_{-\sigma }}[{f_2}]\, {{({d_{-\sigma }})}^*}\, {d_{-\sigma }}+  
\\ &&
\hspace{2.em} {{{a_{\sigma }}[{f_2}]}^*}\, {{{a_{\sigma }}[{f_1}]}^*}\, {a_{\sigma }}[{f_1}]\, {a_{\sigma }}[{f_2}]\, {{{a_{-\sigma }}[{f_1}]}^*}\, {a_{-\sigma
}}[{f_2}]\, {{({d_{\sigma }})}^*}\, {d_{\sigma }}+  
\\ &&
\hspace{2.em} {{{a_{\sigma }}[{f_2}]}^*}\, {{{a_{\sigma }}[{f_1}]}^*}\, {a_{\sigma }}[{f_1}]\, {a_{\sigma }}[{f_2}]\, {{{a_{-\sigma }}[{f_2}]}^*}\, {a_{-\sigma
}}[{f_1}]\, {{({d_{\sigma }})}^*}\, {d_{\sigma }}+  
\\ &&
\hspace{2.em} {{{a_{\sigma }}[{f_2}]}^*}\, {a_{\sigma }}[{f_1}]\, {{{a_{-\sigma }}[{f_2}]}^*}\,   
\\ &&
\hspace{3.em} {{{a_{-\sigma }}[{f_1}]}^*}\, {a_{-\sigma }}[{f_1}]\, {a_{-\sigma }}[{f_2}]\, {{({d_{-\sigma }})}^*}\, {d_{-\sigma }}
}
\dispSFPrintmath{
{R_{43}}  &=& -{{{a_{-\sigma }}[{f_2}]}^*}\, {{{a_{-\sigma }}[{f_1}]}^*}\, {a_{-\sigma }}[{f_1}]\, {a_{-\sigma }}[{f_2}]\, {{({d_{\sigma }})}^*}\, {d_{\sigma
}}\, {{({d_{-\sigma }})}^*}\, {d_{-\sigma }}+  
\\ &&
\hspace{2.em} {{{a_{\sigma }}[{f_1}]}^*}\, {a_{\sigma }}[{f_2}]\, {{{a_{-\sigma }}[{f_2}]}^*}\, {a_{-\sigma }}[{f_1}]\, {{({d_{\sigma }})}^*}\, {d_{\sigma
}}\, {{({d_{-\sigma }})}^*}\, {d_{-\sigma }}-  
\\ &&
\hspace{2.em} {{{a_{\sigma }}[{f_2}]}^*}\, {{{a_{\sigma }}[{f_1}]}^*}\, {a_{\sigma }}[{f_1}]\, {a_{\sigma }}[{f_2}]\, {{({d_{\sigma }})}^*}\, {d_{\sigma
}}\, {{({d_{-\sigma }})}^*}\, {d_{-\sigma }}+  
\\ &&
\hspace{2.em} {{{a_{\sigma }}[{f_2}]}^*}\, {a_{\sigma }}[{f_1}]\, {{{a_{-\sigma }}[{f_1}]}^*}\, {a_{-\sigma }}[{f_2}]\, {{({d_{\sigma }})}^*}\, {d_{\sigma
}}\, {{({d_{-\sigma }})}^*}\, {d_{-\sigma }}
}
\dispSFPrintmath{
{R_{44}}  &=&  {{{a_{-\sigma }}[{f_2}]}^*}\, {{{a_{-\sigma }}[{f_1}]}^*}\, {a_{-\sigma }}[{f_2}]\, {{({d_{\sigma
}})}^*}\, {d_{\sigma }}\, {d_{-\sigma }}-  
\\ &&
\hspace{2.em} {{{a_{-\sigma }}[{f_2}]}^*}\, {a_{-\sigma }}[{f_1}]\, {a_{-\sigma }}[{f_2}]\, {{({d_{\sigma }})}^*}\, {d_{\sigma }}\, {{({d_{-\sigma }})}^*}+
 \\ &&
\hspace{2.em} {{{a_{\sigma }}[{f_2}]}^*}\, {{{a_{-\sigma }}[{f_1}]}^*}\, {a_{-\sigma }}[{f_2}]\, {d_{\sigma }}\, {{({d_{-\sigma }})}^*}\, {d_{-\sigma
}}-  
\\ &&
\hspace{2.em} {a_{\sigma }}[{f_2}]\, {{{a_{-\sigma }}[{f_2}]}^*}\, {a_{-\sigma }}[{f_1}]\, {{({d_{\sigma }})}^*}\, {{({d_{-\sigma }})}^*}\, {d_{-\sigma
}}
}
\end{eqnarray*}
\begin{eqnarray*}
\dispSFPrintmath{
{R_{45}}  &=& {{{a_{\sigma }}[{f_2}]}^*}\, {{{a_{-\sigma }}[{f_2}]}^*}\, {{{a_{-\sigma }}[{f_1}]}^*}\, {a_{-\sigma }}[{f_1}]\, {a_{-\sigma }}[{f_2}]\, {d_{\sigma
}}\, {{({d_{-\sigma }})}^*}\, {d_{-\sigma }}+  
\\ &&
\hspace{2.em} {{{a_{\sigma }}[{f_2}]}^*}\, {{{a_{\sigma }}[{f_1}]}^*}\, {a_{\sigma }}[{f_1}]\, {a_{\sigma }}[{f_2}]\, {{{a_{-\sigma }}[{f_2}]}^*}\, {{({d_{\sigma
}})}^*}\, {d_{\sigma }}\, {d_{-\sigma }}-  
\\ &&
\hspace{2.em} {{{a_{\sigma }}[{f_2}]}^*}\, {{{a_{\sigma }}[{f_1}]}^*}\, {a_{\sigma }}[{f_1}]\, {a_{\sigma }}[{f_2}]\, {a_{-\sigma }}[{f_2}]\, {{({d_{\sigma
}})}^*}\, {d_{\sigma }}\, {{({d_{-\sigma }})}^*}-  
\\ &&
\hspace{2.em} {a_{\sigma }}[{f_2}]\, {{{a_{-\sigma }}[{f_2}]}^*}\, {{{a_{-\sigma }}[{f_1}]}^*}\, {a_{-\sigma }}[{f_1}]\, {a_{-\sigma }}[{f_2}]\, {{({d_{\sigma
}})}^*}\, {{({d_{-\sigma }})}^*}\, {d_{-\sigma }}
}
\dispSFPrintmath{
{R_{46}}  &=&  {{{a_{\sigma }}[{f_2}]}^*}\, {{{a_{-\sigma }}[{f_2}]}^*}\, {a_{-\sigma }}[{f_1}]\, {d_{\sigma
}}\, {{({d_{-\sigma }})}^*}\, {d_{-\sigma }}-  
\\ &&
\hspace{2.em} {a_{\sigma }}[{f_2}]\, {{{a_{-\sigma }}[{f_1}]}^*}\, {a_{-\sigma }}[{f_2}]\, {{({d_{\sigma }})}^*}\, {{({d_{-\sigma }})}^*}\, {d_{-\sigma
}}
}
\dispSFPrintmath{
{R_{47}}  &=& {{{a_{\sigma }}[{f_1}]}^*}\, {a_{\sigma }}[{f_2}]\, {{{a_{-\sigma }}[{f_2}]}^*}\, {{{a_{-\sigma }}[{f_1}]}^*}\, {a_{-\sigma }}[{f_2}]\, {{({d_{\sigma
}})}^*}\, {d_{\sigma }}\, {d_{-\sigma }}+  
\\ &&
\hspace{2.em} {{{a_{\sigma }}[{f_2}]}^*}\, {{{a_{\sigma }}[{f_1}]}^*}\, {a_{\sigma }}[{f_2}]\, {{{a_{-\sigma }}[{f_1}]}^*}\, {a_{-\sigma }}[{f_2}]\, {d_{\sigma
}}\, {{({d_{-\sigma }})}^*}\, {d_{-\sigma }}-  
\\ &&
\hspace{2.em} {{{a_{\sigma }}[{f_2}]}^*}\, {a_{\sigma }}[{f_1}]\, {{{a_{-\sigma }}[{f_2}]}^*}\, {a_{-\sigma }}[{f_1}]\, {a_{-\sigma }}[{f_2}]\, {{({d_{\sigma
}})}^*}\, {d_{\sigma }}\, {{({d_{-\sigma }})}^*}-  
\\ &&
\hspace{2.em} {{{a_{\sigma }}[{f_2}]}^*}\, {a_{\sigma }}[{f_1}]\, {a_{\sigma }}[{f_2}]\, {{{a_{-\sigma }}[{f_2}]}^*}\, {a_{-\sigma }}[{f_1}]\, {{({d_{\sigma
}})}^*}\, {{({d_{-\sigma }})}^*}\, {d_{-\sigma }}
}
\dispSFPrintmath{
{R_{48}}  &=& -{{{a_{\sigma }}[{f_1}]}^*}\, {a_{\sigma }}[{f_1}]\, {a_{\sigma }}[{f_2}]\, {{{a_{-\sigma }}[{f_2}]}^*}\, {a_{-\sigma }}[{f_2}]\, {{({d_{\sigma
}})}^*}\, {{({d_{-\sigma }})}^*}\, {d_{-\sigma }}+  
\\ &&
\hspace{2.em} {{{a_{\sigma }}[{f_2}]}^*}\, {{{a_{\sigma }}[{f_1}]}^*}\, {a_{\sigma }}[{f_1}]\, {{{a_{-\sigma }}[{f_2}]}^*}\, {a_{-\sigma }}[{f_2}]\, {d_{\sigma
}}\, {{({d_{-\sigma }})}^*}\, {d_{-\sigma }}-  
\\ &&
\hspace{2.em} {{{a_{\sigma }}[{f_2}]}^*}\, {a_{\sigma }}[{f_2}]\, {{{a_{-\sigma }}[{f_1}]}^*}\, {a_{-\sigma }}[{f_1}]\, {a_{-\sigma }}[{f_2}]\, {{({d_{\sigma
}})}^*}\, {d_{\sigma }}\, {{({d_{-\sigma }})}^*}+  
\\ &&
\hspace{2.em} {{{a_{\sigma }}[{f_2}]}^*}\, {a_{\sigma }}[{f_2}]\, {{{a_{-\sigma }}[{f_2}]}^*}\, {{{a_{-\sigma }}[{f_1}]}^*}\, {a_{-\sigma }}[{f_1}]\, {{({d_{\sigma
}})}^*}\, {d_{\sigma }}\, {d_{-\sigma }}
}
\dispSFPrintmath{
{R_{49}}  &=& -{{{a_{\sigma }}[{f_1}]}^*}\, {a_{\sigma }}[{f_1}]\, {{{a_{-\sigma }}[{f_2}]}^*}\, {{{a_{-\sigma }}[{f_1}]}^*}\, {a_{-\sigma }}[{f_2}]\, {{({d_{\sigma
}})}^*}\, {d_{\sigma }}\, {d_{-\sigma }}+  
\\ &&
\hspace{2.em} {{{a_{\sigma }}[{f_1}]}^*}\, {a_{\sigma }}[{f_1}]\, {{{a_{-\sigma }}[{f_2}]}^*}\, {a_{-\sigma }}[{f_1}]\, {a_{-\sigma }}[{f_2}]\, {{({d_{\sigma
}})}^*}\, {d_{\sigma }}\, {{({d_{-\sigma }})}^*}-  
\\ &&
\hspace{2.em} {{{a_{\sigma }}[{f_2}]}^*}\, {{{a_{\sigma }}[{f_1}]}^*}\, {a_{\sigma }}[{f_2}]\, {{{a_{-\sigma }}[{f_1}]}^*}\, {a_{-\sigma }}[{f_1}]\, {d_{\sigma
}}\, {{({d_{-\sigma }})}^*}\, {d_{-\sigma }}+  
\\ &&
\hspace{2.em} {{{a_{\sigma }}[{f_2}]}^*}\, {a_{\sigma }}[{f_1}]\, {a_{\sigma }}[{f_2}]\, {{{a_{-\sigma }}[{f_1}]}^*}\, {a_{-\sigma }}[{f_1}]\, {{({d_{\sigma
}})}^*}\, {{({d_{-\sigma }})}^*}\, {d_{-\sigma }}
}
\dispSFPrintmath{
{R_{50}}  &=&  {{{a_{\sigma }}[{f_2}]}^*}\, {{{a_{\sigma }}[{f_1}]}^*}\, {a_{\sigma }}[{f_1}]\, {a_{\sigma
}}[{f_2}]\, {{{a_{-\sigma }}[{f_2}]}^*}\, {{{a_{-\sigma }}[{f_1}]}^*}\,   
\\ &&
\hspace{3.em} {a_{-\sigma }}[{f_2}]\, {{({d_{\sigma }})}^*}\, {d_{\sigma }}\, {d_{-\sigma }}-{{{a_{\sigma }}[{f_2}]}^*}\, {{{a_{\sigma }}[{f_1}]}^*}\, {a_{\sigma
}}[{f_1}]\,   
\\ &&
\hspace{3.em} {a_{\sigma }}[{f_2}]\, {{{a_{-\sigma }}[{f_2}]}^*}\, {a_{-\sigma }}[{f_1}]\, {a_{-\sigma }}[{f_2}]\, {{({d_{\sigma }})}^*}\, {d_{\sigma
}}\, {{({d_{-\sigma }})}^*}+  
\\ &&
\hspace{2.em} {{{a_{\sigma }}[{f_2}]}^*}\, {{{a_{\sigma }}[{f_1}]}^*}\, {a_{\sigma }}[{f_2}]\, {{{a_{-\sigma }}[{f_2}]}^*}\, {{{a_{-\sigma }}[{f_1}]}^*}\, {a_{-\sigma
}}[{f_1}]\,   
\\ &&
\hspace{3.em} {a_{-\sigma }}[{f_2}]\, {d_{\sigma }}\, {{({d_{-\sigma }})}^*}\, {d_{-\sigma }}-{{{a_{\sigma }}[{f_2}]}^*}\, {a_{\sigma }}[{f_1}]\, {a_{\sigma
}}[{f_2}]\,   
\\ &&
\hspace{3.em} {{{a_{-\sigma }}[{f_2}]}^*}\, {{{a_{-\sigma }}[{f_1}]}^*}\, {a_{-\sigma }}[{f_1}]\, {a_{-\sigma }}[{f_2}]\, {{({d_{\sigma }})}^*}\, {{({d_{-\sigma
}})}^*}\, {d_{-\sigma }}
}
\dispSFPrintmath{
{R_{51}}  &=& -{{{a_{\sigma }}[{f_1}]}^*}\, {a_{\sigma }}[{f_1}]\, {a_{\sigma }}[{f_2}]\, {{{a_{-\sigma }}[{f_2}]}^*}\, {a_{-\sigma }}[{f_1}]\, {{({d_{\sigma
}})}^*}\, {{({d_{-\sigma }})}^*}\, {d_{-\sigma }}+  
\\ &&
\hspace{2.em} {{{a_{\sigma }}[{f_1}]}^*}\, {a_{\sigma }}[{f_2}]\, {{{a_{-\sigma }}[{f_2}]}^*}\, {{{a_{-\sigma }}[{f_1}]}^*}\, {a_{-\sigma }}[{f_1}]\, {{({d_{\sigma
}})}^*}\, {d_{\sigma }}\, {d_{-\sigma }}+  
\\ &&
\hspace{2.em} {{{a_{\sigma }}[{f_2}]}^*}\, {{{a_{\sigma }}[{f_1}]}^*}\, {a_{\sigma }}[{f_1}]\, {{{a_{-\sigma }}[{f_1}]}^*}\, {a_{-\sigma }}[{f_2}]\, {d_{\sigma
}}\, {{({d_{-\sigma }})}^*}\, {d_{-\sigma }}-  
\\ &&
\hspace{2.em} {{{a_{\sigma }}[{f_2}]}^*}\, {a_{\sigma }}[{f_1}]\, {{{a_{-\sigma }}[{f_1}]}^*}\, {a_{-\sigma }}[{f_1}]\, {a_{-\sigma }}[{f_2}]\, {{({d_{\sigma
}})}^*}\, {d_{\sigma }}\, {{({d_{-\sigma }})}^*}
}
\dispSFPrintmath{
{R_{52}}  &=& {{{a_{\sigma }}[{f_1}]}^*}\, {a_{\sigma }}[{f_2}]\, {{{a_{-\sigma }}[{f_2}]}^*}\, {a_{-\sigma }}[{f_1}]\, {a_{-\sigma }}[{f_2}]\, {{({d_{\sigma
}})}^*}\, {d_{\sigma }}\, {{({d_{-\sigma }})}^*}-  
\\ &&
\hspace{2.em} {{{a_{\sigma }}[{f_2}]}^*}\, {{{a_{\sigma }}[{f_1}]}^*}\, {a_{\sigma }}[{f_2}]\, {{{a_{-\sigma }}[{f_2}]}^*}\, {a_{-\sigma }}[{f_1}]\, {d_{\sigma
}}\, {{({d_{-\sigma }})}^*}\, {d_{-\sigma }}-  
\\ &&
\hspace{2.em} {{{a_{\sigma }}[{f_2}]}^*}\, {a_{\sigma }}[{f_1}]\, {{{a_{-\sigma }}[{f_2}]}^*}\, {{{a_{-\sigma }}[{f_1}]}^*}\, {a_{-\sigma }}[{f_2}]\, {{({d_{\sigma
}})}^*}\, {d_{\sigma }}\, {d_{-\sigma }}+  
\\ &&
\hspace{2.em} {{{a_{\sigma }}[{f_2}]}^*}\, {a_{\sigma }}[{f_1}]\, {a_{\sigma }}[{f_2}]\, {{{a_{-\sigma }}[{f_1}]}^*}\, {a_{-\sigma }}[{f_2}]\, {{({d_{\sigma
}})}^*}\, {{({d_{-\sigma }})}^*}\, {d_{-\sigma }}
}
\dispSFPrintmath{
{R_{53}}  &=&  {{{a_{\sigma }}[{f_2}]}^*}\, {{{a_{-\sigma }}[{f_2}]}^*}\, {a_{-\sigma }}[{f_1}]\, {a_{-\sigma
}}[{f_2}]\, {d_{\sigma }}\, {{({d_{-\sigma }})}^*}+  
\\ &&
\hspace{2.em} {{{a_{\sigma }}[{f_2}]}^*}\, {a_{\sigma }}[{f_2}]\, {{{a_{-\sigma }}[{f_1}]}^*}\, {a_{-\sigma }}[{f_2}]\, {{({d_{-\sigma }})}^*}\, {d_{-\sigma
}}+  
\\ &&
\hspace{2.em} {{{a_{\sigma }}[{f_2}]}^*}\, {a_{\sigma }}[{f_2}]\, {{{a_{-\sigma }}[{f_2}]}^*}\, {a_{-\sigma }}[{f_1}]\, {{({d_{-\sigma }})}^*}\, {d_{-\sigma
}}+  
\\ &&
\hspace{2.em} {a_{\sigma }}[{f_2}]\, {{{a_{-\sigma }}[{f_2}]}^*}\, {{{a_{-\sigma }}[{f_1}]}^*}\, {a_{-\sigma }}[{f_2}]\, {{({d_{\sigma }})}^*}\, {d_{-\sigma
}}
}
\end{eqnarray*}
\begin{eqnarray*}
\dispSFPrintmath{
{R_{54}}  &=& {{{a_{-\sigma }}[{f_2}]}^*}\, {{{a_{-\sigma }}[{f_1}]}^*}\, {a_{-\sigma }}[{f_1}]\, {a_{-\sigma }}[{f_2}]\, {{({d_{\sigma }})}^*}\, {d_{\sigma
}}\, {{({d_{-\sigma }})}^*}\, {d_{-\sigma }}+  
\\ &&
\hspace{2.em} {{{a_{\sigma }}[{f_2}]}^*}\, {{{a_{\sigma }}[{f_1}]}^*}\, {a_{\sigma }}[{f_1}]\, {a_{\sigma }}[{f_2}]\, {{({d_{\sigma }})}^*}\, {d_{\sigma
}}\, {{({d_{-\sigma }})}^*}\, {d_{-\sigma }}
}
\dispSFPrintmath{
{R_{55}}  &=& {{{a_{\sigma }}[{f_2}]}^*}\, {{{a_{\sigma }}[{f_1}]}^*}\, {a_{\sigma }}[{f_1}]\, {a_{\sigma }}[{f_2}]\, {{{a_{-\sigma }}[{f_2}]}^*}\, {a_{-\sigma
}}[{f_2}]\, {{({d_{\sigma }})}^*}\, {d_{\sigma }}+  
\\ &&
\hspace{2.em} {{{a_{\sigma }}[{f_2}]}^*}\, {a_{\sigma }}[{f_2}]\, {{{a_{-\sigma }}[{f_2}]}^*}\,   
\\ &&
\hspace{3.em} {{{a_{-\sigma }}[{f_1}]}^*}\, {a_{-\sigma }}[{f_1}]\, {a_{-\sigma }}[{f_2}]\, {{({d_{-\sigma }})}^*}\, {d_{-\sigma }}
}
\dispSFPrintmath{
{R_{56}}  &=&  {{{a_{\sigma }}[{f_2}]}^*}\, {a_{\sigma }}[{f_2}]\, {{{a_{-\sigma }}[{f_1}]}^*}\, {a_{-\sigma
}}[{f_2}]\, {{({d_{\sigma }})}^*}\, {d_{\sigma }}\, {{({d_{-\sigma }})}^*}\, {d_{-\sigma }}+  
\\ &&
\hspace{2.em} {{{a_{\sigma }}[{f_2}]}^*}\, {a_{\sigma }}[{f_2}]\, {{{a_{-\sigma }}[{f_2}]}^*}\, {a_{-\sigma }}[{f_1}]\, {{({d_{\sigma }})}^*}\, {d_{\sigma
}}\, {{({d_{-\sigma }})}^*}\, {d_{-\sigma }}
}
\dispSFPrintmath{
{R_{57}}  &=& {{{a_{\sigma }}[{f_2}]}^*}\, {{{a_{\sigma }}[{f_1}]}^*}\, {a_{\sigma }}[{f_1}]\, {a_{\sigma }}[{f_2}]\, {{{a_{-\sigma }}[{f_2}]}^*}\, {a_{-\sigma
}}[{f_2}]\, {{({d_{-\sigma }})}^*}\, {d_{-\sigma }}+  
\\ &&
\hspace{2.em} {{{a_{\sigma }}[{f_2}]}^*}\, {a_{\sigma }}[{f_2}]\, {{{a_{-\sigma }}[{f_2}]}^*}\, {{{a_{-\sigma }}[{f_1}]}^*}\, {a_{-\sigma }}[{f_1}]\, {a_{-\sigma
}}[{f_2}]\, {{({d_{\sigma }})}^*}\, {d_{\sigma }}
}
\dispSFPrintmath{
{R_{58}}  &=&  {{{a_{\sigma }}[{f_1}]}^*}\, {a_{\sigma }}[{f_1}]\, {{{a_{-\sigma }}[{f_2}]}^*}\, {a_{-\sigma
}}[{f_2}]\, {{({d_{\sigma }})}^*}\, {d_{\sigma }}\, {{({d_{-\sigma }})}^*}\, {d_{-\sigma }}+  
\\ &&
\hspace{2.em} {{{a_{\sigma }}[{f_2}]}^*}\, {a_{\sigma }}[{f_2}]\, {{{a_{-\sigma }}[{f_1}]}^*}\, {a_{-\sigma }}[{f_1}]\, {{({d_{\sigma }})}^*}\, {d_{\sigma
}}\, {{({d_{-\sigma }})}^*}\, {d_{-\sigma }}
}
\dispSFPrintmath{
{R_{59}}  &=&  {{{a_{\sigma }}[{f_1}]}^*}\, {a_{\sigma }}[{f_1}]\, {{{a_{-\sigma }}[{f_1}]}^*}\, {a_{-\sigma
}}[{f_2}]\, {{({d_{\sigma }})}^*}\, {d_{\sigma }}\, {{({d_{-\sigma }})}^*}\, {d_{-\sigma }}+  
\\ &&
\hspace{2.em} {{{a_{\sigma }}[{f_1}]}^*}\, {a_{\sigma }}[{f_1}]\, {{{a_{-\sigma }}[{f_2}]}^*}\, {a_{-\sigma }}[{f_1}]\, {{({d_{\sigma }})}^*}\, {d_{\sigma
}}\, {{({d_{-\sigma }})}^*}\, {d_{-\sigma }}+  
\\ &&
\hspace{2.em} {{{a_{\sigma }}[{f_1}]}^*}\, {a_{\sigma }}[{f_2}]\, {{{a_{-\sigma }}[{f_1}]}^*}\, {a_{-\sigma }}[{f_1}]\, {{({d_{\sigma }})}^*}\, {d_{\sigma
}}\, {{({d_{-\sigma }})}^*}\, {d_{-\sigma }}+  
\\ &&
\hspace{2.em} {{{a_{\sigma }}[{f_2}]}^*}\, {a_{\sigma }}[{f_1}]\, {{{a_{-\sigma }}[{f_1}]}^*}\, {a_{-\sigma }}[{f_1}]\, {{({d_{\sigma }})}^*}\, {d_{\sigma
}}\, {{({d_{-\sigma }})}^*}\, {d_{-\sigma }}
}
\dispSFPrintmath{
{R_{60}}  &=&  {{{a_{\sigma }}[{f_1}]}^*}\, {a_{\sigma }}[{f_2}]\, {{{a_{-\sigma }}[{f_2}]}^*}\, {{{a_{-\sigma
}}[{f_1}]}^*}\, {a_{-\sigma }}[{f_1}]\, {a_{-\sigma }}[{f_2}]\,   
\\ &&
\hspace{3.em} {{({d_{\sigma }})}^*}\, {d_{\sigma }}\, {{({d_{-\sigma }})}^*}\, {d_{-\sigma }}+{{{a_{\sigma }}[{f_2}]}^*}\, {{{a_{\sigma }}[{f_1}]}^*}\, {a_{\sigma
}}[{f_1}]\,   
\\ &&
\hspace{3.em} {a_{\sigma }}[{f_2}]\, {{{a_{-\sigma }}[{f_1}]}^*}\, {a_{-\sigma }}[{f_2}]\, {{({d_{\sigma }})}^*}\, {d_{\sigma }}\, {{({d_{-\sigma }})}^*}\, {d_{-\sigma
}}+  
\\ &&
\hspace{2.em} {{{a_{\sigma }}[{f_2}]}^*}\, {{{a_{\sigma }}[{f_1}]}^*}\, {a_{\sigma }}[{f_1}]\, {a_{\sigma }}[{f_2}]\, {{{a_{-\sigma }}[{f_2}]}^*}\, {a_{-\sigma
}}[{f_1}]\,   
\\ &&
\hspace{3.em} {{({d_{\sigma }})}^*}\, {d_{\sigma }}\, {{({d_{-\sigma }})}^*}\, {d_{-\sigma }}+{{{a_{\sigma }}[{f_2}]}^*}\, {a_{\sigma }}[{f_1}]\, {{{a_{-\sigma
}}[{f_2}]}^*}\,   
\\ &&
\hspace{3.em} {{{a_{-\sigma }}[{f_1}]}^*}\, {a_{-\sigma }}[{f_1}]\, {a_{-\sigma }}[{f_2}]\, {{({d_{\sigma }})}^*}\, {d_{\sigma }}\, {{({d_{-\sigma }})}^*}\, {d_{-\sigma
}}
}
\dispSFPrintmath{
{R_{61}}  &=&  {{{a_{\sigma }}[{f_2}]}^*}\, {{{a_{\sigma }}[{f_1}]}^*}\, {a_{\sigma }}[{f_1}]\, {a_{\sigma
}}[{f_2}]\, {{{a_{-\sigma }}[{f_2}]}^*}\, {a_{-\sigma }}[{f_2}]\,   
\\ &&
\hspace{3.em} {{({d_{\sigma }})}^*}\, {d_{\sigma }}\, {{({d_{-\sigma }})}^*}\, {d_{-\sigma }}+{{{a_{\sigma }}[{f_2}]}^*}\, {a_{\sigma }}[{f_2}]\, {{{a_{-\sigma
}}[{f_2}]}^*}\,   
\\ &&
\hspace{3.em} {{{a_{-\sigma }}[{f_1}]}^*}\, {a_{-\sigma }}[{f_1}]\, {a_{-\sigma }}[{f_2}]\, {{({d_{\sigma }})}^*}\, {d_{\sigma }}\, {{({d_{-\sigma }})}^*}\, {d_{-\sigma
}}
}
\dispSFPrintmath{
{R_{62}}  &=& {{{a_{\sigma }}[{f_2}]}^*}\, {{{a_{\sigma }}[{f_1}]}^*}\, {a_{\sigma }}[{f_2}]\, {{{a_{-\sigma }}[{f_2}]}^*}\, {a_{-\sigma }}[{f_1}]\, {a_{-\sigma
}}[{f_2}]\, {d_{\sigma }}\, {{({d_{-\sigma }})}^*}+  
\\ &&
\hspace{2.em} {{{a_{\sigma }}[{f_2}]}^*}\, {a_{\sigma }}[{f_1}]\, {a_{\sigma }}[{f_2}]\, {{{a_{-\sigma }}[{f_2}]}^*}\, {{{a_{-\sigma }}[{f_1}]}^*}\, {a_{-\sigma
}}[{f_2}]\, {{({d_{\sigma }})}^*}\, {d_{-\sigma }}
}
\dispSFPrintmath{
{R_{63}}  &=& {{{a_{\sigma }}[{f_2}]}^*}\, {a_{\sigma }}[{f_2}]\, {{{a_{-\sigma }}[{f_2}]}^*}\, {{{a_{-\sigma }}[{f_1}]}^*}\, {a_{-\sigma }}[{f_2}]\, {{({d_{\sigma
}})}^*}\, {d_{\sigma }}\, {d_{-\sigma }}-  
\\ &&
\hspace{2.em} {{{a_{\sigma }}[{f_2}]}^*}\, {a_{\sigma }}[{f_2}]\, {{{a_{-\sigma }}[{f_2}]}^*}\, {a_{-\sigma }}[{f_1}]\, {a_{-\sigma }}[{f_2}]\, {{({d_{\sigma
}})}^*}\, {d_{\sigma }}\, {{({d_{-\sigma }})}^*}
}
\dispSFPrintmath{
{R_{64}}  &=& -{{{a_{\sigma }}[{f_1}]}^*}\, {{{a_{-\sigma }}[{f_1}]}^*}\, {a_{-\sigma }}[{f_1}]\, {d_{\sigma }}+{a_{\sigma }}[{f_1}]\, {{{a_{-\sigma
}}[{f_1}]}^*}\, {a_{-\sigma }}[{f_1}]\, {{({d_{\sigma }})}^*}
}
\dispSFPrintmath{{R_{65}}  &=&  {{{a_{\sigma }}[{f_1}]}^*}\, {{{a_{-\sigma }}[{f_1}]}^*}\, {d_{\sigma }}\, {d_{-\sigma
}}+{a_{\sigma }}[{f_1}]\, {a_{-\sigma }}[{f_1}]\, {{({d_{\sigma }})}^*}\, {{({d_{-\sigma }})}^*}}
\dispSFPrintmath{{R_{66}}  &=&  {{{a_{-\sigma }}[{f_1}]}^*}\, {a_{-\sigma }}[{f_1}]\, {{({d_{\sigma }})}^*}\, {d_{\sigma
}}\, {{({d_{-\sigma }})}^*}\, {d_{-\sigma }}}
\dispSFPrintmath{
{R_{67}}  &=&  {{{a_{\sigma }}[{f_1}]}^*}\, {{{a_{-\sigma }}[{f_1}]}^*}\, {a_{-\sigma }}[{f_1}]\, {a_{-\sigma
}}[{f_2}]\, {d_{\sigma }}\, {{({d_{-\sigma }})}^*}+  
\\ &&
\hspace{2.em} {{{a_{\sigma }}[{f_1}]}^*}\, {a_{\sigma }}[{f_2}]\, {{{a_{-\sigma }}[{f_1}]}^*}\, {a_{-\sigma }}[{f_1}]\, {{({d_{-\sigma }})}^*}\, {d_{-\sigma
}}+  
\\ &&
\hspace{2.em} {{{a_{\sigma }}[{f_2}]}^*}\, {a_{\sigma }}[{f_1}]\, {{{a_{-\sigma }}[{f_1}]}^*}\, {a_{-\sigma }}[{f_1}]\, {{({d_{-\sigma }})}^*}\, {d_{-\sigma
}}+  
\\ &&
\hspace{2.em} {a_{\sigma }}[{f_1}]\, {{{a_{-\sigma }}[{f_2}]}^*}\, {{{a_{-\sigma }}[{f_1}]}^*}\, {a_{-\sigma }}[{f_1}]\, {{({d_{\sigma }})}^*}\, {d_{-\sigma
}}
}
\dispSFPrintmath{
{R_{68}}  &=&  {{{a_{\sigma }}[{f_1}]}^*}\, {{{a_{-\sigma }}[{f_2}]}^*}\, {{{a_{-\sigma }}[{f_1}]}^*}\, {a_{-\sigma
}}[{f_1}]\, {d_{\sigma }}\, {d_{-\sigma }}+  
\\ &&
\hspace{2.em} {a_{\sigma }}[{f_1}]\, {{{a_{-\sigma }}[{f_1}]}^*}\, {a_{-\sigma }}[{f_1}]\, {a_{-\sigma }}[{f_2}]\, {{({d_{\sigma }})}^*}\, {{({d_{-\sigma
}})}^*}
}
\end{eqnarray*}
\begin{eqnarray*}
\dispSFPrintmath{
{R_{69}}  &=&  -{{{a_{\sigma }}[{f_1}]}^*}\, {{{a_{-\sigma }}[{f_1}]}^*}\, {a_{-\sigma }}[{f_2}]\, {d_{\sigma
}}-  
\\ &&
\hspace{2.em} {{{a_{\sigma }}[{f_1}]}^*}\, {a_{\sigma }}[{f_2}]\, {{{a_{-\sigma }}[{f_1}]}^*}\, {d_{-\sigma }}+{{{a_{\sigma }}[{f_2}]}^*}\, {a_{\sigma
}}[{f_1}]\, {a_{-\sigma }}[{f_1}]\, {{({d_{-\sigma }})}^*}+  
\\ &&
\hspace{2.em} {a_{\sigma }}[{f_1}]\, {{{a_{-\sigma }}[{f_2}]}^*}\, {a_{-\sigma }}[{f_1}]\, {{({d_{\sigma }})}^*}
}
\dispSFPrintmath{
{R_{70}}  &=&  -{{{a_{-\sigma }}[{f_1}]}^*}\, {a_{-\sigma }}[{f_1}]\, {a_{-\sigma }}[{f_2}]\, {{({d_{\sigma
}})}^*}\, {d_{\sigma }}\, {{({d_{-\sigma }})}^*}+  
\\ &&
\hspace{2.em} {{{a_{-\sigma }}[{f_2}]}^*}\, {{{a_{-\sigma }}[{f_1}]}^*}\, {a_{-\sigma }}[{f_1}]\, {{({d_{\sigma }})}^*}\, {d_{\sigma }}\, {d_{-\sigma
}}+  
\\ &&
\hspace{2.em} {{{a_{\sigma }}[{f_2}]}^*}\, {{{a_{-\sigma }}[{f_1}]}^*}\, {a_{-\sigma }}[{f_1}]\, {d_{\sigma }}\, {{({d_{-\sigma }})}^*}\, {d_{-\sigma
}}-  
\\ &&
\hspace{2.em} {a_{\sigma }}[{f_2}]\, {{{a_{-\sigma }}[{f_1}]}^*}\, {a_{-\sigma }}[{f_1}]\, {{({d_{\sigma }})}^*}\, {{({d_{-\sigma }})}^*}\, {d_{-\sigma
}}
}
\dispSFPrintmath{
{R_{71}}  &=&  {{{a_{\sigma }}[{f_1}]}^*}\, {{{a_{-\sigma }}[{f_1}]}^*}\, {a_{-\sigma }}[{f_1}]\, {d_{\sigma
}}\, {{({d_{-\sigma }})}^*}\, {d_{-\sigma }}-  
\\ &&
\hspace{2.em} {a_{\sigma }}[{f_1}]\, {{{a_{-\sigma }}[{f_1}]}^*}\, {a_{-\sigma }}[{f_1}]\, {{({d_{\sigma }})}^*}\, {{({d_{-\sigma }})}^*}\, {d_{-\sigma
}}
}
\dispSFPrintmath{
{R_{72}}  &=&  {{{a_{\sigma }}[{f_1}]}^*}\, {{{a_{-\sigma }}[{f_1}]}^*}\, {a_{-\sigma }}[{f_2}]\, {d_{\sigma
}}\, {{({d_{-\sigma }})}^*}\, {d_{-\sigma }}-  
\\ &&
\hspace{2.em} {a_{\sigma }}[{f_1}]\, {{{a_{-\sigma }}[{f_2}]}^*}\, {a_{-\sigma }}[{f_1}]\, {{({d_{\sigma }})}^*}\, {{({d_{-\sigma }})}^*}\, {d_{-\sigma
}}
}
\dispSFPrintmath{
{R_{73}}  &=& -{{{a_{\sigma }}[{f_1}]}^*}\, {a_{\sigma }}[{f_2}]\, {{{a_{-\sigma }}[{f_1}]}^*}\, {a_{-\sigma }}[{f_1}]\, {a_{-\sigma }}[{f_2}]\, {{({d_{\sigma
}})}^*}\, {d_{\sigma }}\, {{({d_{-\sigma }})}^*}+  
\\ &&
\hspace{2.em} {{{a_{\sigma }}[{f_2}]}^*}\, {a_{\sigma }}[{f_1}]\, {{{a_{-\sigma }}[{f_2}]}^*}\, {{{a_{-\sigma }}[{f_1}]}^*}\, {a_{-\sigma }}[{f_1}]\, {{({d_{\sigma
}})}^*}\, {d_{\sigma }}\, {d_{-\sigma }}
}
\dispSFPrintmath{
{R_{74}}  &=& {{{a_{\sigma }}[{f_1}]}^*}\, {{{a_{-\sigma }}[{f_2}]}^*}\, {{{a_{-\sigma }}[{f_1}]}^*}\, {a_{-\sigma }}[{f_1}]\, {a_{-\sigma }}[{f_2}]\, {d_{\sigma
}}\, {{({d_{-\sigma }})}^*}\, {d_{-\sigma }}+  
\\ &&
\hspace{2.em} {{{a_{\sigma }}[{f_1}]}^*}\, {a_{\sigma }}[{f_2}]\, {{{a_{-\sigma }}[{f_2}]}^*}\, {{{a_{-\sigma }}[{f_1}]}^*}\, {a_{-\sigma }}[{f_1}]\, {{({d_{\sigma
}})}^*}\, {d_{\sigma }}\, {d_{-\sigma }}-  
\\ &&
\hspace{2.em} {{{a_{\sigma }}[{f_2}]}^*}\, {a_{\sigma }}[{f_1}]\, {{{a_{-\sigma }}[{f_1}]}^*}\, {a_{-\sigma }}[{f_1}]\, {a_{-\sigma }}[{f_2}]\, {{({d_{\sigma
}})}^*}\, {d_{\sigma }}\, {{({d_{-\sigma }})}^*}-  
\\ &&
\hspace{2.em} {a_{\sigma }}[{f_1}]\, {{{a_{-\sigma }}[{f_2}]}^*}\, {{{a_{-\sigma }}[{f_1}]}^*}\, {a_{-\sigma }}[{f_1}]\, {a_{-\sigma }}[{f_2}]\, {{({d_{\sigma
}})}^*}\, {{({d_{-\sigma }})}^*}\, {d_{-\sigma }}
}
\dispSFPrintmath{
{R_{75}}  &=&  -{{{a_{\sigma }}[{f_1}]}^*}\, {{{a_{-\sigma }}[{f_2}]}^*}\, {{{a_{-\sigma }}[{f_1}]}^*}\, {a_{-\sigma
}}[{f_1}]\, {a_{-\sigma }}[{f_2}]\, {d_{\sigma }}-  
\\ &&
\hspace{2.em} {{{a_{\sigma }}[{f_1}]}^*}\, {a_{\sigma }}[{f_2}]\, {{{a_{-\sigma }}[{f_2}]}^*}\, {{{a_{-\sigma }}[{f_1}]}^*}\, {a_{-\sigma }}[{f_1}]\, {d_{-\sigma
}}+  
\\ &&
\hspace{2.em} {{{a_{\sigma }}[{f_2}]}^*}\, {a_{\sigma }}[{f_1}]\, {{{a_{-\sigma }}[{f_1}]}^*}\, {a_{-\sigma }}[{f_1}]\, {a_{-\sigma }}[{f_2}]\, {{({d_{-\sigma
}})}^*}+  
\\ &&
\hspace{2.em} {a_{\sigma }}[{f_1}]\, {{{a_{-\sigma }}[{f_2}]}^*}\, {{{a_{-\sigma }}[{f_1}]}^*}\, {a_{-\sigma }}[{f_1}]\, {a_{-\sigma }}[{f_2}]\, {{({d_{\sigma
}})}^*}
}
\dispSFPrintmath{
{R_{76}}  &=&  {{{a_{\sigma }}[{f_1}]}^*}\, {{{a_{-\sigma }}[{f_2}]}^*}\, {{{a_{-\sigma }}[{f_1}]}^*}\, {a_{-\sigma
}}[{f_2}]\, {d_{\sigma }}\, {d_{-\sigma }}+  
\\ &&
\hspace{2.em} {{{a_{\sigma }}[{f_2}]}^*}\, {{{a_{\sigma }}[{f_1}]}^*}\, {a_{\sigma }}[{f_2}]\, {{{a_{-\sigma }}[{f_1}]}^*}\, {d_{\sigma }}\, {d_{-\sigma
}}+  
\\ &&
\hspace{2.em} {{{a_{\sigma }}[{f_2}]}^*}\, {a_{\sigma }}[{f_1}]\, {a_{\sigma }}[{f_2}]\, {a_{-\sigma }}[{f_1}]\, {{({d_{\sigma }})}^*}\, {{({d_{-\sigma
}})}^*}+  
\\ &&
\hspace{2.em} {a_{\sigma }}[{f_1}]\, {{{a_{-\sigma }}[{f_2}]}^*}\, {a_{-\sigma }}[{f_1}]\, {a_{-\sigma }}[{f_2}]\, {{({d_{\sigma }})}^*}\, {{({d_{-\sigma
}})}^*}
}
\dispSFPrintmath{
{R_{77}}  &=&  {{{a_{-\sigma }}[{f_2}]}^*}\, {{{a_{-\sigma }}[{f_1}]}^*}\, {a_{-\sigma }}[{f_1}]\, {a_{-\sigma
}}[{f_2}]\, {{({d_{\sigma }})}^*}\, {d_{\sigma }}+  
\\ &&
\hspace{2.em} {{{a_{\sigma }}[{f_2}]}^*}\, {{{a_{-\sigma }}[{f_1}]}^*}\, {a_{-\sigma }}[{f_1}]\, {a_{-\sigma }}[{f_2}]\, {d_{\sigma }}\, {{({d_{-\sigma
}})}^*}+  
\\ &&
\hspace{2.em} {{{a_{\sigma }}[{f_2}]}^*}\, {a_{\sigma }}[{f_2}]\, {{{a_{-\sigma }}[{f_1}]}^*}\, {a_{-\sigma }}[{f_1}]\, {{({d_{-\sigma }})}^*}\, {d_{-\sigma
}}+  
\\ &&
\hspace{2.em} {a_{\sigma }}[{f_2}]\, {{{a_{-\sigma }}[{f_2}]}^*}\, {{{a_{-\sigma }}[{f_1}]}^*}\, {a_{-\sigma }}[{f_1}]\, {{({d_{\sigma }})}^*}\, {d_{-\sigma
}}
}
\dispSFPrintmath{
{R_{78}}  &=& {{{a_{\sigma }}[{f_2}]}^*}\, {{{a_{-\sigma }}[{f_2}]}^*}\, {{{a_{-\sigma }}[{f_1}]}^*}\, {a_{-\sigma }}[{f_1}]\, {a_{-\sigma }}[{f_2}]\, {d_{\sigma
}}\, {{({d_{-\sigma }})}^*}\, {d_{-\sigma }}-  
\\ &&
\hspace{2.em} {a_{\sigma }}[{f_2}]\, {{{a_{-\sigma }}[{f_2}]}^*}\, {{{a_{-\sigma }}[{f_1}]}^*}\, {a_{-\sigma }}[{f_1}]\, {a_{-\sigma }}[{f_2}]\, {{({d_{\sigma
}})}^*}\, {{({d_{-\sigma }})}^*}\, {d_{-\sigma }}
}
\dispSFPrintmath{
{R_{79}}  &=& -{{{a_{\sigma }}[{f_2}]}^*}\, {a_{\sigma }}[{f_2}]\, {{{a_{-\sigma }}[{f_1}]}^*}\, {a_{-\sigma }}[{f_1}]\, {a_{-\sigma }}[{f_2}]\, {{({d_{\sigma
}})}^*}\, {d_{\sigma }}\, {{({d_{-\sigma }})}^*}+  
\\ &&
\hspace{2.em} {{{a_{\sigma }}[{f_2}]}^*}\, {a_{\sigma }}[{f_2}]\, {{{a_{-\sigma }}[{f_2}]}^*}\, {{{a_{-\sigma }}[{f_1}]}^*}\, {a_{-\sigma }}[{f_1}]\, {{({d_{\sigma
}})}^*}\, {d_{\sigma }}\, {d_{-\sigma }}
}
\dispSFPrintmath{
{R_{80}}  &=&  {{{a_{\sigma }}[{f_1}]}^*}\, {a_{\sigma }}[{f_2}]\, {{{a_{-\sigma }}[{f_1}]}^*}\, {a_{-\sigma
}}[{f_2}]\, {{({d_{-\sigma }})}^*}\, {d_{-\sigma }}+  
\\ &&
\hspace{2.em} {{{a_{\sigma }}[{f_2}]}^*}\, {a_{\sigma }}[{f_1}]\, {{{a_{-\sigma }}[{f_2}]}^*}\, {a_{-\sigma }}[{f_1}]\, {{({d_{-\sigma }})}^*}\, {d_{-\sigma
}}
}
\dispSFPrintmath{
{R_{81}}  &=& {{{a_{\sigma }}[{f_2}]}^*}\, {{{a_{\sigma }}[{f_1}]}^*}\, {a_{\sigma }}[{f_2}]\, {{{a_{-\sigma }}[{f_1}]}^*}\, {a_{-\sigma }}[{f_1}]\, {a_{-\sigma
}}[{f_2}]\, {d_{\sigma }}\, {{({d_{-\sigma }})}^*}+  
\\ &&
\hspace{2.em} {{{a_{\sigma }}[{f_2}]}^*}\, {a_{\sigma }}[{f_1}]\, {a_{\sigma }}[{f_2}]\, {{{a_{-\sigma }}[{f_2}]}^*}\, {{{a_{-\sigma }}[{f_1}]}^*}\, {a_{-\sigma
}}[{f_1}]\, {{({d_{\sigma }})}^*}\, {d_{-\sigma }}
}
\end{eqnarray*}
\begin{eqnarray*}
\dispSFPrintmath{
{R_{82}}  &=&  {{{a_{\sigma }}[{f_1}]}^*}\, {a_{\sigma }}[{f_2}]\, {{{a_{-\sigma }}[{f_2}]}^*}\, {{{a_{-\sigma }}[{f_1}]}^*}\, {a_{-\sigma }}[{f_1}]\, {a_{-\sigma
}}[{f_2}]\, {{({d_{\sigma }})}^*}\, {d_{\sigma }}+  
\\ &&
\hspace{2.em} {{{a_{\sigma }}[{f_2}]}^*}\, {a_{\sigma }}[{f_1}]\, {{{a_{-\sigma }}[{f_2}]}^*}\, {{{a_{-\sigma }}[{f_1}]}^*}\, {a_{-\sigma }}[{f_1}]\, {a_{-\sigma
}}[{f_2}]\, {{({d_{\sigma }})}^*}\, {d_{\sigma }}
}
\dispSFPrintmath{
{R_{83}}  &=&  {{{a_{\sigma }}[{f_1}]}^*}\, {a_{\sigma }}[{f_2}]\, {{{a_{-\sigma }}[{f_1}]}^*}\, {a_{-\sigma
}}[{f_1}]\, {{({d_{\sigma }})}^*}\, {d_{\sigma }}\, {{({d_{-\sigma }})}^*}\, {d_{-\sigma }}+  
\\ &&
\hspace{2.em} {{{a_{\sigma }}[{f_2}]}^*}\, {a_{\sigma }}[{f_1}]\, {{{a_{-\sigma }}[{f_1}]}^*}\, {a_{-\sigma }}[{f_1}]\, {{({d_{\sigma }})}^*}\, {d_{\sigma
}}\, {{({d_{-\sigma }})}^*}\, {d_{-\sigma }}
}
\dispSFPrintmath{
{R_{84}}  &=&  {{{a_{\sigma }}[{f_1}]}^*}\, {a_{\sigma }}[{f_2}]\, {{{a_{-\sigma }}[{f_2}]}^*}\, {{{a_{-\sigma
}}[{f_1}]}^*}\,   
\\ &&
\hspace{3.em} {a_{-\sigma }}[{f_1}]\, {a_{-\sigma }}[{f_2}]\, {{({d_{-\sigma }})}^*}\, {d_{-\sigma }}+{{{a_{\sigma }}[{f_2}]}^*}\, {a_{\sigma }}[{f_1}]\, 
 \\ &&
\hspace{3.em} {{{a_{-\sigma }}[{f_2}]}^*}\, {{{a_{-\sigma }}[{f_1}]}^*}\, {a_{-\sigma }}[{f_1}]\, {a_{-\sigma }}[{f_2}]\, {{({d_{-\sigma }})}^*}\, {d_{-\sigma
}}
}
\dispSFPrintmath{
{R_{85}}  &=& -{{{a_{-\sigma }}[{f_2}]}^*}\, {{{a_{-\sigma }}[{f_1}]}^*}\, {a_{-\sigma }}[{f_1}]\, {a_{-\sigma }}[{f_2}]\, {{({d_{\sigma }})}^*}\, {d_{\sigma
}}\, {{({d_{-\sigma }})}^*}\, {d_{-\sigma }}+  
\\ &&
\hspace{2.em} {{{a_{\sigma }}[{f_2}]}^*}\, {{{a_{\sigma }}[{f_1}]}^*}\, {a_{\sigma }}[{f_1}]\, {a_{\sigma }}[{f_2}]\, {{({d_{\sigma }})}^*}\, {d_{\sigma
}}\, {{({d_{-\sigma }})}^*}\, {d_{-\sigma }}
}
\dispSFPrintmath{
{R_{86}}  &=&  {{{a_{\sigma }}[{f_1}]}^*}\, {{{a_{-\sigma }}[{f_2}]}^*}\, {{{a_{-\sigma }}[{f_1}]}^*}\, {a_{-\sigma
}}[{f_2}]\, {d_{\sigma }}\, {d_{-\sigma }}+  
\\ &&
\hspace{2.em} {a_{\sigma }}[{f_1}]\, {{{a_{-\sigma }}[{f_2}]}^*}\, {a_{-\sigma }}[{f_1}]\, {a_{-\sigma }}[{f_2}]\, {{({d_{\sigma }})}^*}\, {{({d_{-\sigma
}})}^*}
}
\dispSFPrintmath{
{R_{87}}  &=& {{{a_{\sigma }}[{f_2}]}^*}\, {{{a_{\sigma }}[{f_1}]}^*}\, {a_{\sigma }}[{f_2}]\, {{{a_{-\sigma }}[{f_2}]}^*}\, {{{a_{-\sigma }}[{f_1}]}^*}\, {a_{-\sigma
}}[{f_1}]\, {d_{\sigma }}\, {d_{-\sigma }}+  
\\ &&
\hspace{2.em} {{{a_{\sigma }}[{f_2}]}^*}\, {a_{\sigma }}[{f_1}]\, {a_{\sigma }}[{f_2}]\, {{{a_{-\sigma }}[{f_1}]}^*}\,   
\\ &&
\hspace{3.em} {a_{-\sigma }}[{f_1}]\, {a_{-\sigma }}[{f_2}]\, {{({d_{\sigma }})}^*}\, {{({d_{-\sigma }})}^*}
}
\dispSFPrintmath{
{R_{88}}  &=&  -{{{a_{\sigma }}[{f_1}]}^*}\, {a_{\sigma }}[{f_2}]\, {{{a_{-\sigma }}[{f_2}]}^*}\, {{{a_{-\sigma
}}[{f_1}]}^*}\, {a_{-\sigma }}[{f_2}]\, {d_{-\sigma }}-  
\\ &&
\hspace{2.em} {{{a_{\sigma }}[{f_2}]}^*}\, {{{a_{\sigma }}[{f_1}]}^*}\, {a_{\sigma }}[{f_2}]\, {{{a_{-\sigma }}[{f_1}]}^*}\, {a_{-\sigma }}[{f_2}]\, {d_{\sigma
}}+  
\\ &&
\hspace{2.em} {{{a_{\sigma }}[{f_2}]}^*}\, {a_{\sigma }}[{f_1}]\, {{{a_{-\sigma }}[{f_2}]}^*}\, {a_{-\sigma }}[{f_1}]\, {a_{-\sigma }}[{f_2}]\, {{({d_{-\sigma
}})}^*}+  
\\ &&
\hspace{2.em} {{{a_{\sigma }}[{f_2}]}^*}\, {a_{\sigma }}[{f_1}]\, {a_{\sigma }}[{f_2}]\, {{{a_{-\sigma }}[{f_2}]}^*}\, {a_{-\sigma }}[{f_1}]\, {{({d_{\sigma
}})}^*}
}
\dispSFPrintmath{
{R_{89}}  &=& {{{a_{-\sigma }}[{f_2}]}^*}\, {{{a_{-\sigma }}[{f_1}]}^*}\, {a_{-\sigma }}[{f_1}]\, {a_{-\sigma }}[{f_2}]\, {{({d_{\sigma }})}^*}\, {d_{\sigma
}}\, {{({d_{-\sigma }})}^*}\, {d_{-\sigma }}-  
\\ &&
\hspace{2.em} {{{a_{\sigma }}[{f_2}]}^*}\, {a_{\sigma }}[{f_2}]\, {{{a_{-\sigma }}[{f_2}]}^*}\,   
\\ &&
\hspace{3.em} {{{a_{-\sigma }}[{f_1}]}^*}\, {a_{-\sigma }}[{f_1}]\, {a_{-\sigma }}[{f_2}]\, {{({d_{-\sigma }})}^*}\, {d_{-\sigma }}
}
\dispSFPrintmath{
{R_{90}}  &=&  -{{{a_{\sigma }}[{f_2}]}^*}\, {a_{\sigma }}[{f_2}]\, {{{a_{-\sigma }}[{f_1}]}^*}\, {a_{-\sigma
}}[{f_1}]\, {{({d_{\sigma }})}^*}\, {d_{\sigma }}\, {{({d_{-\sigma }})}^*}\, {d_{-\sigma }}+  
\\ &&
\hspace{2.em} {{{a_{\sigma }}[{f_2}]}^*}\, {a_{\sigma }}[{f_2}]\, {{{a_{-\sigma }}[{f_2}]}^*}\, {{{a_{-\sigma }}[{f_1}]}^*}\, {a_{-\sigma }}[{f_1}]\, {a_{-\sigma
}}[{f_2}]\, {{({d_{\sigma }})}^*}\, {d_{\sigma }}
}
\dispSFPrintmath{
{R_{91}}  &=& -{{{a_{\sigma }}[{f_1}]}^*}\, {a_{\sigma }}[{f_2}]\, {{{a_{-\sigma }}[{f_2}]}^*}\, {{{a_{-\sigma }}[{f_1}]}^*}\, {a_{-\sigma }}[{f_2}]\, {{({d_{\sigma
}})}^*}\, {d_{\sigma }}\, {d_{-\sigma }}+  
\\ &&
\hspace{2.em} {{{a_{\sigma }}[{f_2}]}^*}\, {a_{\sigma }}[{f_1}]\, {{{a_{-\sigma }}[{f_2}]}^*}\, {a_{-\sigma }}[{f_1}]\, {a_{-\sigma }}[{f_2}]\, {{({d_{\sigma
}})}^*}\, {d_{\sigma }}\, {{({d_{-\sigma }})}^*}
}
\dispSFPrintmath{
{R_{92}}  &=& {{{a_{\sigma }}[{f_2}]}^*}\, {{{a_{\sigma }}[{f_1}]}^*}\, {a_{\sigma }}[{f_2}]\, {{{a_{-\sigma }}[{f_1}]}^*}\, {a_{-\sigma }}[{f_1}]\, {d_{\sigma
}}\, {{({d_{-\sigma }})}^*}\, {d_{-\sigma }}-  
\\ &&
\hspace{2.em} {{{a_{\sigma }}[{f_2}]}^*}\, {a_{\sigma }}[{f_1}]\, {a_{\sigma }}[{f_2}]\, {{{a_{-\sigma }}[{f_1}]}^*}\, {a_{-\sigma }}[{f_1}]\, {{({d_{\sigma
}})}^*}\, {{({d_{-\sigma }})}^*}\, {d_{-\sigma }}
}
\dispSFPrintmath{
{R_{93}}  &=&  {{{a_{\sigma }}[{f_2}]}^*}\, {{{a_{\sigma }}[{f_1}]}^*}\, {a_{\sigma }}[{f_2}]\, {{{a_{-\sigma
}}[{f_2}]}^*}\, {{{a_{-\sigma }}[{f_1}]}^*}\, {a_{-\sigma }}[{f_1}]\,   
\\ &&
\hspace{3.em} {a_{-\sigma }}[{f_2}]\, {d_{\sigma }}\, {{({d_{-\sigma }})}^*}\, {d_{-\sigma }}-{{{a_{\sigma }}[{f_2}]}^*}\, {a_{\sigma }}[{f_1}]\, {a_{\sigma
}}[{f_2}]\,   
\\ &&
\hspace{3.em} {{{a_{-\sigma }}[{f_2}]}^*}\, {{{a_{-\sigma }}[{f_1}]}^*}\, {a_{-\sigma }}[{f_1}]\, {a_{-\sigma }}[{f_2}]\, {{({d_{\sigma }})}^*}\, {{({d_{-\sigma
}})}^*}\, {d_{-\sigma }}
}
\dispSFPrintmath{
{R_{94}}  &=&  -{{{a_{\sigma }}[{f_1}]}^*}\, {a_{\sigma }}[{f_2}]\, {{{a_{-\sigma }}[{f_2}]}^*}\, {{{a_{-\sigma }}[{f_1}]}^*}\, {a_{-\sigma }}[{f_1}]\, {{({d_{\sigma
}})}^*}\, {d_{\sigma }}\, {d_{-\sigma }}+  
\\ &&
\hspace{2.em} {{{a_{\sigma }}[{f_2}]}^*}\, {a_{\sigma }}[{f_1}]\, {{{a_{-\sigma }}[{f_1}]}^*}\, {a_{-\sigma }}[{f_1}]\, {a_{-\sigma }}[{f_2}]\, {{({d_{\sigma
}})}^*}\, {d_{\sigma }}\, {{({d_{-\sigma }})}^*}
}
\dispSFPrintmath{
{R_{95}}  &=& -{{{a_{\sigma }}[{f_1}]}^*}\, {a_{\sigma }}[{f_2}]\, {{{a_{-\sigma }}[{f_2}]}^*}\, {a_{-\sigma }}[{f_1}]\, {a_{-\sigma }}[{f_2}]\, {{({d_{\sigma
}})}^*}\, {d_{\sigma }}\, {{({d_{-\sigma }})}^*}+  
\\ &&
\hspace{2.em} {{{a_{\sigma }}[{f_2}]}^*}\, {a_{\sigma }}[{f_1}]\, {{{a_{-\sigma }}[{f_2}]}^*}\, {{{a_{-\sigma }}[{f_1}]}^*}\, {a_{-\sigma }}[{f_2}]\, {{({d_{\sigma
}})}^*}\, {d_{\sigma }}\, {d_{-\sigma }}
}
\dispSFPrintmath{{R_{96}}  &=&  {{{a_{\sigma }}[{f_2}]}^*}\, {a_{\sigma }}[{f_2}]\, {{{a_{-\sigma }}[{f_1}]}^*}\, {a_{-\sigma
}}[{f_1}]\, {{({d_{\sigma }})}^*}\, {d_{\sigma }}\, {{({d_{-\sigma }})}^*}\, {d_{-\sigma }}}
\end{eqnarray*}
\begin{eqnarray*}
\dispSFPrintmath{
{R_{97}}  &=&  {{{a_{\sigma }}[{f_1}]}^*}\, {a_{\sigma }}[{f_2}]\, {{{a_{-\sigma }}[{f_2}]}^*}\, {{{a_{-\sigma
}}[{f_1}]}^*}\, {a_{-\sigma }}[{f_1}]\, {a_{-\sigma }}[{f_2}]\,   
\\ &&
\hspace{3.em} {{({d_{\sigma }})}^*}\, {d_{\sigma }}\, {{({d_{-\sigma }})}^*}\, {d_{-\sigma }}+{{{a_{\sigma }}[{f_2}]}^*}\, {a_{\sigma }}[{f_1}]\, {{{a_{-\sigma
}}[{f_2}]}^*}\,   
\\ &&
\hspace{3.em} {{{a_{-\sigma }}[{f_1}]}^*}\, {a_{-\sigma }}[{f_1}]\, {a_{-\sigma }}[{f_2}]\, {{({d_{\sigma }})}^*}\, {d_{\sigma }}\, {{({d_{-\sigma }})}^*}\, {d_{-\sigma
}}
}
\dispSFPrintmath{
{R_{98}}  &=&  {{{a_{\sigma }}[{f_2}]}^*}\, {a_{\sigma }}[{f_2}]\, {{{a_{-\sigma }}[{f_2}]}^*}\,   
\\ &&
\hspace{2.em} {{{a_{-\sigma }}[{f_1}]}^*}\, {a_{-\sigma }}[{f_1}]\, {a_{-\sigma }}[{f_2}]\, {{({d_{\sigma }})}^*}\, {d_{\sigma }}\, {{({d_{-\sigma }})}^*}\, {d_{-\sigma
}}
}
\dispSFPrintmath{
{R_{99}}  &=&  -{{{a_{\sigma }}[{f_1}]}^*}\, {a_{\sigma }}[{f_2}]\, {{{a_{-\sigma }}[{f_2}]}^*}\, {{{a_{-\sigma
}}[{f_1}]}^*}\, {a_{-\sigma }}[{f_2}]\, {d_{-\sigma }}+  
\\ &&
\hspace{2.em} {{{a_{\sigma }}[{f_2}]}^*}\, {a_{\sigma }}[{f_1}]\, {{{a_{-\sigma }}[{f_2}]}^*}\, {a_{-\sigma }}[{f_1}]\, {a_{-\sigma }}[{f_2}]\, {{({d_{-\sigma
}})}^*}
}
\dispSFPrintmath{
{R_{100}}  &=&  {{{a_{\sigma }}[{f_2}]}^*}\, {{{a_{\sigma }}[{f_1}]}^*}\, {a_{\sigma }}[{f_2}]\, {{{a_{-\sigma
}}[{f_2}]}^*}\,   
\\ &&
\hspace{3.em} {{{a_{-\sigma }}[{f_1}]}^*}\, {a_{-\sigma }}[{f_1}]\, {a_{-\sigma }}[{f_2}]\, {d_{\sigma }}-{{{a_{\sigma }}[{f_2}]}^*}\, {a_{\sigma }}[{f_1}]\, 
 \\ &&
\hspace{3.em} {a_{\sigma }}[{f_2}]\, {{{a_{-\sigma }}[{f_2}]}^*}\, {{{a_{-\sigma }}[{f_1}]}^*}\, {a_{-\sigma }}[{f_1}]\, {a_{-\sigma }}[{f_2}]\, {{({d_{\sigma
}})}^*}
}
\dispSFPrintmath{
{R_{101}}  &=& {{{a_{\sigma }}[{f_2}]}^*}\, {{{a_{\sigma }}[{f_1}]}^*}\, {a_{\sigma }}[{f_2}]\, {{{a_{-\sigma }}[{f_2}]}^*}\, {{{a_{-\sigma }}[{f_1}]}^*}\, {a_{-\sigma
}}[{f_2}]\, {d_{\sigma }}\, {d_{-\sigma }}+  
\\ &&
\hspace{2.em} {{{a_{\sigma }}[{f_2}]}^*}\, {a_{\sigma }}[{f_1}]\, {a_{\sigma }}[{f_2}]\, {{{a_{-\sigma }}[{f_2}]}^*}\,   
\\ &&
\hspace{3.em} {a_{-\sigma }}[{f_1}]\, {a_{-\sigma }}[{f_2}]\, {{({d_{\sigma }})}^*}\, {{({d_{-\sigma }})}^*}
}
\dispSFPrintmath{{R_{102}}  &=&  {{{a_{\sigma }}[{f_1}]}^*}\, {a_{-\sigma }}[{f_1}]\, {d_{\sigma }}\, {{({d_{-\sigma
}})}^*}+{a_{\sigma }}[{f_1}]\, {{{a_{-\sigma }}[{f_1}]}^*}\, {{({d_{\sigma }})}^*}\, {d_{-\sigma }}}
\dispSFPrintmath{{R_{103}}  &=&  {{{a_{-\sigma }}[{f_1}]}^*}\, {{({d_{\sigma }})}^*}\, {d_{\sigma }}\, {d_{-\sigma
}}-{a_{-\sigma }}[{f_1}]\, {{({d_{\sigma }})}^*}\, {d_{\sigma }}\, {{({d_{-\sigma }})}^*}}
\dispSFPrintmath{
{R_{104}}  &=&  {{{a_{\sigma }}[{f_1}]}^*}\, {{{a_{-\sigma }}[{f_2}]}^*}\, {a_{-\sigma }}[{f_1}]\, {d_{\sigma
}}\, {{({d_{-\sigma }})}^*}\, {d_{-\sigma }}-  
\\ &&
\hspace{2.em} {{{a_{\sigma }}[{f_1}]}^*}\, {a_{\sigma }}[{f_2}]\, {a_{-\sigma }}[{f_1}]\, {{({d_{\sigma }})}^*}\, {d_{\sigma }}\, {{({d_{-\sigma }})}^*}+
 \\ &&
\hspace{2.em} {{{a_{\sigma }}[{f_2}]}^*}\, {a_{\sigma }}[{f_1}]\, {{{a_{-\sigma }}[{f_1}]}^*}\, {{({d_{\sigma }})}^*}\, {d_{\sigma }}\, {d_{-\sigma }}-
 \\ &&
\hspace{2.em} {a_{\sigma }}[{f_1}]\, {{{a_{-\sigma }}[{f_1}]}^*}\, {a_{-\sigma }}[{f_2}]\, {{({d_{\sigma }})}^*}\, {{({d_{-\sigma }})}^*}\, {d_{-\sigma
}}
}
\dispSFPrintmath{
{R_{105}}  &=&  {{{a_{-\sigma }}[{f_1}]}^*}\, {a_{-\sigma }}[{f_2}]\, {{({d_{\sigma }})}^*}\, {d_{\sigma }}+{{{a_{-\sigma
}}[{f_2}]}^*}\, {a_{-\sigma }}[{f_1}]\, {{({d_{\sigma }})}^*}\, {d_{\sigma }}+  
\\ &&
\hspace{2.em} {{{a_{\sigma }}[{f_2}]}^*}\, {a_{-\sigma }}[{f_1}]\, {d_{\sigma }}\, {{({d_{-\sigma }})}^*}+{a_{\sigma }}[{f_2}]\, {{{a_{-\sigma }}[{f_1}]}^*}\, {{({d_{\sigma
}})}^*}\, {d_{-\sigma }}
}
\dispSFPrintmath{
{R_{106}}  &=&  {{{a_{\sigma }}[{f_1}]}^*}\, {{{a_{-\sigma }}[{f_2}]}^*}\, {a_{-\sigma }}[{f_1}]\, {a_{-\sigma
}}[{f_2}]\, {d_{\sigma }}\, {{({d_{-\sigma }})}^*}+  
\\ &&
\hspace{2.em} {{{a_{\sigma }}[{f_1}]}^*}\, {a_{\sigma }}[{f_2}]\, {{{a_{-\sigma }}[{f_2}]}^*}\, {a_{-\sigma }}[{f_1}]\, {{({d_{-\sigma }})}^*}\, {d_{-\sigma
}}+  
\\ &&
\hspace{2.em} {{{a_{\sigma }}[{f_1}]}^*}\, {a_{\sigma }}[{f_2}]\, {{{a_{-\sigma }}[{f_2}]}^*}\, {a_{-\sigma }}[{f_1}]\, {{({d_{\sigma }})}^*}\, {d_{\sigma
}}+  
\\ &&
\hspace{2.em} {{{a_{\sigma }}[{f_2}]}^*}\, {{{a_{\sigma }}[{f_1}]}^*}\, {a_{\sigma }}[{f_2}]\, {a_{-\sigma }}[{f_1}]\, {d_{\sigma }}\, {{({d_{-\sigma
}})}^*}+  
\\ &&
\hspace{2.em} {{{a_{\sigma }}[{f_2}]}^*}\, {a_{\sigma }}[{f_1}]\, {{{a_{-\sigma }}[{f_1}]}^*}\, {a_{-\sigma }}[{f_2}]\, {{({d_{-\sigma }})}^*}\, {d_{-\sigma
}}+  
\\ &&
\hspace{2.em} {{{a_{\sigma }}[{f_2}]}^*}\, {a_{\sigma }}[{f_1}]\, {{{a_{-\sigma }}[{f_1}]}^*}\, {a_{-\sigma }}[{f_2}]\, {{({d_{\sigma }})}^*}\, {d_{\sigma
}}+  
\\ &&
\hspace{2.em} {{{a_{\sigma }}[{f_2}]}^*}\, {a_{\sigma }}[{f_1}]\, {a_{\sigma }}[{f_2}]\, {{{a_{-\sigma }}[{f_1}]}^*}\, {{({d_{\sigma }})}^*}\, {d_{-\sigma
}}+  
\\ &&
\hspace{2.em} {a_{\sigma }}[{f_1}]\, {{{a_{-\sigma }}[{f_2}]}^*}\, {{{a_{-\sigma }}[{f_1}]}^*}\, {a_{-\sigma }}[{f_2}]\, {{({d_{\sigma }})}^*}\, {d_{-\sigma
}}
}
\dispSFPrintmath{
{R_{107}}  &=&  {{{a_{\sigma }}[{f_2}]}^*}\, {{{a_{-\sigma }}[{f_1}]}^*}\, {a_{-\sigma }}[{f_2}]\, {d_{\sigma
}}\, {{({d_{-\sigma }})}^*}\, {d_{-\sigma }}-  
\\ &&
\hspace{2.em} {a_{\sigma }}[{f_2}]\, {{{a_{-\sigma }}[{f_2}]}^*}\, {a_{-\sigma }}[{f_1}]\, {{({d_{\sigma }})}^*}\, {{({d_{-\sigma }})}^*}\, {d_{-\sigma
}}
}
\dispSFPrintmath{
{R_{108}}  &=&  {{{a_{\sigma }}[{f_2}]}^*}\, {a_{\sigma }}[{f_2}]\, {{{a_{-\sigma }}[{f_1}]}^*}\, {{({d_{\sigma
}})}^*}\, {d_{\sigma }}\, {d_{-\sigma }}-  
\\ &&
\hspace{2.em} {{{a_{\sigma }}[{f_2}]}^*}\, {a_{\sigma }}[{f_2}]\, {a_{-\sigma }}[{f_1}]\, {{({d_{\sigma }})}^*}\, {d_{\sigma }}\, {{({d_{-\sigma }})}^*}
}
\dispSFPrintmath{
{R_{109}}  &=&  {{{a_{\sigma }}[{f_2}]}^*}\, {a_{\sigma }}[{f_2}]\, {{{a_{-\sigma }}[{f_1}]}^*}\, {a_{-\sigma
}}[{f_2}]\, {{({d_{-\sigma }})}^*}\, {d_{-\sigma }}+  
\\ &&
\hspace{2.em} {{{a_{\sigma }}[{f_2}]}^*}\, {a_{\sigma }}[{f_2}]\, {{{a_{-\sigma }}[{f_2}]}^*}\, {a_{-\sigma }}[{f_1}]\, {{({d_{-\sigma }})}^*}\, {d_{-\sigma
}}
}
\dispSFPrintmath{
{R_{110}}  &=&  {{{a_{\sigma }}[{f_2}]}^*}\, {a_{\sigma }}[{f_2}]\, {{{a_{-\sigma }}[{f_1}]}^*}\, {a_{-\sigma
}}[{f_2}]\, {{({d_{\sigma }})}^*}\, {d_{\sigma }}+  
\\ &&
\hspace{2.em} {{{a_{\sigma }}[{f_2}]}^*}\, {a_{\sigma }}[{f_2}]\, {{{a_{-\sigma }}[{f_2}]}^*}\, {a_{-\sigma }}[{f_1}]\, {{({d_{\sigma }})}^*}\, {d_{\sigma
}}
}
\end{eqnarray*}
\begin{eqnarray*}
\dispSFPrintmath{{R_{111}}  &=&  {{{a_{\sigma }}[{f_1}]}^*}\, {a_{\sigma }}[{f_1}]\, {{({d_{-\sigma }})}^*}\, {d_{-\sigma
}}}
\dispSFPrintmath{
{R_{112}}  &=&  {{{a_{\sigma }}[{f_1}]}^*}\, {a_{\sigma }}[{f_1}]\, {{{a_{-\sigma }}[{f_2}]}^*}\, {{({d_{\sigma
}})}^*}\, {d_{\sigma }}\, {d_{-\sigma }}-  
\\ &&
\hspace{2.em} {{{a_{\sigma }}[{f_1}]}^*}\, {a_{\sigma }}[{f_1}]\, {a_{-\sigma }}[{f_2}]\, {{({d_{\sigma }})}^*}\, {d_{\sigma }}\, {{({d_{-\sigma }})}^*}
}
\dispSFPrintmath{
{R_{113}}  &=&  {{{a_{\sigma }}[{f_1}]}^*}\, {a_{\sigma }}[{f_1}]\, {a_{\sigma }}[{f_2}]\, {{{a_{-\sigma }}[{f_2}]}^*}\, {{({d_{\sigma
}})}^*}\, {d_{-\sigma }}+  
\\ &&
\hspace{2.em} {{{a_{\sigma }}[{f_2}]}^*}\, {{{a_{\sigma }}[{f_1}]}^*}\, {a_{\sigma }}[{f_1}]\, {a_{-\sigma }}[{f_2}]\, {d_{\sigma }}\, {{({d_{-\sigma
}})}^*}
}
\dispSFPrintmath{
{R_{114}}  &=&  {{{a_{\sigma }}[{f_1}]}^*}\, {a_{\sigma }}[{f_1}]\, {{({d_{\sigma }})}^*}\, {d_{\sigma }}\, {{({d_{-\sigma
}})}^*}\, {d_{-\sigma }}-  
\\ &&
\hspace{2.em} {{{a_{\sigma }}[{f_1}]}^*}\, {a_{\sigma }}[{f_1}]\, {{{a_{-\sigma }}[{f_2}]}^*}\, {a_{-\sigma }}[{f_2}]\, {{({d_{\sigma }})}^*}\, {d_{\sigma
}}
}
\dispSFPrintmath{{R_{115}}  &=&  \int {{{a_{\sigma }}[{{\delta }_k}]}^*}\, {a_{\sigma }}[{{\delta }_k}]
\epsilon [k] \, dk}
\dispSFPrintmath{{R_{116}}  &=&  -{{{a_{\sigma }}[\epsilon [{f_2}]]}^*}\, {d_{\sigma }}\, {{({d_{-\sigma
}})}^*}\, {d_{-\sigma }}+{a_{\sigma }}[\epsilon [{f_2}]]\, {{({d_{\sigma }})}^*}\, {{({d_{-\sigma }})}^*}\, {d_{-\sigma }}}
\dispSFPrintmath{
{R_{117}}  &=& {{{a_{\sigma }}[{f_2}]}^*}\, {a_{\sigma }}[\epsilon [{f_2}]]\, {{({d_{-\sigma }})}^*}\, {d_{-\sigma }}+{{{a_{\sigma }}[\epsilon [{f_2}]]}^*}\, {a_{-\sigma
}}[{f_2}]\, {d_{\sigma }}\, {{({d_{-\sigma }})}^*}+  
\\ &&
\hspace{2.em} {{{a_{\sigma }}[\epsilon [{f_2}]]}^*}\, {a_{\sigma }}[{f_2}]\, {{({d_{-\sigma }})}^*}\, {d_{-\sigma }}+{a_{\sigma }}[\epsilon [{f_2}]]\, {{{a_{-\sigma
}}[{f_2}]}^*}\, {{({d_{\sigma }})}^*}\, {d_{-\sigma }}
}
\dispSFPrintmath{
{R_{118}}  &=&  {{{a_{\sigma }}[{f_2}]}^*}\, {a_{\sigma }}[\epsilon [{f_2}]]\, {{{a_{-\sigma }}[{f_2}]}^*}\, {{({d_{\sigma
}})}^*}\, {d_{\sigma }}\, {d_{-\sigma }}-  
\\ &&
\hspace{2.em} {{{a_{\sigma }}[\epsilon [{f_2}]]}^*}\, {a_{\sigma }}[{f_2}]\, {a_{-\sigma }}[{f_2}]\, {{({d_{\sigma }})}^*}\, {d_{\sigma }}\, {{({d_{-\sigma
}})}^*}
}
\dispSFPrintmath{
{R_{119}}  &=&  -{{{a_{\sigma }}[{f_2}]}^*}\, {a_{\sigma }}[\epsilon [{f_2}]]\, {a_{-\sigma }}[{f_2}]\, {{({d_{\sigma
}})}^*}\, {d_{\sigma }}\, {{({d_{-\sigma }})}^*}+  
\\ &&
\hspace{2.em} {{{a_{\sigma }}[\epsilon [{f_2}]]}^*}\, {{{a_{-\sigma }}[{f_2}]}^*}\, {a_{-\sigma }}[{f_2}]\, {d_{\sigma }}\, {{({d_{-\sigma }})}^*}\, {d_{-\sigma
}}+  
\\ &&
\hspace{2.em} {{{a_{\sigma }}[\epsilon [{f_2}]]}^*}\, {a_{\sigma }}[{f_2}]\, {{{a_{-\sigma }}[{f_2}]}^*}\, {{({d_{\sigma }})}^*}\, {d_{\sigma }}\, {d_{-\sigma
}}-  
\\ &&
\hspace{2.em} {a_{\sigma }}[\epsilon [{f_2}]]\, {{{a_{-\sigma }}[{f_2}]}^*}\, {a_{-\sigma }}[{f_2}]\, {{({d_{\sigma }})}^*}\, {{({d_{-\sigma }})}^*}\, {d_{-\sigma
}}
}
\dispSFPrintmath{
{R_{120}}  &=&  {{{a_{\sigma }}[{f_2}]}^*}\, {{{a_{\sigma }}[\epsilon [{f_2}]]}^*}\, {a_{\sigma }}[{f_2}]\, {a_{-\sigma
}}[{f_2}]\, {d_{\sigma }}\, {{({d_{-\sigma }})}^*}+  
\\ &&
\hspace{2.em} {{{a_{\sigma }}[{f_2}]}^*}\, {a_{\sigma }}[\epsilon [{f_2}]]\, {a_{\sigma }}[{f_2}]\, {{{a_{-\sigma }}[{f_2}]}^*}\, {{({d_{\sigma }})}^*}\, {d_{-\sigma
}}
}
\dispSFPrintmath{
{R_{121}}  &=&  -{{{a_{\sigma }}[{f_2}]}^*}\, {a_{\sigma }}[\epsilon [{f_2}]]\, {{({d_{\sigma }})}^*}\, {d_{\sigma
}}\, {{({d_{-\sigma }})}^*}\, {d_{-\sigma }}+  
\\ &&
\hspace{2.em} {{{a_{\sigma }}[{f_2}]}^*}\, {a_{\sigma }}[\epsilon [{f_2}]]\, {{{a_{-\sigma }}[{f_2}]}^*}\, {a_{-\sigma }}[{f_2}]\, {{({d_{\sigma }})}^*}\, {d_{\sigma
}}-  
\\ &&
\hspace{2.em} {{{a_{\sigma }}[\epsilon [{f_2}]]}^*}\, {a_{\sigma }}[{f_2}]\, {{({d_{\sigma }})}^*}\, {d_{\sigma }}\, {{({d_{-\sigma }})}^*}\, {d_{-\sigma
}}+  
\\ &&
\hspace{2.em} {{{a_{\sigma }}[\epsilon [{f_2}]]}^*}\, {a_{\sigma }}[{f_2}]\, {{{a_{-\sigma }}[{f_2}]}^*}\, {a_{-\sigma }}[{f_2}]\, {{({d_{\sigma }})}^*}\, {d_{\sigma
}}
}
\dispSFPrintmath{
{R_{122}}  &=&  {{{a_{\sigma }}[{f_2}]}^*}\, {a_{\sigma }}[\epsilon [{f_2}]]\, {{{a_{-\sigma }}[{f_2}]}^*}\, {a_{-\sigma
}}[{f_2}]\, {{({d_{-\sigma }})}^*}\, {d_{-\sigma }}+  
\\ &&
\hspace{2.em} {{{a_{\sigma }}[\epsilon [{f_2}]]}^*}\, {a_{\sigma }}[{f_2}]\, {{{a_{-\sigma }}[{f_2}]}^*}\, {a_{-\sigma }}[{f_2}]\, {{({d_{-\sigma }})}^*}\, {d_{-\sigma
}}
}
\dispSFPrintmath{
{R_{123}}  &=&  -{{{a_{\sigma }}[{f_2}]}^*}\, {{{a_{\sigma }}[\epsilon [{f_2}]]}^*}\, {a_{\sigma }}[{f_2}]\, {d_{\sigma
}}\, {{({d_{-\sigma }})}^*}\, {d_{-\sigma }}+  
\\ &&
\hspace{2.em} {{{a_{\sigma }}[{f_2}]}^*}\, {a_{\sigma }}[\epsilon [{f_2}]]\, {a_{\sigma }}[{f_2}]\, {{({d_{\sigma }})}^*}\, {{({d_{-\sigma }})}^*}\, {d_{-\sigma
}}+  
\\ &&
\hspace{2.em} {{{a_{\sigma }}[{f_2}]}^*}\, {{{a_{\sigma }}[\epsilon [{f_2}]]}^*}\, {a_{\sigma }}[{f_2}]\, {{{a_{-\sigma }}[{f_2}]}^*}\, {a_{-\sigma }}[{f_2}]\, {d_{\sigma
}}\, {{({d_{-\sigma }})}^*}\, {d_{-\sigma }}-  
\\ &&
\hspace{2.em} {{{a_{\sigma }}[{f_2}]}^*}\, {a_{\sigma }}[\epsilon [{f_2}]]\, {a_{\sigma }}[{f_2}]\,   
\\ &&
\hspace{3.em} {{{a_{-\sigma }}[{f_2}]}^*}\, {a_{-\sigma }}[{f_2}]\, {{({d_{\sigma }})}^*}\, {{({d_{-\sigma }})}^*}\, {d_{-\sigma }}
}
\dispSFPrintmath{
{R_{124}}  &=&  {{{a_{\sigma }}[{f_2}]}^*}\, {a_{\sigma }}[\epsilon [{f_2}]]\, {a_{-\sigma }}[{f_2}]\, {{({d_{\sigma
}})}^*}\, {d_{\sigma }}\, {{({d_{-\sigma }})}^*}-  
\\ &&
\hspace{2.em} {{{a_{\sigma }}[\epsilon [{f_2}]]}^*}\, {a_{\sigma }}[{f_2}]\, {{{a_{-\sigma }}[{f_2}]}^*}\, {{({d_{\sigma }})}^*}\, {d_{\sigma }}\, {d_{-\sigma
}}
}
\dispSFPrintmath{
{R_{125}}  &=& {{{a_{\sigma }}[{f_2}]}^*}\, {a_{\sigma }}[\epsilon [{f_2}]]\, {{{a_{-\sigma }}[{f_2}]}^*}\, {a_{-\sigma }}[{f_2}]\, {{({d_{\sigma }})}^*}\, {d_{\sigma
}}\, {{({d_{-\sigma }})}^*}\, {d_{-\sigma }}+  
\\ &&
\hspace{2.em} {{{a_{\sigma }}[\epsilon [{f_2}]]}^*}\, {a_{\sigma }}[{f_2}]\, {{{a_{-\sigma }}[{f_2}]}^*}\, {a_{-\sigma }}[{f_2}]\, {{({d_{\sigma }})}^*}\, {d_{\sigma
}}\, {{({d_{-\sigma }})}^*}\, {d_{-\sigma }}
}
\dispSFPrintmath{
{R_{126}}  &=&  {{{a_{\sigma }}[{f_2}]}^*}\, {a_{\sigma }}[\epsilon [{f_2}]]\, {{({d_{\sigma }})}^*}\, {d_{\sigma
}}\, {{({d_{-\sigma }})}^*}\, {d_{-\sigma }}+  
\\ &&
\hspace{2.em} {{{a_{\sigma }}[\epsilon [{f_2}]]}^*}\, {a_{\sigma }}[{f_2}]\, {{({d_{\sigma }})}^*}\, {d_{\sigma }}\, {{({d_{-\sigma }})}^*}\, {d_{-\sigma
}}
}
\end{eqnarray*}
\begin{eqnarray*}
\dispSFPrintmath{
{R_{127}}  &=& {{{a_{\sigma }}[{f_2}]}^*}\, {{{a_{\sigma }}[\epsilon [{f_2}]]}^*}\, {a_{\sigma }}[{f_2}]\, {{{a_{-\sigma }}[{f_2}]}^*}\, {a_{-\sigma }}[{f_2}]\, {d_{\sigma
}}\, {{({d_{-\sigma }})}^*}\, {d_{-\sigma }}-  
\\ &&
\hspace{2.em} {{{a_{\sigma }}[{f_2}]}^*}\, {a_{\sigma }}[\epsilon [{f_2}]]\, {a_{\sigma }}[{f_2}]\,   
\\ &&
\hspace{3.em} {{{a_{-\sigma }}[{f_2}]}^*}\, {a_{-\sigma }}[{f_2}]\, {{({d_{\sigma }})}^*}\, {{({d_{-\sigma }})}^*}\, {d_{-\sigma }}
}
R_{128} &=& a_{\sigma}[f_{1}]^{*} \, d_{\sigma} - a_{\sigma}[f_{1}] \, 
  (d_{\sigma})^{*} \\
R_{129} &=& a_{\sigma}[f_{1}]^{*} \, a_{-\sigma}[f_{2}]^{*} \, d_{\sigma} \, 
  d_{-\sigma} + a_{\sigma}[f_{1}] \, a_{-\sigma}[f_{2}] \, (d_{\sigma})^{*} \, 
  (d_{-\sigma})^{*} \\
R_{130} &=& a_{\sigma}[f_{2}]^{*} \, a_{\sigma}[f_{1}] \, a_{-\sigma}[f_{2}]
  \, (d_{-\sigma})^{*} - a_{\sigma}[f_{1}]^{*} \, a_{-\sigma}[f_{2}]^{*} \, 
  a_{-\sigma}[f_{2}] \, d_{\sigma} \\ &&
\quad
  - a_{\sigma}[f_{1}]^{*} \, a_{\sigma}[f_{2}] \, a_{-\sigma}[f_{2}]^{*}
  \, d_{-\sigma} + a_{\sigma}[f_{1}] \, a_{-\sigma}[f_{2}]^{*} \, 
  a_{-\sigma}[f_{2}] \, (d_{\sigma})^{*} \\ &&
\quad 
  + a_{\sigma}[f_{1}]^{*} \, a_{-\sigma}[f_{2}]^{*} \, a_{-\sigma}[f_{2}] \,
  d_{\sigma} \, (d_{-\sigma})^{*} \, d_{-\sigma} 
  - a_{\sigma}[f_{1}] \, a_{-\sigma}[f_{2}]^{*} \, a_{-\sigma}[f_{2}] \,
  (d_{\sigma})^{*} \, (d_{-\sigma})^{*} \, d_{-\sigma} \\ 
R_{131} &=& a_{\sigma}[f_{2}]^{*} \, a_{\sigma}[f_{1}] \,
  (d_{\sigma})^{*} \, d_{\sigma} \, (d_{-\sigma})^{*} \, d_{-\sigma} 
  + a_{\sigma}[f_{1}]^{*} \, a_{\sigma}[f_{2}] \,
  (d_{\sigma})^{*} \, d_{\sigma} \, (d_{-\sigma})^{*} \, d_{-\sigma} \\
R_{132} &=& -a_{\sigma}[f_{2}]^{*} \, a_{\sigma}[f_{1}]^{*} \,
  a_{\sigma}[f_{2}] \, a_{-\sigma}[f_{2}]^{*} \, a_{-\sigma}[f_{2}] \, 
  d_{\sigma} + a_{\sigma}[f_{2}]^{*} \, a_{\sigma}[f_{1}] \,
  a_{\sigma}[f_{2}] \, a_{-\sigma}[f_{2}]^{*} \, a_{-\sigma}[f_{2}] \, 
  (d_{\sigma})^{*} \\
R_{133} &=& a_{\sigma}[f_{2}]^{*} \, a_{\sigma}[f_{1}]^{*} \,
  a_{\sigma}[f_{2}] \, a_{-\sigma}[f_{2}]^{*} \, d_{\sigma} \, d_{-\sigma} 
  + a_{\sigma}[f_{2}]^{*} \, a_{\sigma}[f_{1}] \,
  a_{\sigma}[f_{2}] \, a_{-\sigma}[f_{2}] \, (d_{\sigma})^{*} \,
  (d_{-\sigma})^{*}  
\end{eqnarray*}
%
%
\begin{eqnarray*}
\dispSFPrintmath{\Mstring{$\partial_{t}$}
 {R_1} &=&  -{R_2} {{\omega }_{12}}}
\dispSFPrintmath{\Mstring{$\partial_{t}$}
 {R_2}  &=&  ({R_3}-2 {P_7} {{\omega }_{12}})}
\dispSFPrintmath{\Mstring{$\partial_{t}$}
 {R_3}  &=&  (-2 {R_2}+2 {R_4}+{R_5}+(-{P_5}+{P_6}-{P_{11}}) {{\omega }_{12}}+{P_{12}} {{\omega }_{12}})}
\dispSFPrintmath{\Mstring{$\partial_{t}$}
 {R_4}  &=&  ({R_6}+{R_7})}
\dispSFPrintmath{\Mstring{$\partial_{t}$}
 {R_5}  &=&  ({R_6}+{R_8}-2 {P_{26}} {{\omega }_{12}})}
\dispSFPrintmath{\Mstring{$\partial_{t}$}
 {R_6}  &=&  (-2 {R_4}+2 {R_9}+{R_{10}}+{R_{11}})}
\dispSFPrintmath{\Mstring{$\partial_{t}$}
 {R_7}  &=&  (-{R_4}+{R_{10}}+{R_{11}}-({P_{20}}-{P_{22}}) {{\omega }_{12}})}
\dispSFPrintmath{\Mstring{$\partial_{t}$}
 {R_8}  &=&  (-{R_5}-{R_9}-({P_{21}}-{P_{23}}) {{\omega }_{12}})}
\dispSFPrintmath{\Mstring{$\partial_{t}$}
 {R_9} &=&  -{R_6}}
\dispSFPrintmath{\Mstring{$\partial_{t}$}
 {R_{10}}  &=&  (-{R_6}-{R_7}+{R_{12}}-2 {P_{24}} {{\omega }_{12}})}
\dispSFPrintmath{\Mstring{$\partial_{t}$}
 {R_{11}}  &=&  (-{R_{13}}+2 {R_{14}} {{\omega }_{12}})}
\dispSFPrintmath{\Mstring{$\partial_{t}$}
 {R_{12}}  &=&  (-{R_4}+{R_9}+{R_{11}})}
\dispSFPrintmath{\Mstring{$\partial_{t}$}
 {R_{13}}  &=&  {R_{11}}}
\dispSFPrintmath{\Mstring{$\partial_{t}$}
 {R_{14}} &=& 0}
\dispSFPrintmath{\Mstring{$\partial_{t}$}
 {R_{15}}  &=&  (-{R_2} {{\Gamma }_{12}}-{R_{16}} {{\omega }_{12}})}
\dispSFPrintmath{\Mstring{$\partial_{t}$}
 {R_{16}}  &=&  ({R_{17}}-2 {P_7} {{\Gamma }_{12}})}
\dispSFPrintmath{\Mstring{$\partial_{t}$}
 {R_{17}}  &=&  (-2 {R_{16}}+2 {R_{18}}+{R_{19}}+(-{P_5}+{P_6}-{P_{11}}+{P_{12}}) {{\Gamma }_{12}})}
\dispSFPrintmath{\Mstring{$\partial_{t}$}
 {R_{18}}  &=&  ({R_{20}}-{R_{21}}+{R_{22}})}
\dispSFPrintmath{\Mstring{$\partial_{t}$}
 {R_{19}}  &=&  (-{R_{21}}+{R_{22}}+{R_{23}}-2 {P_{26}} {{\Gamma }_{12}})}
\dispSFPrintmath{\Mstring{$\partial_{t}$}
 {R_{20}}  &=&  (-{R_{18}}+{R_{24}}+{R_{25}}-({P_{20}}-{P_{22}}) {{\Gamma }_{12}})}
\dispSFPrintmath{\Mstring{$\partial_{t}$}
 {R_{21}}  &=&  ({R_{18}}+{R_{19}}-{R_{24}}-{R_{26}})}
\dispSFPrintmath{\Mstring{$\partial_{t}$}
 {R_{22}}  &=&  (-{R_{18}}-{R_{19}}+{R_{25}}+{R_{26}})}
\dispSFPrintmath{\Mstring{$\partial_{t}$}
 {R_{23}}  &=&  (-{R_{26}}-({P_{21}}-{P_{23}}) {{\Gamma }_{12}})}
\dispSFPrintmath{\Mstring{$\partial_{t}$}
 {R_{24}}  &=&  (-{R_{20}}+{R_{21}}-2 {P_{24}} {{\Gamma }_{12}})}
\dispSFPrintmath{\Mstring{$\partial_{t}$}
 {R_{25}}  &=&  (-{R_{27}}+2 {R_{14}} {{\Gamma }_{12}})}
\end{eqnarray*}
\begin{eqnarray*}
\dispSFPrintmath{\Mstring{$\partial_{t}$}
 {R_{26}}  &=&  ({R_{23}}-2 {P_{26}} {{\Gamma }_{12}})}
\dispSFPrintmath{\Mstring{$\partial_{t}$}
 {R_{27}}  &=&  {R_{25}}}
\dispSFPrintmath{\Mstring{$\partial_{t}$}
 {R_{28}}  &=&  {R_{29}} {{\omega }_{12}}}
\dispSFPrintmath{\Mstring{$\partial_{t}$}
 {R_{29}}  &=&  (-{R_{31}}+2 {R_{30}} {{\omega }_{12}})}
\dispSFPrintmath{\Mstring{$\partial_{t}$}
 {R_{30}}  &=&  {R_{32}}}
\dispSFPrintmath{\Mstring{$\partial_{t}$}
 {R_{31}}  &=&  (2 {R_{29}}-2 {R_{34}}+{R_{35}}-{R_{32}} {{\omega }_{12}}+{R_{33}} {{\omega }_{12}})}
\dispSFPrintmath{\Mstring{$\partial_{t}$}
 {R_{32}}  &=&  (-4 {R_{30}}+2 {R_{36}}-{R_6} {{\omega }_{12}}+{R_{12}} {{\omega }_{12}}+{R_{37}} {{\omega }_{12}})}
\dispSFPrintmath{\Mstring{$\partial_{t}$}
 {R_{33}}  &=&  ({R_{38}}+{R_6} {{\omega }_{12}}-{R_{12}} {{\omega }_{12}}-{R_{37}} {{\omega }_{12}})}
\dispSFPrintmath{\Mstring{$\partial_{t}$}
 {R_{34}}  &=&  (-{R_{39}}+{R_{40}}-2 {R_{41}} {{\omega }_{12}})}
\dispSFPrintmath{\Mstring{$\partial_{t}$}
 {R_{35}}  &=&  (-{R_{40}}+{R_{42}}+2 {R_{43}} {{\omega }_{12}})}
\dispSFPrintmath{\Mstring{$\partial_{t}$}
 {R_{36}}  &=&  (-2 {R_{32}}+2 {R_{45}}+2 {R_{48}}-{R_9} {{\omega }_{12}}-{R_{10}} {{\omega }_{12}}+{R_{44}} {{\omega
}_{12}})}
\dispSFPrintmath{\Mstring{$\partial_{t}$}
 {R_{37}}  &=&  ({R_{44}}+{R_{46}})}
\dispSFPrintmath{\Mstring{$\partial_{t}$}
 {R_{38}}  &=&  (-{R_{33}}+{R_{47}}-{R_4} {{\omega }_{12}}-{R_{46}} {{\omega }_{12}})}
\dispSFPrintmath{\Mstring{$\partial_{t}$}
 {R_{39}}  &=&  ({R_{34}}+{R_{49}}+{R_{50}}+{R_{47}} {{\omega }_{12}}+{R_{48}} {{\omega }_{12}})}
\dispSFPrintmath{\Mstring{$\partial_{t}$}
 {R_{40}}  &=&  (-2 {R_{34}}-{R_{49}}-{R_{50}}+2 {R_{51}}-{R_{47}} {{\omega }_{12}}+{R_{52}} {{\omega }_{12}})}
\dispSFPrintmath{\Mstring{$\partial_{t}$}
 {R_{41}}  &=&  {R_{47}}}
\dispSFPrintmath{\Mstring{$\partial_{t}$}
 {R_{42}}  &=&  (-{R_{35}}+{R_{51}}-{R_{45}} {{\omega }_{12}})}
\dispSFPrintmath{\Mstring{$\partial_{t}$}
 {R_{43}}  &=&  (-{R_{45}}-{R_{52}})}
\dispSFPrintmath{\Mstring{$\partial_{t}$}
 {R_{44}}  &=&  (-2 {R_{37}}+{R_{53}}+2 {P_{26}} {{\omega }_{12}})}
\dispSFPrintmath{\Mstring{$\partial_{t}$}
 {R_{45}}  &=&  (-2 {R_{54}}+2 {R_{55}})}
\dispSFPrintmath{\Mstring{$\partial_{t}$}
 {R_{46}}  &=&  (-{R_{37}}+{R_{53}})}
\dispSFPrintmath{\Mstring{$\partial_{t}$}
 {R_{47}}  &=&  (-2 {R_{41}}+{R_{13}} {{\omega }_{12}}+{R_{56}} {{\omega }_{12}})}
\dispSFPrintmath{\Mstring{$\partial_{t}$}
 {R_{48}}  &=&  (2 {R_{57}}-2 {R_{58}}+{R_{13}} {{\omega }_{12}}+{R_{56}} {{\omega }_{12}})}
\dispSFPrintmath{\Mstring{$\partial_{t}$}
 {R_{49}}  &=&  (-{R_{39}}+{R_{59}}-2 {R_{58}} {{\omega }_{12}})}
\dispSFPrintmath{\Mstring{$\partial_{t}$}
 {R_{50}}  &=&  (-{R_{60}}+2 {R_{61}} {{\omega }_{12}})}
\dispSFPrintmath{\Mstring{$\partial_{t}$}
 {R_{51}}  &=&  (-{R_{40}}+2 {R_{43}} {{\omega }_{12}}+2 {R_{54}} {{\omega }_{12}})}
\dispSFPrintmath{\Mstring{$\partial_{t}$}
 {R_{52}}  &=&  (2 {R_{43}}+2 {R_{54}}-2 {R_{62}}-{R_{13}} {{\omega }_{12}}-{R_{56}} {{\omega }_{12}})}
\dispSFPrintmath{\Mstring{$\partial_{t}$}
 {R_{53}}  &=&  (-{R_{44}}-2 {R_{46}}+2 {R_{63}}+({P_{21}}-{P_{23}}) {{\omega }_{12}})}
\dispSFPrintmath{\Mstring{$\partial_{t}$}
 {R_{54}}  &=&  {R_{45}}}
\dispSFPrintmath{\Mstring{$\partial_{t}$}
 {R_{55}} &=&  -{R_{45}}}
\dispSFPrintmath{\Mstring{$\partial_{t}$}
 {R_{56}}  &=&  {R_{63}}}
\dispSFPrintmath{\Mstring{$\partial_{t}$}
 {R_{57}}  &=&  (-{R_{48}}+{R_{11}} {{\omega }_{12}}+{R_{63}} {{\omega }_{12}})}
\dispSFPrintmath{\Mstring{$\partial_{t}$}
 {R_{58}}  &=&  {R_{48}}}
\dispSFPrintmath{\Mstring{$\partial_{t}$}
 {R_{59}}  &=&  (-{R_{34}}-{R_{49}}+{R_{51}})}
\dispSFPrintmath{\Mstring{$\partial_{t}$}
 {R_{60}}  &=&  {R_{50}}}
\dispSFPrintmath{\Mstring{$\partial_{t}$}
 {R_{61}} &=& 0}
\dispSFPrintmath{\Mstring{$\partial_{t}$}
 {R_{62}}  &=&  ({R_{52}}+{R_{11}} {{\omega }_{12}}+{R_{63}} {{\omega }_{12}})}
\end{eqnarray*}
\begin{eqnarray*}
\dispSFPrintmath{\Mstring{$\partial_{t}$}
 {R_{63}}  &=&  (-{R_{56}}+2 {R_{14}} {{\omega }_{12}})}
\dispSFPrintmath{\Mstring{$\partial_{t}$}
 {R_{64}}  &=&  (-{R_{67}}-{R_{68}}-{R_{65}} {{\omega }_{12}}+2 {R_{66}} {{\omega }_{12}})}
\dispSFPrintmath{\Mstring{$\partial_{t}$}
 {R_{65}}  &=&  ({R_{33}}+{R_{69}})}
\dispSFPrintmath{\Mstring{$\partial_{t}$}
 {R_{66}}  &=&  {R_{70}}}
\dispSFPrintmath{\Mstring{$\partial_{t}$}
 {R_{67}}  &=&  (-2 {R_{71}}+2 {R_{73}}+{R_{74}}-{R_{70}} {{\omega }_{12}}+{R_{72}} {{\omega }_{12}})}
\dispSFPrintmath{\Mstring{$\partial_{t}$}
 {R_{68}}  &=&  ({R_{74}}+{R_{75}})}
\dispSFPrintmath{\Mstring{$\partial_{t}$}
 {R_{69}}  &=&  (-{R_{38}}-2 {R_{65}}-{R_{76}}-{R_6} {{\omega }_{12}}+{R_{12}} {{\omega }_{12}}+{R_{37}} {{\omega }_{12}})}
\dispSFPrintmath{\Mstring{$\partial_{t}$}
 {R_{70}}  &=&  (-4 {R_{66}}+2 {R_{77}}+{R_{37}} {{\omega }_{12}})}
\dispSFPrintmath{\Mstring{$\partial_{t}$}
 {R_{71}}  &=&  ({R_{67}}-2 {R_{66}} {{\omega }_{12}})}
\dispSFPrintmath{\Mstring{$\partial_{t}$}
 {R_{72}}  &=&  ({R_{80}}-{R_{37}} {{\omega }_{12}})}
\dispSFPrintmath{\Mstring{$\partial_{t}$}
 {R_{73}}  &=&  ({R_{81}}+{R_{82}}-{R_{83}}+{R_{41}} {{\omega }_{12}})}
\dispSFPrintmath{\Mstring{$\partial_{t}$}
 {R_{74}}  &=&  ({R_{82}}-{R_{83}}+{R_{84}}+{R_{43}} {{\omega }_{12}}+{R_{85}} {{\omega }_{12}})}
\dispSFPrintmath{\Mstring{$\partial_{t}$}
 {R_{75}}  &=&  (-2 {R_{68}}-{R_{82}}+{R_{83}}-{R_{84}}-{R_{87}}-{R_{43}} {{\omega }_{12}}-{R_{85}} {{\omega }_{12}}+{R_{86}}
{{\omega }_{12}})}
\dispSFPrintmath{\Mstring{$\partial_{t}$}
 {R_{76}}  &=&  ({R_{47}}+{R_{88}})}
\dispSFPrintmath{\Mstring{$\partial_{t}$}
 {R_{77}}  &=&  (-2 {R_{70}}+2 {R_{78}}+2 {R_{79}}+{R_{44}} {{\omega }_{12}})}
\dispSFPrintmath{\Mstring{$\partial_{t}$}
 {R_{78}} &=&  -2  {R_{89}}}
\dispSFPrintmath{\Mstring{$\partial_{t}$}
 {R_{79}}  &=&  (2 {R_{90}}+{R_{56}} {{\omega }_{12}})}
\dispSFPrintmath{\Mstring{$\partial_{t}$}
 {R_{80}}  &=&  (-{R_{72}}-{R_{91}}-{R_{46}} {{\omega }_{12}})}
\dispSFPrintmath{\Mstring{$\partial_{t}$}
 {R_{81}}  &=&  (-{R_{73}}-{R_{92}}+{R_{93}}+{R_{47}} {{\omega }_{12}}+{R_{79}} {{\omega }_{12}}+{R_{91}} {{\omega }_{12}})}
\dispSFPrintmath{\Mstring{$\partial_{t}$}
 {R_{82}}  &=&  (-{R_{73}}+{R_{93}}+{R_{94}}-{R_{91}} {{\omega }_{12}}+{R_{95}} {{\omega }_{12}})}
\dispSFPrintmath{\Mstring{$\partial_{t}$}
 {R_{83}}  &=&  ({R_{73}}+{R_{92}}-{R_{94}})}
\dispSFPrintmath{\Mstring{$\partial_{t}$}
 {R_{84}}  &=&  (-{R_{74}}-{R_{94}}-{R_{78}} {{\omega }_{12}})}
\dispSFPrintmath{\Mstring{$\partial_{t}$}
 {R_{85}}  &=&  ({R_{45}}-2 {R_{78}})}
\dispSFPrintmath{\Mstring{$\partial_{t}$}
 {R_{86}}  &=&  (-{R_{91}}+{R_{99}})}
\dispSFPrintmath{\Mstring{$\partial_{t}$}
 {R_{87}}  &=&  ({R_{93}}-{R_{100}})}
\dispSFPrintmath{\Mstring{$\partial_{t}$}
 {R_{88}}  &=&  (2 {R_{41}}-{R_{76}}-2 {R_{101}}-{R_{13}} {{\omega }_{12}}-{R_{56}} {{\omega }_{12}})}
\dispSFPrintmath{\Mstring{$\partial_{t}$}
 {R_{89}} &=&  2  {R_{78}}}
\dispSFPrintmath{\Mstring{$\partial_{t}$}
 {R_{90}}  &=&  (-2 {R_{79}}+{R_{63}} {{\omega }_{12}})}
\dispSFPrintmath{\Mstring{$\partial_{t}$}
 {R_{91}}  &=&  ({R_{41}}-{R_{13}} {{\omega }_{12}})}
\dispSFPrintmath{\Mstring{$\partial_{t}$}
 {R_{92}}  &=&  ({R_{81}}-{R_{83}}+2 {R_{96}} {{\omega }_{12}})}
\dispSFPrintmath{\Mstring{$\partial_{t}$}
 {R_{93}}  &=&  (-{R_{97}}+2 {R_{98}} {{\omega }_{12}})}
\dispSFPrintmath{\Mstring{$\partial_{t}$}
 {R_{94}}  &=&  (-{R_{82}}+{R_{83}}-{R_{43}} {{\omega }_{12}}-{R_{54}} {{\omega }_{12}})}
\dispSFPrintmath{\Mstring{$\partial_{t}$}
 {R_{95}}  &=&  (-{R_{43}}-{R_{54}}+{R_{62}}+{R_{13}} {{\omega }_{12}})}
\dispSFPrintmath{\Mstring{$\partial_{t}$}
 {R_{96}}  &=&  {R_{79}}}
\dispSFPrintmath{\Mstring{$\partial_{t}$}
 {R_{97}}  &=&  {R_{93}}}
\dispSFPrintmath{\Mstring{$\partial_{t}$}
 {R_{98}} &=& 0}
\dispSFPrintmath{\Mstring{$\partial_{t}$}
 {R_{99}}  &=&  ({R_{41}}-{R_{86}}-{R_{101}}-{R_{13}} {{\omega }_{12}})}
\end{eqnarray*}
\begin{eqnarray*}
\dispSFPrintmath{\Mstring{$\partial_{t}$}
 {R_{100}}  &=&  ({R_{87}}-{R_{97}}+2 {R_{98}} {{\omega }_{12}}-{R_{101}} {{\omega }_{12}})}
\dispSFPrintmath{\Mstring{$\partial_{t}$}
 {R_{101}} &=& 0}
\dispSFPrintmath{\Mstring{$\partial_{t}$}
 {R_{102}}  &=&  ({R_{104}}-{R_2} {{\omega }_{12}}-{R_{103}} {{\omega }_{12}})}
\dispSFPrintmath{\Mstring{$\partial_{t}$}
 {R_{103}}  &=&  ({R_{105}}-2 {P_7} {{\omega }_{12}})}
\dispSFPrintmath{\Mstring{$\partial_{t}$}
 {R_{104}}  &=&  ({R_{106}}+{R_6} {{\omega }_{12}}-{R_{12}} {{\omega }_{12}}-{R_{37}} {{\omega }_{12}})}
\dispSFPrintmath{\Mstring{$\partial_{t}$}
 {R_{105}}  &=&  (2 {R_{46}}-2 {R_{103}}+{R_{107}}+{R_{108}}+(-{P_5}+{P_6}-{P_{11}}+{P_{12}}) {{\omega }_{12}})}
\dispSFPrintmath{\Mstring{$\partial_{t}$}
 {R_{106}}  &=&  (-2 {R_{52}}-2 {R_{104}}+{R_5} {{\omega }_{12}}+2 {R_9} {{\omega }_{12}}+{R_{10}} {{\omega }_{12}}-{R_{44}}
{{\omega }_{12}}+{R_{108}} {{\omega }_{12}})}
\dispSFPrintmath{\Mstring{$\partial_{t}$}
 {R_{107}}  &=&  (-{R_{37}}+{R_{109}})}
\dispSFPrintmath{\Mstring{$\partial_{t}$}
 {R_{108}}  &=&  ({R_{110}}-2 {P_{24}} {{\omega }_{12}})}
\dispSFPrintmath{\Mstring{$\partial_{t}$}
 {R_{109}}  &=&  (-{R_{46}}+{R_{63}}-{R_{107}})}
\dispSFPrintmath{\Mstring{$\partial_{t}$}
 {R_{110}}  &=&  (-{R_{108}}+({P_{20}}-{P_{22}}) {{\omega }_{12}})}
\dispSFPrintmath{\Mstring{$\partial_{t}$}
 {R_{111}}  &=&  ({R_{112}}-{R_2} {{\omega }_{12}})}
\dispSFPrintmath{\Mstring{$\partial_{t}$}
 {R_{112}}  &=&  ({R_{113}}-2 {R_{114}})}
\dispSFPrintmath{\Mstring{$\partial_{t}$}
 {R_{113}}  &=&  (-{R_{32}}+{R_{45}}+{R_{48}}+{R_{70}}-{R_{78}}-{R_{79}}-{R_9} {{\omega }_{12}})}
\dispSFPrintmath{\Mstring{$\partial_{t}$}
 {R_{114}}  &=&  ({R_{32}}-{R_{48}}-{R_{70}}+{R_{79}}+{R_{112}})}
\dispSFPrintmath{\Mstring{$\partial_{t}$}
 {R_{115}}  &=&  {R_{116}}}
\dispSFPrintmath{\Mstring{$\partial_{t}$}
 {R_{116}}  &=&  (-{R_{117}}+2 {P_7} {{\Gamma }_2})}
\dispSFPrintmath{\Mstring{$\partial_{t}$}
 {R_{117}}  &=&  (2 {R_{116}}+2 {R_{118}}+{R_{119}}+(-{P_5}+{P_6}-{P_{11}}+{P_{12}}) {{\Gamma }_2})}
\dispSFPrintmath{\Mstring{$\partial_{t}$}
 {R_{118}}  &=&  ({R_{120}}+{R_{121}})}
\dispSFPrintmath{\Mstring{$\partial_{t}$}
 {R_{119}}  &=&  ({R_{121}}+{R_{122}}-2 {P_{26}} {{\Gamma }_2})}
\dispSFPrintmath{\Mstring{$\partial_{t}$}
 {R_{120}}  &=&  (-{R_{118}}+{R_{123}}+(-{P_{20}}+{P_{22}}) {{\Gamma }_2})}
\dispSFPrintmath{\Mstring{$\partial_{t}$}
 {R_{121}}  &=&  (-2 {R_{118}}+{R_{123}}+2 {R_{124}})}
\dispSFPrintmath{\Mstring{$\partial_{t}$}
 {R_{122}}  &=&  (-{R_{119}}-{R_{124}}+(-{P_{21}}+{P_{23}}) {{\Gamma }_2})}
\dispSFPrintmath{\Mstring{$\partial_{t}$}
 {R_{123}}  &=&  (-{R_{120}}-{R_{125}}+{R_{126}}-2 ({P_{24}} {{\Gamma }_2}-{R_{14}} {{\Gamma }_2}))}
\dispSFPrintmath{\Mstring{$\partial_{t}$}
 {R_{124}} &=&  - {R_{121}}}
\dispSFPrintmath{\Mstring{$\partial_{t}$}
 {R_{125}}  &=&  {R_{127}}}
\dispSFPrintmath{\Mstring{$\partial_{t}$}
 {R_{126}}  &=&  ({R_{118}}-{R_{123}}-{R_{124}}+{R_{127}})}
\dispSFPrintmath{\Mstring{$\partial_{t}$}
 {R_{127}}  &=&  (-{R_{125}}+2 {R_{14}} {{\Gamma }_2})}
\partial_{t} R_{128} &=& R_{3} + R_{129} -2 P_{7} \omega_{12} \\
\partial_{t} R_{129} &=& -R_{9} + R_{130} \\
\partial_{t} R_{130} &=& -R_{12} -2 R_{129} + R_{131} - R_{133} \\
\partial_{t} R_{131} &=& R_{4} - R_{9} -R_{10} \\
\partial_{t} R_{132} &=& R_{13} - R_{133} - 2 R_{14} \omega_{12}\\
\partial_{t} R_{133} &=& R_{11} + R_{132}
\end{eqnarray*}


\end{document}